\documentclass[twocolumn]{aastex61}

\usepackage{amsmath,bm}
\hypersetup{breaklinks=true,colorlinks=true,citecolor=blue,linkcolor=blue,urlcolor=blue}

\shorttitle{TDE rate \& host-galaxy global properties}
\shortauthors{Graur et al.}

\begin{document}

\title{A dependence of the tidal disruption event rate on global stellar surface mass density and stellar velocity dispersion}

\correspondingauthor{Or Graur}
\email{or.graur@cfa.harvard.edu}

\author{Or Graur}
\altaffiliation{NSF Astronomy and Astrophysics Postdoctoral Fellow}
\affiliation{Harvard-Smithsonian Center for Astrophysics, 60 Garden St., Cambridge, MA 02138, USA}
\affiliation{Department of Astrophysics, American Museum of Natural History, New York, NY 10024, USA}

\author{K. Decker French}
\altaffiliation{Hubble Fellow}
\affiliation{Steward Observatory, University of Arizona, 933 North Cherry Avenue, Tucson AZ 85721, USA}
\affiliation{The Observatories of the Carnegie Institution for Science, 813 Santa Barbara Street, Pasadena CA 91101}

\author{H. Jabran Zahid}
\affiliation{Harvard-Smithsonian Center for Astrophysics, 60 Garden St., Cambridge, MA 02138, USA}

\author{James Guillochon}
\affiliation{Harvard-Smithsonian Center for Astrophysics, 60 Garden St., Cambridge, MA 02138, USA}

\author{Kaisey S. Mandel}
\affiliation{Harvard-Smithsonian Center for Astrophysics, 60 Garden St., Cambridge, MA 02138, USA}
\affiliation{Institute of Astronomy and Kavli Institute for Cosmology, Madingley Road, Cambridge, CB3 0HA, UK}
\affiliation{Statistical Laboratory, DPMMS, University of Cambridge, Wilberforce Road, Cambridge, CB3 0WB, UK}

\author{Katie Auchettl}
\affiliation{Center for Cosmology and Astro-Particle Physics, The Ohio State University, 191 West Woodruff Avenue, Columbus, OH 43210, USA}
\affiliation{Department of Physics, The Ohio State University, 191 W. Woodruff Avenue, Columbus, OH 43210, USA}

\author{Ann I. Zabludoff}
\affiliation{Steward Observatory, University of Arizona, 933 North Cherry Avenue, Tucson AZ 85721, USA}


\begin{abstract}
\noindent The rate of tidal disruption events (TDEs), $R_\text{TDE}$, is predicted to depend on stellar conditions near the super-massive black hole (SMBH), which are on difficult-to-measure sub-parsec scales. We test whether $R_\text{TDE}$ depends on kpc-scale global galaxy properties, which are observable. We concentrate on stellar surface mass density, $\Sigma_{M_\star}$, and velocity dispersion, $\sigma_v$, which correlate with the stellar density and velocity dispersion of the stars around the SMBH. We consider 35 TDE candidates, with and without known X-ray emission. The hosts range from star-forming to quiescent to quiescent with strong Balmer absorption lines. The last (often with post-starburst spectra) are overrepresented in our sample by a factor of $35^{+21}_{-17}$ or $18^{+8}_{-7}$, depending on the strength of the H$\delta$ absorption line. For a subsample of hosts with homogeneous measurements, $\Sigma_{M_\star}=10^9$--$10^{10}~{\rm M_\sun / kpc^2}$, higher on average than for a volume-weighted control sample of Sloan Digital Sky Survey galaxies with similar redshifts and stellar masses. This is because: (1) most of the TDE hosts here are quiescent galaxies, which tend to have higher $\Sigma_{M_\star}$ than the star-forming galaxies that dominate the control, and (2) the star-forming hosts have higher average $\Sigma_{M_\star}$ than the star-forming control. There is also a weak suggestion that TDE hosts have lower $\sigma_v$ than for the quiescent control. Assuming that $R_{\rm TDE}\propto \Sigma_{M_\star}^\alpha \times \sigma_v^\beta$, and applying a statistical model to the TDE hosts and control sample, we estimate $\hat{\alpha}=0.9\pm0.2$ and $\hat{\beta}=-1.0\pm0.6$. This is broadly consistent with $R_\text{TDE}$ being tied to the dynamical relaxation of stars surrounding the SMBH.
\end{abstract}



\section{Introduction}
\label{sec:intro}

Tidal disruption events (TDEs) are luminous flares predicted to occur when a super-massive black hole (SMBH), generally found in the centers of galaxies (see, e.g., \citealt{2016ASSL..418..263G}), disrupts and accretes a star that orbits it within the tidal radius of the SMBH \citep{1975Natur.254..295H,1988Natur.333..523R}. Theory suggests that disruption takes place when the tidal radius is larger than the Schwarzshild radius, which is satisfied when the mass of the SMBH, $M_\bullet$, together with the mass and radius of the disrupted star, $M_\star$ and $R_\star$, follow $M_\bullet \lesssim 10^8~{\rm M_\sun}~(R_\star/R_\sun)^{3/2} (M_\star/{\rm M_\sun})^{-1/2}$ \citep{2012PhRvD..85b4037K}. Roughly half of the disrupted star is unbound and leaves the system, while the rest of the material falls onto the SMBH and forms an accretion disk. It is the circularization of the accretion disk, or collisions between the infalling streams of matter, that is thought to cause the observed flare (e.g., \citealt{1982ApJ...262..120L,1989ApJ...346L..13E,1989Natur.340..595P}).

TDE candidates were first detected in archival searches in the soft X-ray part of the spectrum (e.g., \citealt{1995A&A...299L...5G,1996A&A...309L..35B,1999A&A...343..775K,1999AA...349L..45K}), where they are expected to peak \citep{1999ApJ...514..180U}. In recent years, most new TDE candidates have been discovered in optical, real-time transient surveys, such as the All-Sky Automated Survey for SuperNovae (ASASSN; \citealt{2017MNRAS.464.2672H}), Palomar Transient Factory (PTF; \citealt{2009PASP..121.1334R}), and the Panoramic Survey Telescope and Rapid Response System (PanSTARRS; \citealt{2002SPIE.4836..154K}). Several TDE candidates were discovered in archival data from the Sloan Digital Sky Survey (SDSS; \citealt{2000AJ....120.1579Y}), in both imaging \citep{2011ApJ...741...73V} and spectroscopy (due to the presence of high-ionization coronal lines; e.g., \citealt{2009ApJ...701..105K,2011ApJ...740...85W,2012ApJ...749..115W}). 

The Open TDE Catalog currently lists $\sim 66$ TDE candidates.\footnote{\url{https://tde.space/}} However, it is still unclear what observational features define a TDE as opposed to, e.g., a flaring active galactic nucleus (AGN) or a supernova (see, for example, \citealt{2017arXiv170306141A} or the discussion surrounding ASASSN-15lh; e.g., \citealt{2016Sci...351..257D,2016NatAs...1E...2L,2017ApJ...845...85L,2017ApJ...836...25M}). Moreover, most objects do not have multiwavelength coverage, and it is unclear if the objects we collectively call ``TDE candidates'' all belong to one monolithic class. For example, while many TDE candidates are observed in X-rays, as expected, some are detected in the UV/optical but show no signs of X-ray emission (e.g., \citealt{Gezari2012}). We refer to all TDE candidates in this paper as ``TDEs.''

Recently, \citet[hereafter A17]{2017ApJ...838..149A} conducted a comprehensive analysis of the X-ray emission of all the TDE candidates in the literature and theorized that, rather than two separate groups of astronomical phenomena, the division between X-ray- and optically/UV-bright TDEs was due to the latter type of objects originating in systems with enhanced column densities. This leads to significant reprocessing of the X-rays emitted by these ``veiled'' TDEs, so that they are not observed in this part of the spectrum.

The TDE rate is expected to depend on the mass of the SMBH along with the density and velocity dispersion of stars in orbit around it. Theoretical formulations predict a TDE rate of $\sim 10^{-4}~{\rm yr^{-1}}$ per galaxy \citep{1999MNRAS.309..447M,2004ApJ...600..149W,2016MNRAS.455..859S}. While some observational studies have estimated similar rates (e.g., \citealt{2008A&A...489..543E,2017arXiv170306141A}), others imply a rate an order of magnitude lower (e.g., \citealt{2002AJ....124.1308D,2008ApJ...676..944G,2014ApJ...792...53V,2016MNRAS.455.2918H}).

For a star to be disrupted by the SMBH, it needs to enter its ``loss cone,'' i.e., the star's trajectory and velocity have to be such that it will enter the space between the  Schwarzshild and tidal radii of the SMBH. The most straightforward way for this to happen is through collisional two-body relaxation, although other dynamical mechanisms have been suggested (see \citealt{2012EPJWC..3905001A} for a review). Thus, it is natural to expect that the TDE rate would depend on the mass of the SMBH and the density and velocity dispersion of the stars in the loss cone. Regrettably, as the size of the loss cone is on sub-pc scales, these properties cannot be measured directly.

Some local galaxy properties, however, are correlated with global properties on kpc scales, which can be measured. Based on integral field observations, \citet{2006MNRAS.366.1126C} showed that the central stellar velocity dispersions of galaxies are correlated over galactic scales and the dependence of the central stellar velocity dispersion on aperture size is very weak. There exists a well known correlation between the stellar velocity dispersion, $\sigma_v$, and the black hole mass (e.g., \citealt{1995ARA&A..33..581K,1998AJ....115.2285M,2000ApJ...539L..13G,2002ApJ...574..740T,2013ApJ...764..184M}). Thus, the central stellar velocity dispersion is correlated with properties of the central SMBH. Similarly, galaxy light profiles of massive galaxies are smooth and thus the stellar surface mass density, $\Sigma_{M_\star}$, measured within the half-light radius is correlated with the central stellar surface mass density. Thus, while we are not able to probe the host galaxy on the pc-scales corresponding to the loss cone of the SMBH, we can examine global galaxy properties on kpc scales that correlate with galaxy properties on small scales.

\citet{2014ApJ...793...38A}, \citet[hereafter F16]{2016ApJ...818L..21F}, and \citet{2017ApJ...835..176F} analyzed a sample of eight TDEs that fall into this second class of events. They had all been discovered by optical surveys, had no detectable emission in X-ray observations,\footnote{Except for ASASSN14li, which was observed to have soft X-ray emission \citep{2016MNRAS.455.2918H}. \citet{2014ApJ...793...38A} detected X-ray emission at the location of PTF09axc, but were unable to determine if it was directly related to the TDE candidate.} and had optical spectra that showed wide H$\alpha$ and He~II emission lines. A majority of these TDEs were hosted by quiescent, Balmer-strong galaxies characterized by weak or no emission lines coupled with deep Balmer absorption features. These features hint that, although current star formation is either low or nonexistent (hence the lack of emission lines), the galaxy experienced a large burst of star formation in the last Gyr or so, leaving behind A stars that are characterized by their Balmer absorption features. This type of star-formation history is often seen in post-starburst galaxies, as they evolve from late-type, star-forming galaxies to early-type, quiescent galaxies (e.g., \citealt{1983ApJ...270....7D,1996ApJ...466..104Z,2016ApJ...831..146Z}). As this is a transitional phase, post-starburst galaxies represent only $\approx0.1$--$2.0$\% of all galaxies. Based on this TDE sample, F16 concluded that the TDE rate in quiescent Balmer-strong galaxies was boosted by a factor of $\sim 30$--$200$. \citet{2016MNRAS.455..859S} have suggested that the overabundance of TDEs in these rare galaxies could be due to the high density of stars in the cores of post-merger galaxies \citep{2004ApJ...607..258Y,2006ApJ...646L..33Y,2008ApJ...688..945Y,2012MNRAS.420..672S,2016ApJ...825L..14S,2016ApJ...831..146Z}. 

Here, we use a sample of 35 TDEs and a volume-weighted control sample of galaxies (described in Section~\ref{sec:data}) to formulate an empirical relation between the TDE rate and global galaxy properties. In Section~\ref{sec:props}, we show that TDEs preferentially occur in galaxies that have higher stellar surface mass densities than the galaxies in the control sample. This effect is due to the star-forming TDE hosts, which have higher $\Sigma_{M_\star}$ values on average than the star-forming subsample of control galaxies. We also find a suggestion that quiescent TDE hosts may have lower stellar velocity dispersions than the quiescent-galaxy control sample. Also, we confirm that quiescent Balmer-strong galaxies are overrepresented among TDE hosts. This overrepresentation may arise, at least in part, because these galaxies also exhibit high stellar densities. In Section~\ref{sec:rate}, we statistically model the TDE rate. Consistent with theory, our TDE sample suggests a direct (and significant) dependence of the TDE rate on stellar surface mass density and an inverse (but not significantly different than zero) dependence on velocity dispersion. We discuss our results in Section~\ref{sec:discuss} and summarize them in Section~\ref{sec:summary}.

Throughout this paper we assume a $\Lambda$-cold-dark-matter cosmological model with parameters $\Omega_{\Lambda} = 0.7$, $\Omega_m = 0.3$, and $H_0 = 70$~km~s$^{-1}$~Mpc$^{-1}$.


\section{Data}
\label{sec:data}

We describe our TDE sample in Section~\ref{subsec:TDEs} and summarize its properties in Table~\ref{table:host_props}. The control galaxy sample is described in Section~\ref{subsec:gals}.

\subsection{TDE sample}
\label{subsec:TDEs}

Our TDE sample comprises 35 events, most of which are taken from A17. The TDE candidates were classified according to a combination of their available X-ray data, followed by data in other wavelengths. Accordingly, the TDEs are classified into four groups: X-ray TDEs, likely X-ray TDEs, possible X-ray TDEs, and veiled TDEs. The criteria for each group are defined in section 3.1 of A17. All four groups share the following basic criteria:
\begin{enumerate}
 \item The candidate's location is consistent with the nucleus of its host galaxy.
 \item There is no evidence of an AGN.\footnote{Despite this criterion, we keep the host galaxies of ASASSN14li, PS16dtm, F01004, SDSS J0952, and SDSS J1350, which are known to host AGNs (e.g., \citealt{2016ApJ...830L..32P}), as the TDEs do not resemble AGN flares (see, e.g., \citealt{2016Sci...351...62V,2017ApJ...843..106B}). See Section~\ref{subsec:agn} for further details.}
 \item A supernova or gamma-ray burst classification of the candidate has been ruled out.
\end{enumerate}
The X-ray TDEs, likely X-ray TDEs, and possible X-ray TDEs also have X-ray emission, where requirements regarding the shape and quality of an event's X-ray light curve, as well as its peak brightness, are gradually relaxed for each successive group. Finally, veiled TDEs are candidates that have no known X-ray emission but have been classified as TDEs according to their UV/optical data (we refer the reader to A17 for a detailed account of the X-ray observations of each TDE in our sample). The TDEs used here, grouped according to this classification scheme, are presented in Table~\ref{table:host_props}.

\floattable
\startlongtable
\begin{deluxetable}{lCCDCl}
 \tablecaption{TDE host-galaxy properties \label{table:host_props}}
 \tablehead{
 \colhead{TDE Host} & \colhead{H$\alpha$ EW\tablenotemark{a}}  & \colhead{H$\delta_{\rm A}$} & \twocolhead{Redshift} & \colhead{Slit Width}     & \colhead{Spectrum source} \\
 \colhead{}         & \colhead{($\rm \AA{}$)}                  & \colhead{($\rm \AA{}$)}     & \twocolhead{}         & \colhead{(arcsec (kpc))} & \colhead{}
 }
 \decimals
 \startdata
  \multicolumn{7}{c}{X-ray TDEs} \\
  ASASSN14li\tablenotemark{b,c}  & -0.6 \pm 0.5 & 5.7 \pm 0.6 & 0.02058 & 3.0~(1.3) & SDSS \\
  Swift J1644                    & -2.5 \pm 0.8 & 4.7 \pm 1.1 & 0.3534  & 1.0~(9.1) & \citet{2011Sci...333..199L} \\
  XMM J0740                      & -0.3 \pm 0.6 & 0.4 \pm 0.4 & 0.0173  & 4.8~(1.7) & \citet{2017AA...598A..29S} \\
  \hline
  \multicolumn{7}{c}{Likely X-ray TDEs} \\
  SDSS J1201                   &   0.7 \pm 0.3 & -1.1 \pm 2.4 & 0.146   & \cdots     & \citet{2012AA...541A.106S} \\
  2MASX J0249\tablenotemark{d} & - 5.7 \pm 0.6 &  0.4 \pm 0.5 & 0.0186  & \cdots     & \citet{2007AA...462L..49E} \\
  PTF10iya                     & -20.5 \pm 0.6 &  2.9 \pm 0.9 & 0.22405 & 1.0~(5.4)  & \citet{2012MNRAS.420.2684C} \\
  SDSS J1311                   & - 2.1 \pm 1.5 &  3.6 \pm 1.1 & 0.195   & 2.0~(9.2)  & \citet{2010ApJ...722.1035M} \\
  SDSS J1323                   & - 0.2 \pm 0.5 & -1.2 \pm 1.5 & 0.0875  & 3.0~(5.8)  & SDSS \\
  3XMM J1521                   &   0.8 \pm 1.1 & -1.5 \pm 2.4 & 0.17901 & 2.0~(8.4)  & BOSS \\
  3XMM J1500                   & -45.1 \pm 0.8 &  1.4 \pm 1.7 & 0.145   & 2.0~(6.7)  & BOSS \\
  \hline
  \multicolumn{7}{c}{Possible X-ray TDEs} \\
  ASASSN15oi\tablenotemark{c}  & 0.1 \pm 0.3   & 1.9 \pm 0.7  & 0.0484  & 1.65~(1.7) & \citet{2016MNRAS.463.3813H} \\
  RX J1242-A\tablenotemark{d}  &   1.1 \pm 0.8 &  0.9 \pm 1.2 & 0.05    & 1.5~(1.6)  & \citet{1999AA...349L..45K} \\
  RX J1242-B\tablenotemark{d}  & - 0.9 \pm 0.9 & -0.4 \pm 2.7 & 0.05    & 1.5~(1.6)  & \citet{1999AA...349L..45K} \\
  RX J1420-A                   & - 0.2 \pm 0.9 & -2.8 \pm 2.0 & 0.148   & 3.0~(10.2) & SDSS \\
  RX J1420-B                   & -71.6 \pm 1.7 &  5.8 \pm 3.4 & 0.147   & 3.0~(10.1) & SDSS \\  
  SDSS J0159                   & -19.9 \pm 0.8 &  1.7 \pm 1.0 & 0.31167 & 3.0~(23.6) & SDSS \\
  RBS 1032                     & - 0.5 \pm 0.4 &  4.1 \pm 0.4 & 0.02604 & 3.0~(1.7)  & SDSS \\
  RX J1624                     &   0.6 \pm 1.3 & -1.1 \pm 2.1 & 0.0636  & 1.7~(2.4)  & \citet{1999AA...350L..31G} \\
  NGC 5905                     & -28.4 \pm 0.1 & \cdots       & 0.01131 & 2.0~(0.4)  & \citet{1995ApJS...98..477H} \\
  \hline
  \multicolumn{7}{c}{Veiled TDEs} \\
  iPTF16fnl\tablenotemark{c}       &   0.8 \pm 0.6 &  5.8 \pm 0.3 & 0.0163  & 1.2~(0.4)  & \citet{2018MNRAS.473.1130B} \\
  PS16dtm                          & -31.8 \pm 0.4 & -0.2 \pm 1.1 & 0.0804  & 3.0~(5.3)  & SDSS \\
  F01004                           & -47.0 \pm 0.2 & -0.2 \pm 0.8 & 0.1178  & 1.5~(4.0)  & \citet{2009MNRAS.400.1139R} \\
  D23H-1                           & -13.3 \pm 0.9 &  4.3 \pm 1.5 & 0.1855  & 2.0~(8.7)  & BOSS \\
  PS1-11af                         &   0.7 \pm 1.2 &  1.5 \pm 1.4 & 0.4046  & 1.0~(10.7) & \citet{2014ApJ...780...44C} \\
  SDSS J0952\tablenotemark{e}      & -27.8 \pm 0.4 & -2.0 \pm 1.1 & 0.0789  & 3.0~(5.2)  & SDSS \\
  SDSS J1342\tablenotemark{e}      & -15.1 \pm 0.5 & -1.0 \pm 1.3 & 0.0366  & 3.0~(2.3)  & SDSS \\
  SDSS J1350\tablenotemark{e}      & -20.3 \pm 0.4 &  0.9 \pm 1.3 & 0.0777  & 3.0~(5.1)  & SDSS \\
  TDE1                             &   1.2 \pm 1.0 & -1.3 \pm 1.3 & 0.1359  & 2.0~(6.2)  & BOSS \\
  TDE2\tablenotemark{c}            & - 4.5 \pm 0.5 &  3.7 \pm 0.6 & 0.2515  & 1.5~(9.2)  & \citet{2011ApJ...741...73V} \\
  SDSS J0748\tablenotemark{b,c,e}  & -11.4 \pm 1.0 & 1.2 \pm 0.8 & 0.0615   & 1.0~(1.3)  & \citet{2013ApJ...774...46Y} \\
  ASASSN14ae\tablenotemark{b,c}    & - 0.7 \pm 0.4 & 3.4 \pm 0.8 & 0.0436   & 3.0~(2.8)  & SDSS \\
  PTF09axc\tablenotemark{b,c}      & - 1.1 \pm 0.7 & 4.9 \pm 0.4 & 0.1146   & 1.0~(2.6)  & \citet{2014ApJ...793...38A} \\
  PTF09djl\tablenotemark{b,c}      & - 0.3 \pm 0.7 & 4.7 \pm 0.5 & 0.184    & 1.0~(4.3)  & \citet{2014ApJ...793...38A} \\
  PTF09ge\tablenotemark{b,c}       & - 1.7 \pm 0.8 & 0.3 \pm 0.7 & 0.064    & 0.7~(1.0)  & \citet{2014ApJ...793...38A} \\
  PS1-10jh\tablenotemark{b,c}      & - 0.5 \pm 0.7 & 1.7 \pm 0.8 & 0.1696   & 1.0~(4.0)  & \citet{2014ApJ...793...38A} \\
  iPTF15af\tablenotemark{b,c}       & - 1.7 \pm 0.3 & 1.3 \pm 1.9 & 0.079    & 3.0~(5.2)  & SDSS \\
  iPTF16axa                        & - 1.1 \pm 1.7 & 0.2 \pm 1.5 & 0.108    & 0.5~(1.2)  & \citet{2017MNRAS.471.1694W} \\
 \enddata
 \tablenotetext{a}{Negative values indicate emission. H$\alpha$ EW values have been corrected for stellar absorption.}
 \tablenotetext{b}{Analyzed by F16.}
 \tablenotetext{c}{TDEs with optical spectra that exhibit broad H or He features.}
 \tablenotetext{d}{Extracted with WebPlotDigitizer.}
 \tablenotetext{e}{TDEs with optical spectra that exhibit high-ionization coronal lines.}
\end{deluxetable}

The A17 classification scheme includes a measure of the reliability of a given event being a bona-fide TDE (i.e., ``likely'' TDEs are more reliable candidates than ``possible'' TDEs). Throughout this work, we test whether our results are biased by the inclusion of less-reliable TDE candidates by repeating the statistical analyses using different TDE subsamples: (1) all TDEs excluding ``possible'' TDEs; (2) only TDEs with known X-ray emission; (3) TDEs with known X-ray emission but excluding ``possible'' X-ray TDEs; (4) veiled TDEs with no known X-ray emission (including those objects with coronal lines or broad H or He features); (5) TDEs with spectra that exhibit broad H or He lines; and (6) the eight TDEs analyzed by F16. The results of these tests, which find no significant dependence on the choice of TDE subsample, are collected in Table~\ref{table:subsamples}.

We also include 3XMM J1500 \citep{2017NatAs...1E..33L}, F01004 \citep{2017NatAs...1E..61T}, PS16dtm \citep{2017ApJ...843..106B}, and iPTF16axa \citep{2017ApJ...842...29H}, which were published after A17 had been submitted. According to the A17 criteria, we treat these events as a likely X-ray TDE and three veiled TDEs, respectively. TDE1 \citep{2011ApJ...741...73V} was not classified by A17 because there were no X-ray observations overlapping the position of the TDE. Based on its optical emission, we treat this event as a veiled TDE. iPTF15af, which was included in the F16 sample, was not detected in X-rays and so is classified here as a veiled TDE with broad H/He features (Blagorodnova et al., in prep., and private communication).

The objects in our sample were selected according to the A17 classification (i.e., we did not include TDEs that were rejected by A17) and by the availability of host-galaxy spectra that covered the H$\alpha$ and H$\delta$ features, necessary for our analysis in Section~\ref{sec:props}, below. The latter requirement biases our sample against TDEs in high-redshift galaxies. However, this leads to the exclusion of only one TDE, Swift J2058, at $z\approx1.2$.

Where available, we collect the host-galaxy spectra of our TDEs from the literature, the SDSS, or the Baryon Oscillation Spectroscopic Survey (BOSS; \citealt{Dawson2012BOSS}). When a spectrum was published as a plot in a paper but the data were unavailable (namely, 2MASX J0249 and the two host-galaxy candidates of RX J1242; \citealt{2007AA...462L..49E,1999AA...349L..45K}), we use the {\sc WebPlotDigitizer}\footnote{\url{http://arohatgi.info/WebPlotDigitizer/}} to digitize the figure. 

Two of the TDEs, namely RX J1420 \citep{2000A&A...362L..25G} and RX J1242 \citep{1999AA...349L..45K,2004MNRAS.349L...1V}, each have two potential host galaxies identified by their discoverers. We analyze all four galaxies in this work, and in Section~\ref{subsec:rxj1420} pay special attention to RX J1420.

Our TDE sample is heterogeneous, as it includes TDEs collected and classified by different surveys. As noted by \citet{2017ApJ...837..120G}, heterogeneous samples should be treated with care, as they propagate the different classification, survey depth, and selection biases of the surveys they draw from. F16 attempted to allay this problem by limiting their sample to flares that were discovered by optical surveys and had spectra that showed broad H$\alpha$ or He II features. While this insured a more homogeneous sample, it also limited its size. As we note in Section~\ref{sec:intro}, it is still unclear whether or not veiled TDEs and X-ray TDEs form two separate classes of objects. As such, we test whether the overabundance of veiled TDEs in quiescent Balmer-strong galaxies is unique to those objects, or whether other types of TDEs also occure in this rare type of galaxy. By relying on the systematic classification of A17 and analyzing all host-galaxy spectra in the same manner, we alleviate the worst biases inherent in a heterogeneous sample. The results we report in Section~\ref{sec:props}, below, bolster our confidence in these measures.

\subsection{Control galaxy sample}
\label{subsec:gals}

In the following sections, we compare the host galaxies of the TDE sample in Table~\ref{table:host_props} to a representative sample of SDSS galaxies. We select galaxies from the 12th data release \citep{2015ApJS..219...12A}\footnote{\url{http://www.sdss.org/dr12}} SDSS Main Galaxy Sample consisting of $\sim900,000$ galaxies observed over $\sim10,000$ deg$^2$ to a limiting magnitude of $r^\prime<17.77$ mag. The spectral range of the SDSS observations is $3800-9200~\mathrm{\AA}$ at a resolution of $R\sim1500$ at 5000~$\mathrm{\AA}$ \citep{2013AJ....146...32S}. 

\begin{figure*}
 \includegraphics[width=0.94\textwidth]{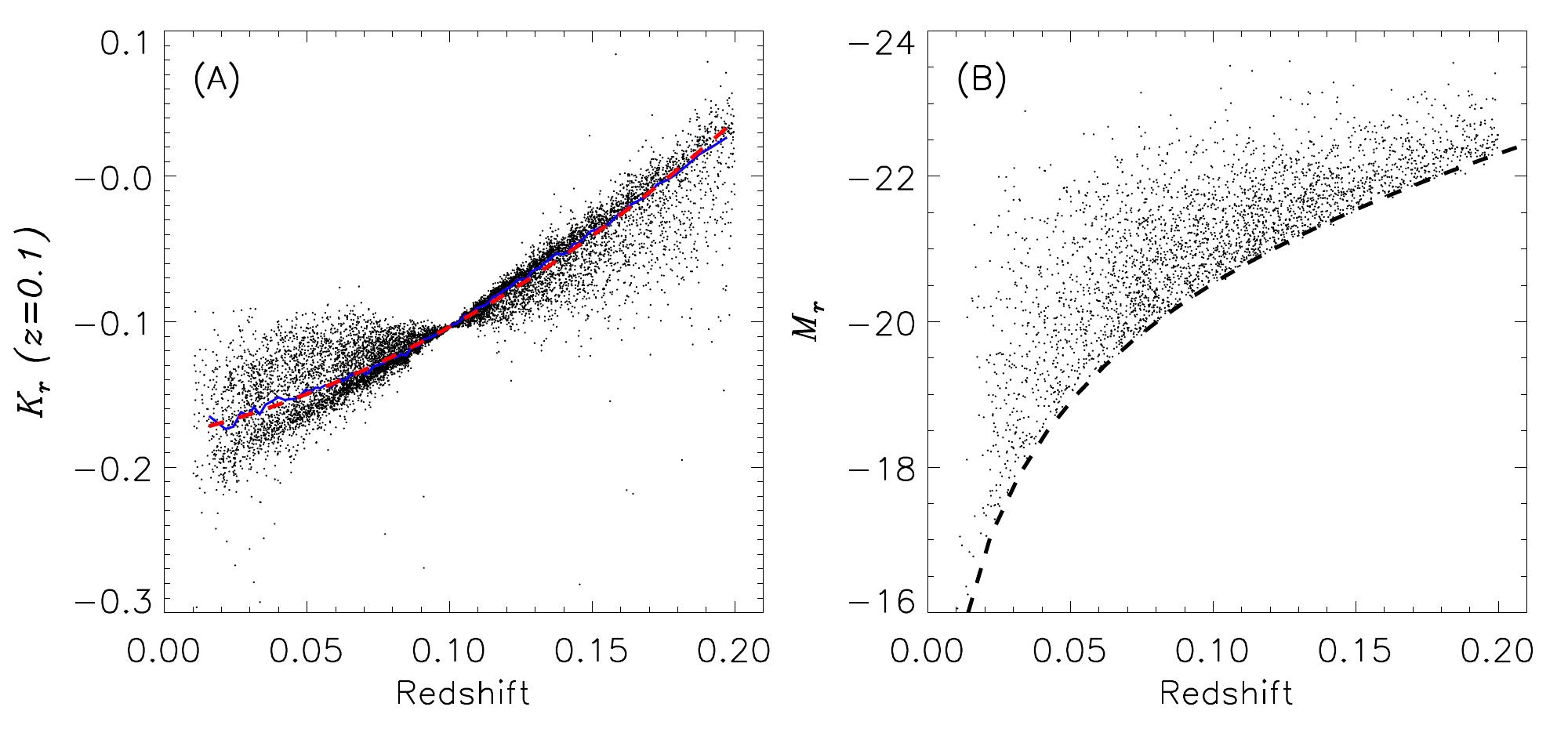}
 \caption{(A) The $z=0.1$ $r^\prime-$band $K-$correction as a function of redshift. Black points are individual galaxies, the blue curve is the median correction in bins of redshift and the red dashed line is a polynomial fit of the correction as a function of redshift. We apply the fit of the correction to the magnitude limit plotted in (B). (B) $K-$corrected absolute $r^\prime-$band magnitude as a function of redshift. Black points are individual galaxies and dashed line delineates the magnitude limit as a function of redshift. We use the magnitude limit and absolute magnitude to determine the volume over which each galaxy may be observed. For clarity, we plot a random subset of the full sample.}
 \label{fig:kcorr}
\end{figure*}

We limit the comparison sample to galaxies in the redshift range $0.01<z<0.2$, which covers $\sim 86$\% of the redshifts of the TDE host galaxies in our sample.\footnote{Five TDE hosts have redshifts $>0.2$: PTF10iya ($z=0.22405$), TDE2 ($z=0.2515$), SDSS J0159 ($z=0.31167$), Swift J1644 ($z=0.3534$), and PS1-11af ($z=0.4046$).} The lower redshift limit ensures that distance estimates are not severely affected by peculiar velocities. The upper redshift limit means the sample contains bright, intrinsically rare galaxies. A two-sided Kolmogorov-Smirnov test does not reject the null hypothesis that the redshift distribution of the TDE hosts is drawn from the same distribution as the redshifts of the control sample, at a $p$-value of $0.01$. Furthermore, varying the redshift limit of the sample, e.g., to $0.01<z<0.1$, has no significant effect on our results.  

The galaxies in our control sample range in stellar mass from $\text{log}(M_\star/{\rm M_\sun})=6.5$ to $12.3$. The TDE hosts for which we measure stellar masses, on the other hand, are limited to a range of $\text{log}(M_\star/{\rm M_\sun})=9.2$--$10.8$. In Sections~\ref{subsec:compactness} and \ref{subsec:estimate}, we control for this narrower stellar mass range and show that applying different stellar mass cuts to the control sample has very little effect on our results.

The SDSS survey is magnitude limited and thus the same objects are not observed throughout the survey volume; only nearby fainter objects are included in the sample. Our aim is to compare the TDE host galaxy properties to the properties of the general galaxy population. To account for the magnitude limit of the survey, we derive a volume weight for each galaxy in the sample. This volume weight corresponds to the survey volume over which a particular galaxy may be observed given the magnitude limit of the SDSS. This approach is analogous to the volume weighting first proposed to account for incompleteness when deriving the quasar luminosity function \citep{1968ApJ...151..393S}.

The sample is observed over a large redshift range, so we must account for the shift in the photometric bands. We $K-$correct the individual measurements to $z=0.1$ and derive a mean $K-$correction for the sample as a function of redshift, which we apply to the magnitude limit. Figure \ref{fig:kcorr}A shows the $K-$correction as a function of redshift. The median correction in bins of redshift is
\begin{equation}
K_{r^\prime}(z) = 1.09z_{0.1} + 3.31z_{0.1}^2 - 2.5\mathrm{log}_{10}(1.1),
\end{equation}
where $z_{0.1} = z - 0.1$. We apply the best-fit correction to the magnitude limit shown in Figure \ref{fig:kcorr}B.

We calculate the volume weight from the absolute magnitude limit plotted in Figure \ref{fig:kcorr}B. The limit is 
\begin{equation}
M_{r^\prime,lim}(z)  = 17.77 - 5[\mathrm{log}(L_D) - 1] - K_{r^\prime}(z) - A_{r^\prime}
\end{equation}
where $L_D$ is the luminosity distance and $A_{r^\prime} = 0.086$ mag is the average $r^\prime-$band extinction. We invert this relation to determine the maximum redshift, $z_{max}$, over which an object of particular absolute magnitude may be observed. For objects with $z_{max} > 0.2$, we set $z_{max} = 0.2$. The volume weight is
\begin{equation}
w_{vol} = \delta_{sky} \times \frac{4\pi}{3} \times (r_{max}^{3} - r_{min}^{3}).
\end{equation}
Here $\delta_{sky}$ is the fraction of sky observed by SDSS and $r_{max}$ and $r_{min}$ are the maximum and minimum comoving distance, respectively. We determine $r_{max}$ from $z_{max}$ and $r_{min}$ is the comoving distance at $z=0.01$.

In this work, we concentrate on several global galaxy properties: stellar mass, $M_\star$; size (by way of the S\'{e}rsic half-light radius, $r_{50}$, as measured in the $r^\prime$ band); stellar velocity dispersion, $\sigma_v$; and surface mass density, 
\begin{equation}
\Sigma_{M_\star} = \frac{M_\star}{r_{50}^2},
\end{equation}
where, following \citet{2012ApJ...760..131C}, we have emitted the constant $2\pi$ in the denominator.

We derive stellar masses with the {\sc Lephare} code \citep{1999MNRAS.310..540A,2009ApJ...690.1236I} and SDSS $ugriz$ c-model magnitudes \citep{2002AJ....123..485S,2010AJ....139.1628D}, corrected for line-of-sight extinction via the \citet{2011ApJ...737..103S} reddening maps. $r^\prime$-band half-light radii, measured through a S\'{e}rsic profile fit, were taken from the Value Added Galaxy Catalog produced by the New York University group \citep{2005AJ....129.2562B}. Velocity dispersions were calculated by the Portsmouth Group.\footnote{\url{http://www.sdss.org/dr12/spectro/galaxy_portsmouth/}}

We apply several quality cuts to the galaxy sample. As detailed below, we exclude galaxies with:
\begin{enumerate}
 \item Bad H$\alpha$ fits (i.e., $\chi^2\leq0$; 0.8\% of the sample).
 \item Bad velocity dispersion measurements ($\sigma_v=0$, $\sigma(\sigma_v)=0$, or $\sigma(\sigma_v)>=\sigma_v$; 5.1\% of the sample).
 \item Bad stellar mass measurements ($M_\star<0$ or $\sigma(M_\star)\leq 0$; $0.9$\% of the sample).
 \item Bad diameter measurements or values smaller than the SDSS seeing of $\sim1.2^{\prime\prime}$ ($0.6$\% of the sample).
 \item Evidence of AGN: we exclude composite galaxies ($6.5$\% of the sample, $1.9$\% after volume-weighting), AGNs ($3.1$\%, $0.7$\%), and low signal-to-noise LINERS ($10.4$\%, $2.1$\%), which correspond to classes 3--5 in the {\sc Galspec} BPT classification \citep{1981PASP...93....5B,2003MNRAS.346.1055K,2004MNRAS.351.1151B}.
\end{enumerate}
After these cuts, we are left with $443,023$ galaxies. While the TDE hosts are not drawn from the SDSS control sample, they satisfy cuts 1--4 (AGN activity is discussed in full in Section~\ref{subsec:agn}).


\section{Global properties of TDE host galaxies}
\label{sec:props}

In this section, we first test whether TDEs are overabundant in quiescent Balmer-strong galaxies, then show that TDE host galaxies are preferentially found in galaxies with higher surface mass densities than the general galaxy population. We also find a hint that quiescent TDE hosts may have lower velocity dispersions than the control quiescent population. Quiescent Balmer-strong galaxies are also known to have high stellar surface mass densities, but we cannot yet tell whether their overabundance is driven solely by this property. Finally, we use the test case of RX J1420 to show that the global galaxy properties we focus on here can be used to identify TDE host galaxies in cases where there is more than one candidate host.

\subsection{TDEs and quiescent Balmer-strong galaxies}
\label{subsec:E+A}

We begin by testing whether the overabundance of TDEs in quiescent Balmer-strong galaxies reported by \citet{2014ApJ...793...38A} and F16 persists in our larger sample of TDE host galaxies (34 objects vs. 8). We repeat the analysis performed by F16 and compare the locations of the TDE hosts in the phase space spanned by H$\alpha$ equivalent width (EW) and Lick H$\delta_{\rm A}$ absorption relative to a control sample of volume-weighted SDSS galaxies. 

To compare our results with those of F16, we analyzed the host-galaxy spectra in the same manner, by fitting for the Lick H$\delta_{\rm A}$ index, which probes stellar absorption from A stars \citep{1997ApJS..111..377W}, and the H$\alpha$ EW, which probes current star formation. Following \citet{1999ApJS..122...51D}, quiescent Balmer-strong galaxies were selected as having deep H$\delta$ absorption features (H$\delta_{\rm A}-\sigma({\rm H}\delta_{\rm A})>4$~\AA) and little ongoing star formation (H$\alpha~{\rm EW}<3$~\AA). The latter value corresponds to a specific star formation rate of $\lesssim10^{-11}~{\rm yr^{-1}}$, below which galaxies are usually considered to be ``quiescent'' (e.g., \citealt{2011A&A...533A.119E,2015MNRAS.450..905G}). H$\delta_{\rm A}$ was not corrected for emission-line filling, but the H$\alpha$ EW was corrected for stellar absorption, since these effects are expected to be negligible and significant, respectively, in quiescent Balmer-strong galaxies. 

\floattable
\begin{splitdeluxetable*}{lCCCCCBCCCCCCCCC}
 \tablecaption{Using different TDE subsamples has little to no effect on the determination of: (1) the overrepresentation of quiescent Balmer-strong galaxies among TDE hosts; (2) the high stellar surface mass densities of TDE hosts and their median velocity dispersion; and (3) the estimates of $\alpha,\beta$ in the statistical model of the TDE rate. \label{table:subsamples}}
 \tablehead{
  \colhead{TDE subsample\tablenotemark{a}} & \colhead{$N_{\rm TDE}$\tablenotemark{b}} & \colhead{$N_{\rm sQBS}$\tablenotemark{c}} & \colhead{Overabundance\tablenotemark{d}} & \colhead{$N_{\rm mQBS}$\tablenotemark{e}} & \colhead{Overabundance} & \colhead{$N_{\Sigma_{M_\star}}$} & \colhead{${\rm log}(\Sigma_{M_\star})_{p50}$\tablenotemark{f}} & \colhead{Significance\tablenotemark{g}} & \colhead{$N_{\sigma_v}$} & \colhead{${\rm log}(\sigma_v)_{p50}$\tablenotemark{h}} & \colhead{Significance\tablenotemark{i}} & \colhead{$N_{\Sigma_{M_\star},\sigma_v}$\tablenotemark{j}} & \colhead{$\hat{\alpha}$} & \colhead{$\hat{\beta}$} \\
 }
 \decimals
 \startdata
  All TDEs                          & 35 & 4 & 35^{+21}_{-17}   & 8 & 18^{+8}_{-7}   & 13 & 9.5^{+0.4}_{-0.1} & 6\sigma~(5\sigma) & 14 & 1.8^{+0.3}_{-0.2} & 3\sigma~(1\sigma) & 10 & 0.9 \pm 0.2 & -1.0 \pm 0.6 \\
  All, except ``possible''          & 28 & 4 & 44^{+28}_{-21}   & 7 & 20^{+10}_{-7}  & 10 & 9.5^{+0.2}_{-0.2} & 6\sigma~(5\sigma) & 11 & 1.9^{+0.2}_{-0.2} & 3\sigma~(1\sigma) & 8  & 0.9 \pm 0.2 & -1.1 \pm 0.7 \\
  Veiled + broad H/He + coronal     & 18 & 3 & 50^{+38}_{-29}   & 4 & 17^{+12}_{-8}  & 7  & 9.5^{+0.1}_{-0.2} & 5\sigma~(3\sigma) & 8  & 1.9^{+0.2}_{-0.3} & 2\sigma~(1\sigma) & 6  & 0.9 \pm 0.3 & -1.0 \pm 0.9 \\
  Broad H/He                        & 12 & 4 & 110^{+80}_{-50}  & 5 & 34^{+22}_{-14} & 6  & 9.5^{+0.3}_{-0.2} & 4\sigma~(3\sigma) & 5  & 1.8^{+0.3}_{-0.2} & 2\sigma~(1\sigma) & 6  & 0.9 \pm 0.2 & -1.1 \pm 0.9 \\
  X-ray + ``likely'' + ``possible'' & 17 & 1 & 18^{+22}_{-18}   & 4 & 18^{+13}_{-9}  & 6  & 9.6^{+0.5}_{-0.1} & 4\sigma~(3\sigma) & 6  & 1.8^{+0.3}_{-0.2} & 2\sigma~(1\sigma) & 4  & 1.0 \pm 0.3 & -1.1 \pm 1.0 \\
  X-ray + ``likely''                & 10 & 1 & 29^{+41}_{-29}   & 3 & 23^{+21}_{-13} & 3  & 9.5^{+0.6}_{-0.0} & 3\sigma~(2\sigma) & 3  & 1.8^{+0.1}_{-0.0} & 1\sigma~(1\sigma) & 2  & 1.0 \pm 0.4 & -1.2 \pm 1.5 \\
  F16 sample                        & 8  & 3 & 110^{+110}_{-70} & 4 & 41^{+31}_{-20} & 5  & 9.5^{+0.4}_{-0.2} & 4\sigma~(3\sigma) & 5  & 1.8^{+0.3}_{-0.1} & 2\sigma~(1\sigma) & 5  & 0.9 \pm 0.3 & -0.9 \pm 0.9 \\
 \enddata
 \tablenotetext{a}{The TDE subsamples contain the following types of TDEs: (1) All TDEs; (2) All TDEs except for ``possible'' X-ray TDEs; (3) All veiled TDEs (including those with broad H/He features or coronal lines); (4) TDEs with broad H or He features; (5) All TDEs with known X-ray emission; (6) X-ray and likely X-ray TDEs; and (6) the eight veiled TDEs with broad H or He features analyzed by F16.}
 \tablenotetext{b}{Number of TDEs in the specific subsample.}
 \tablenotetext{c}{Number of TDE hosts that are quiescent, Balmer-strong (post-starburst) galaxies (sQBS).}
 \tablenotetext{d}{All overabundances, including those for the F16 sample, were calculated assuming quiescent Balmer-strong (sQBS; H$\delta_\text{A}-\sigma({\rm H\delta_{A}})>4~{\rm \AA}$) and quiescent moderately Balmer-strong galaxies (mQBS; H$\delta_{\rm A}-\sigma({\rm H\delta_{A}})>1.31~{\rm \AA}$) account for $0.3$\% and $1.2$\% of all galaxies in the volume-weighted sample, respectively.}
 \tablenotetext{e}{Number of TDE hosts that are quiescent, moderately Balmer-strong galaxies (mQBS, a fraction of which are expected to be starburst galaxies).}
 \tablenotetext{f}{Median of ${\rm log}(\Sigma_{M_\star} / {\rm M_\sun}/{\rm kpc^2})$ values for the TDEs in the specific subsample.}
 \tablenotetext{g}{We used a binomial test to ascertain the statistical significance of the higher stellar surface mass densities of the TDE hosts. For these tests, 88\% and 71\% of the galaxies in the volume-weighted sample have values ${\rm log}(\Sigma_{M_\star})<9.5$ and ${\rm log}(\Sigma_{M_\star})<9$, respectively. These values represent the median of the $\Sigma_{M_\star}$ values and the lower limit of the TDE hosts, respectively. The significance of the test that uses the more conservative limit of ${\rm log}(\Sigma_{M_\star})<9$ is shown in parentheses.}
 \tablenotetext{h}{Median of ${\rm log}(\sigma_{v} / {\rm km~s^{-1}})$ values for the TDEs in the specific subsample.}
 \tablenotetext{i}{We used a binomial test to ascertain the statistical significance of the lower velocity dispersions of the TDE hosts. For these tests, 48\% and 0.09\% of the galaxies in the volume-weighted sample have values ${\rm log}(\sigma_v)>1.8$ and ${\rm log}(\sigma_v)>2.2$, respectively. These values represent the median of the $\sigma_v$ values and the upper limit of the TDE hosts, respectively. The significance of the test that uses the more conservative limit of ${\rm log}(\Sigma_{M_\star})>2.2$ is shown in parentheses.}
 \tablenotetext{j}{Number of TDE hosts with $\Sigma_{M_\star}$ and $\sigma_v$ values measured in the same manner as for the control galaxies.}
\end{splitdeluxetable*}

As in F16, we differentiate between quiescent Balmer-strong galaxies (classified as those galaxies with H$\delta_{\rm A}-\sigma({\rm H}\delta_{\rm A})>4$~\AA) and quiescent, ``moderately'' Balmer-strong galaxies (for which the requirement on H$\delta_{\rm A}-\sigma({\rm H}\delta_{\rm A})$ is relaxed to $>1.31$~\AA). The galaxies in the first category are unambiguously post-starburst galaxies (F16), but galaxies in the second category have spectra that could arise from either post-starburst or truncated star-formation histories. To remain exact, we will continue to refer to these galaxies as ``quiescent Balmer-strong'' throughout the paper. Our control sample includes 324 quiescent Balmer-strong galaxies and 2543 quiescent moderately Balmer-strong galaxies. These account for $0.3$\% and $1.2$\% of the comparison galaxy sample, respectively.\footnote{For comparison, in F16 these fractions were $0.2$\% and $2.3$\%, respectively.} 

Our analysis differs from that of F16 in three ways: 
\begin{enumerate}
 \item Our control sample is volume-weighted, so it accounts for the fact that low-mass, low-luminosity galaxies at high redshifts are missed in magnitude-limited surveys such as the SDSS.
 
 \item We do not limit the control sample to galaxies with spectra with a median signal-to-noise ratio $\geq 10$ (to remove galaxies with spurious H$\alpha$ measurements), as such a cut would remove $\approx16$\% of the galaxy sample (before applying the volume weights) and bias it toward more luminous, massive galaxies. 
 
 \item F16 required H$\delta_{\rm A}>1.31$~\AA\ for the quiescent, moderately Balmer-strong galaxies, but we prefer to use the same criterion (H$\delta_{\rm A}-\sigma({\rm H}\delta_{\rm A})$) for both Balmer-strong and moderately Balmer-strong galaxies.
\end{enumerate}

\begin{figure*}
 \includegraphics[width=0.94\textwidth]{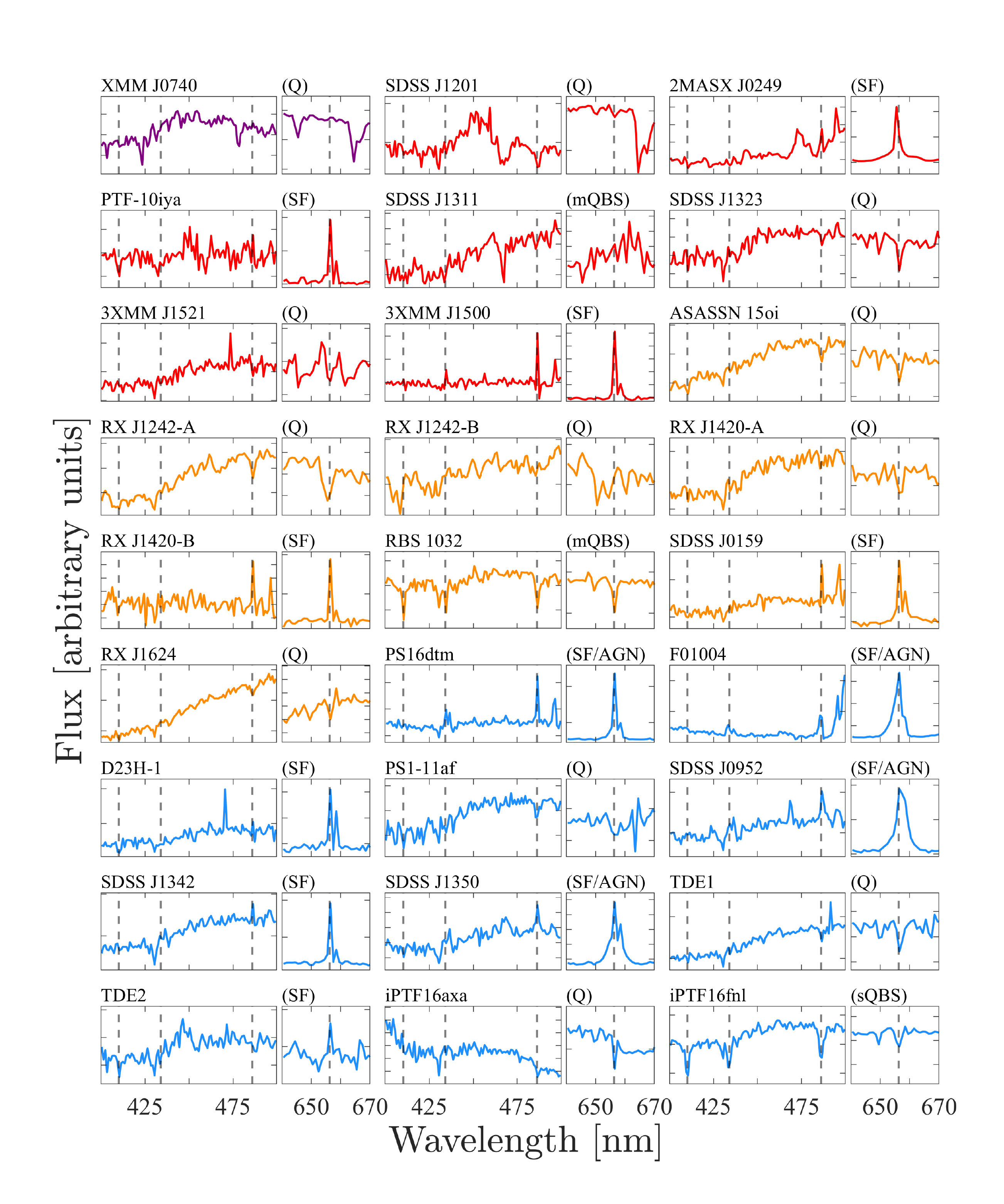}
 \caption{TDE host galaxy spectra used in this work. The spectra of the host galaxies analyzed by F16 are not shown, since they already appear in that paper. The rest-frame spectra have been rebinned into 10~\AA\ bins. The purple, red, orange, and blue colors represent hosts of X-ray TDEs, likely X-ray TDEs, possible X-ray TDEs, and veiled TDEs, respectively. The dashed gray lines denote the wavelengths of the Balmer series: H$\delta$, H$\gamma$, H$\beta$, and H$\alpha$. The galaxy type (``Q'' for quiescent; ``SF'' for star forming; ``SF/AGN'' for galaxies with H$\alpha$ EW$\geq3$~\AA\ with AGN activity (see Section~\ref{subsec:agn}); ``sQBS'' for quiescent Balmer-strong galaxies with H$\delta_{\rm A}-\sigma({\rm H}\delta_{\rm A})>4$~\AA; and ``mQBS'' for quiescent, moderately Balmer-strong galaxies with $1.31~{\rm \AA}<{\rm H}\delta_{\rm A}-\sigma({\rm H}\delta_{\rm A})\leq4$~\AA.) is noted above every panel, along with the name of the TDE. The two panels of each TDE spectrum are not on the same flux scale. The \citet{1995ApJS...98..477H} spectra of NGC 5905 do not cover the H$\delta$ line and are not shown here.}
 \label{fig:spectra}
\end{figure*}

\begin{figure}
 \centering
 \includegraphics[width=0.47\textwidth]{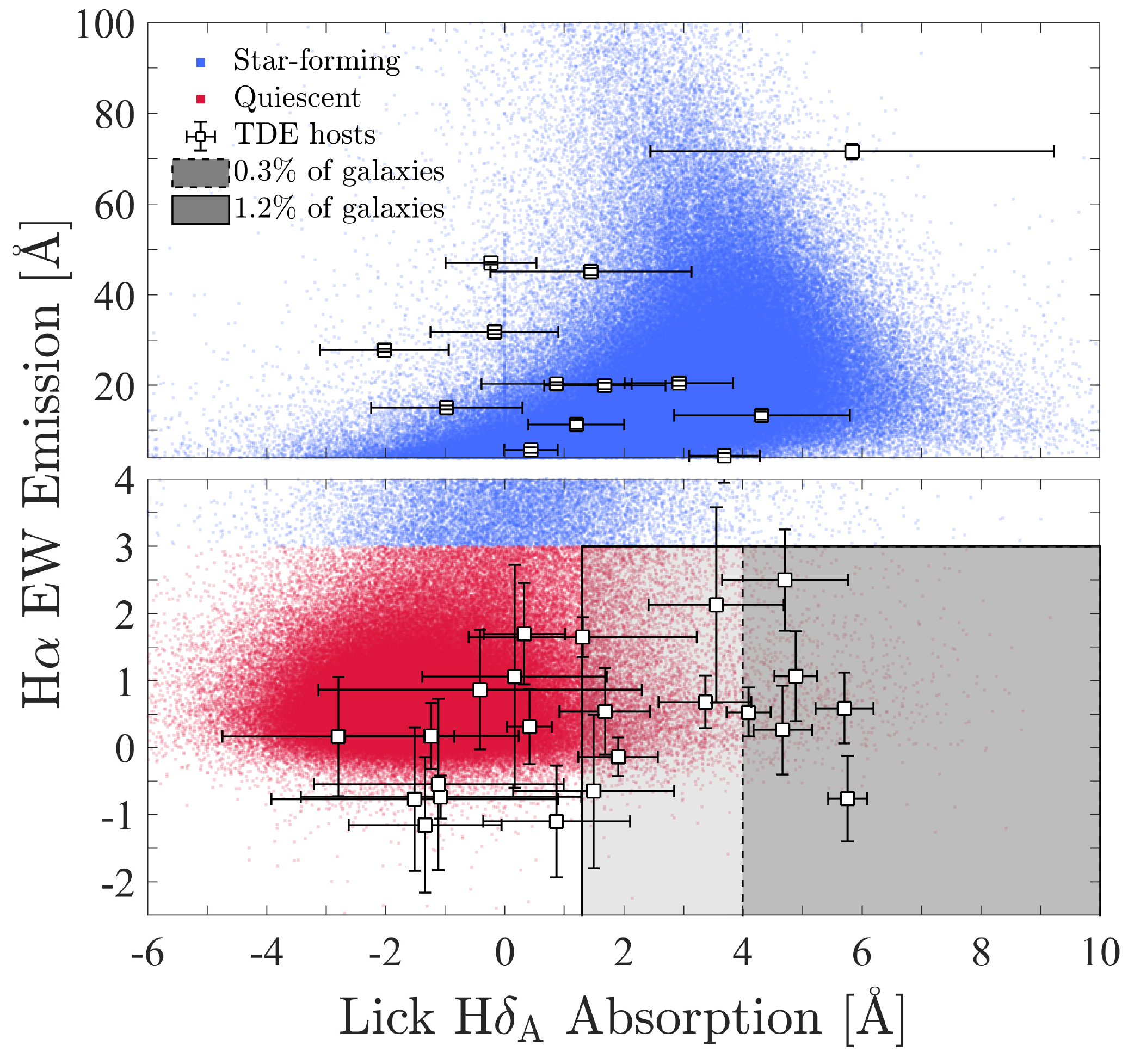}
 \caption{H$\alpha$ EW emission vs. H$\delta_\text{A}$ absorption for an SDSS sample of galaxies at redshifts $0.01<z<0.2$ (blue and red dots, which represent star-forming and quiescent galaxies, respectively) and TDE host galaxies (white squares). For display purposes, the galaxies in this figure have not been volume weighted (see Section~\ref{subsec:gals}), and the upper and lower panels have different linear scalings. The gray patches, delineated by black dashed and solid curves, mark the regions of this phase space inhabited by quiescent Balmer-strong (post-starburst) and moderately Balmer-strong galaxies (some of which are post-starburst). Thirty-five TDEs are represented in this figure (RX J1242 and RX J1420 each have two possible host galaxies). The majority of these broadly follow the global galaxy distribution, with two loci in star-forming and quiescent galaxies. Once the galaxy sample is weighted by volume, quiescent Balmer-strong galaxies account for $0.3$--$1.2$\% of the sample. Between 4--8 of the TDE hosts are quiescent, Balmer-strong galaxies, indicating a lower, but still considerate over-abundance of TDE candidates in this rare type of galaxy.}
 \label{fig:main}
\end{figure}

\begin{figure*}
 \centering
 \includegraphics[width=1\textwidth]{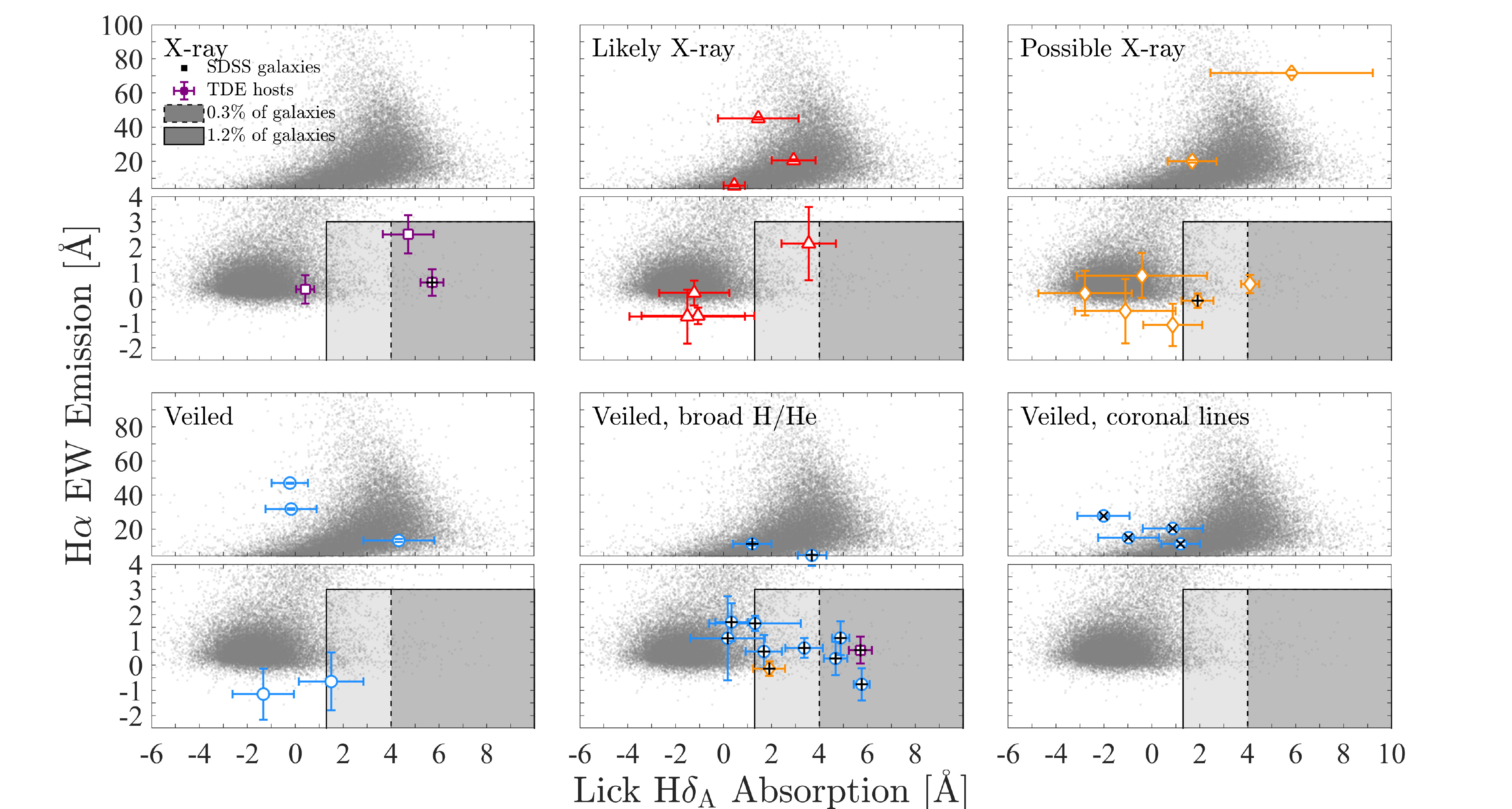}
 \caption{TDE subsamples in H$\alpha$--H$\delta_\text{A}$ phase space. While the small number of events in each panel precludes any statistically significant statement, we see the following: (1) X-ray TDEs are found in quiescent galaxies, including hosts with strong Balmer absorption features; (2) likely and possible X-ray TDEs are found in both quiescent and star-forming galaxies; (3) more than half of the veiled TDEs, including those with broad H or He lines (marked with a `+'), are found in quiescent galaxies, including hosts with strong Balmer absorption lines; (4) veiled TDEs with coronal lines (marked with an `$\times$') lie in star-forming hosts; and (5) overall, TDEs in star-forming galaxies appear to be shifted to lower H$\delta_\text{A}$ absorption values than the general star-forming population. ASASSN14li, ASASSN15oi, and SDSS J0748 appear twice. The first two, marked by a purple square and an orange diamond, respectively, are an X-ray TDE and possible TDE with optical spectra exhibiting broad H and He features. The third event, discovered due to the coronal lines in its spectrum, also exhibits broad H and He features.}
 \label{fig:main_panels}
\end{figure*}

Figure~\ref{fig:spectra} shows the locations of the Balmer-series lines in the TDE host galaxy spectra used here, along with the resulting classifications into star-forming, quiescent, and quiescent Balmer-strong galaxies. We do not show the spectra of those galaxies studied by F16, as they have already been presented in that paper. Table~\ref{table:host_props} lists the TDEs in our sample, along with the properties of their host galaxies and the source of each spectrum.

In Figure~\ref{fig:main}, TDE hosts are found in both star-forming and quiescent galaxies. However, as pointed out by F16, there is a third locus of TDE host galaxies in the region that contains quiescent Balmer-strong galaxies. 

The low fractions of quiescent Balmer-strong galaxies out of the general galaxy population imply an overabundance of TDEs in quiescent Balmer-strong and moderately Balmer-strong galaxies of $4/35/0.003 = 35^{+21}_{-17}$ and $8/35/0.012 = 18^{+8}_{-7}$, respectively. The uncertainties have been calculated by propagating the Poisson uncertainties on the numbers of TDE hosts in the numerator and denominator, and taking the 16th and 84th percentiles of the resulting distribution. Although the overabundance values measured here are formally lower than those measured by F16 ($190^{+115}_{-100}$ and $33^{+7}_{-11}$, respectively), they are consistent within the $2\sigma$ and $1\sigma$ uncertainties, respectively.

Table~\ref{table:subsamples} shows that the choice of TDE subsample has no significant effect on the overrepresentation of quiescent Balmer-strong galaxies. Although formally the F16 and broad H/He subsamples produce overabundance ratios that are $\sim2$--3 times as high as those produced by the other subsamples, all values are consistent within their uncertainties. Importantly, we find the same overabundance values for TDEs with and without known X-ray emission. 

Five of the host galaxies in our sample have redshifts $z>0.2$. Of these, one is a moderately Balmer-Strong galaxy. Removing these galaxies from the sample does not have a significant impact on the overabundance of post-starburst galaxies among the TDE hosts in our sample.

In Figure~\ref{fig:main_panels}, we show each type of TDE on its own in the H$\alpha$--H$\delta_\text{A}$ phase space. Due to the small number of events in each panel, we cannot draw statistically significant conclusions. However, several points stand out:
\begin{enumerate}
 \item The X-ray TDEs ASASSN14li, Swift J1644, and XMM J0740 are all found in quiescent galaxies. Of these, the first two are quiescent Balmer-strong galaxies.
 \item Likely and possible X-ray TDEs are found in both star-forming and quiescent galaxies.
 \item More than half of the veiled TDEs, including those with broad H or He lines, are found in quiescent galaxies, including hosts with strong Balmer absorption lines.
 \item TDEs with coronal line emission are all found in star-forming galaxies. At a given H$\alpha$ emission, these objects also have shallower H$\delta_\text{A}$ absorption features than the background star-forming galaxy population. This effect may be driven by the TDE line emission filling in the H$\delta_\text{A}$ absorption. While the emission line filling of the H$\delta_\text{A}$ absorption from star formation will affect both the TDEs and SDSS galaxies, the additional effect of the on-going TDE observed in the coronal-line TDE host spectra would act to decrease the measured H$\delta_\text{A}$ absorption. In order to quantitatively address this scenario, a fuller picture is required of how the narrow line emission evolves in TDEs with time, and relative to the coronal line emission. Such an analysis is outside the scope of this work, but will be important to understand the underlying host galaxy emission in coronal line emitting TDEs.
 \item Overall, TDEs in star-forming galaxies appear to be shifted to lower H$\delta_\text{A}$ absorption values than the general star-forming population. This shift may be due to: (1) filling of H$\delta_\text{A}$ by emission from the TDE if it was still active when the galaxy spectrum was taken (e.g., SDSS J0748), see above; (2) different stellar mass distributions for the TDE hosts and control galaxy sample (which we tested by cutting on stellar mass and rule out); (3) contamination of the H$\alpha$ line by an AGN or LINER, which affects a few of the TDE hosts (see Section~\ref{subsec:agn}); and (4) a physical difference between the TDE hosts and comparison galaxy sample.
\end{enumerate}

\subsection{Surface stellar mass density and velocity dispersion}
\label{subsec:compactness}

Of the 37 TDE hosts in our sample, 16 appear in both the NYU and Portsmouth value-added catalogs, which allows us to measure their global galaxy properties in the same way as for the control galaxy sample. This allows us to make an ``apples-to-apples'' comparison between the hosts of these TDEs and the general galaxy population. Table~\ref{table:compactness} summarizes the values of these properties for the TDE host galaxies in this subsample.\footnote{The {\sc Lephare} fitting code failed to calculate stellar masses for SDSS J1350 and SDSS J0159. In these instances, we adopted masses calculated by either the Portsmouth or {\sc Galspec} pipelines. The masses from these pipelines were corrected for being 0.13 dex larger, on average, than those produced by {\sc Lephare} (see \citealt{2016ApJ...832..203Z} for the difference between the Portsmouth and {\sc Lephare} values. We calculated the difference between the {\sc Galspec} and {\sc Lephare} masses ourselves.).} 

While this subsample of TDEs accounts for only half of our full sample, it includes all four types of TDE as classified by A17 and spans a wide dynamical range in galaxy properties, including, most importantly, a galaxy stellar mass range of ${\rm log}(M_\star/M_\sun)\approx 8.5$--$11$, over which TDEs are expected to occur. 

To test theoretical estimates of the TDE rate, which find a dependence on the density and velocity dispersion of the stars around the SMBH, we concentrate here on global stellar surface mass density, $\Sigma_{M_\star}$, and velocity dispersion, $\sigma_v$. In the following sections, we will often refer to the base-10 logarithms of these properties. In these cases, the units will be $\text{log}(\Sigma_{M_\star} / (M_\sun / \text{kpc}^2))$ and $\text{log}(\sigma_v / {\rm km~s^{-1}})$. For simplicity, we will omit the units from now on.

\begin{figure*}
 \begin{tabular}{cc}
  \includegraphics[width=0.47\textwidth]{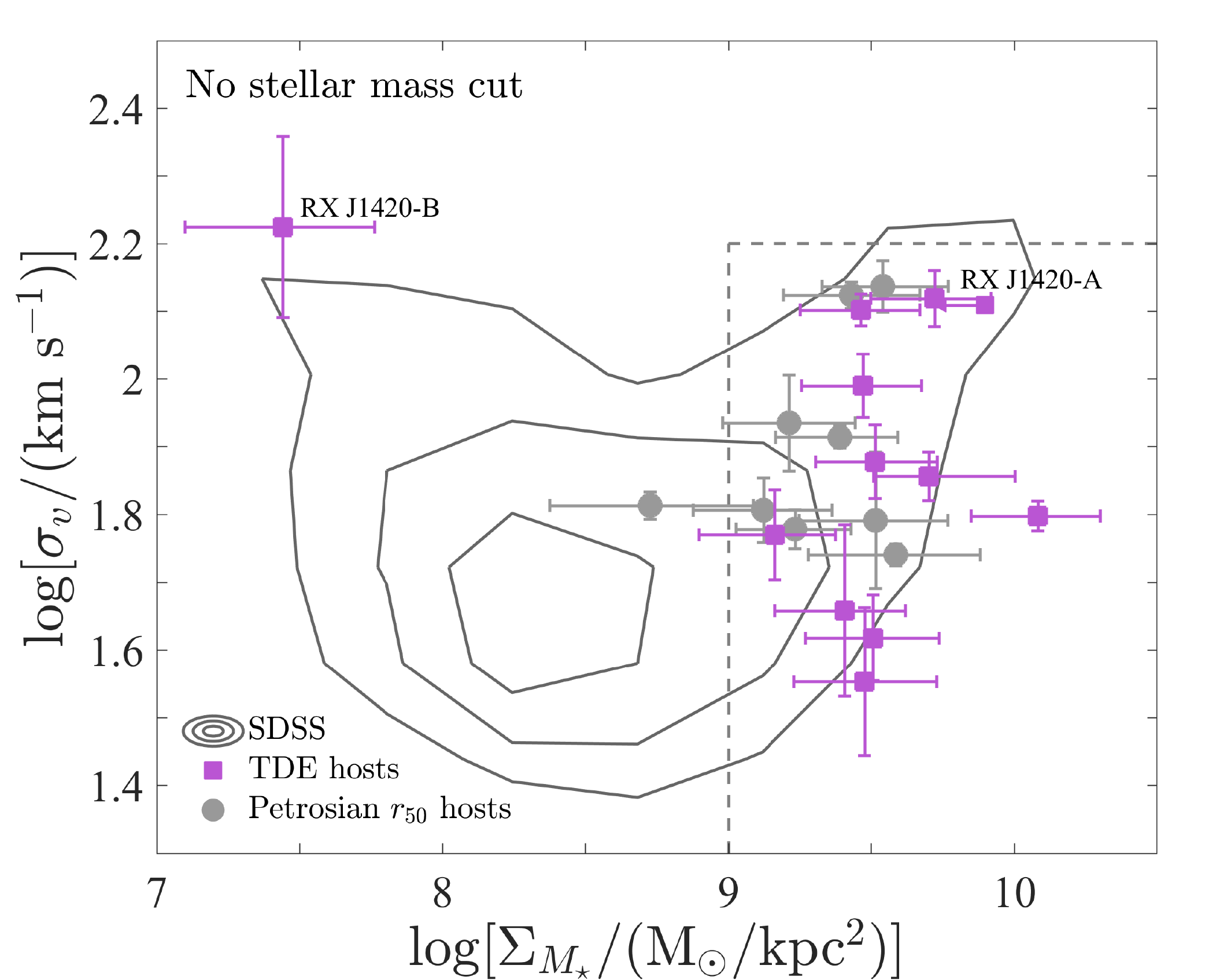} & \includegraphics[width=0.47\textwidth]{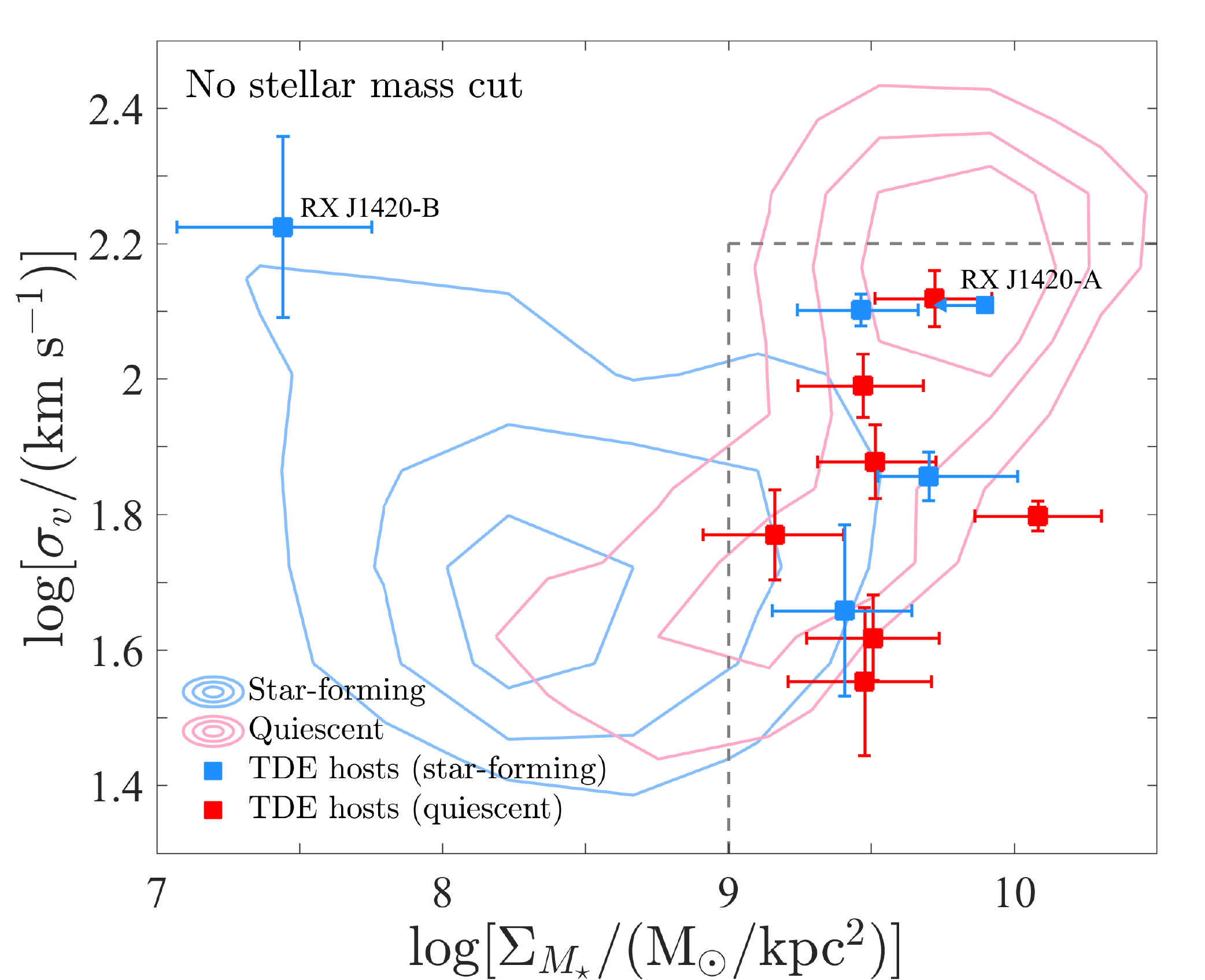} \\
 \end{tabular}
 \caption{Stellar velocity dispersion ($\sigma_v$) vs. stellar surface mass density ($\Sigma_{M_\star}$) of TDE host galaxies. {\it Left}: TDE hosts with galaxy properties measured as for the SDSS control sample are marked with purple squares. Hosts with $\Sigma_{M_\star}$ values based on Petrosian (instead of S\'{e}rsic) half-light radii, most of which have velocity dispersions from \citet{2017MNRAS.471.1694W}, are denoted by gray circles (see Table~\ref{table:compactness}). Contours represent the 25th, 50th, and 84th percentiles of the volume-weighted background galaxy distribution. The phase space delineated by the dashed gray line includes 24\% of the galaxies in the control sample. {\it Right}: The galaxy and TDE host samples are divided into star-forming (blue) and quiescent (red) galaxies. RX J1420-A is found with the rest of the TDE hosts, while RX J1420-B is a stark outlier, hinting that the former galaxy is the real host. Whether we do not apply a stellar mass cut (shown here), or require $\text{log}(M_\star/{\rm M_\sun})>8$, $9$, or $9.5$ (Figure~\ref{fig:appendix}), we find that TDE hosts have significantly higher stellar surface mass densities than the general galaxy population. This trend is driven by star-forming TDE hosts, which have higher $\Sigma_{M_\star}$ values than the star-forming control galaxies. Quiescent TDE hosts have $\Sigma_{M_\star}$ values consistent with those of the quiescent control galaxies. There is also some evidence that quiescent TDE hosts have lower $\sigma_v$ values than the quiescent control population (the significance of this offset depends on the $\sigma_v$ range considered). The gray squares shown in the left panel, though not included in the formal statistical analysis, are consistent with these trends.}
 \label{fig:compact}
\end{figure*}

Thirteen TDE hosts have concrete $\Sigma_{M_\star}$ measurements based on {\sc LePhare} stellar masses and S\'{e}rsic half-light radii,\footnote{Two TDE hosts are unresolved, leading to lower limits on their sizes and upper limits on their surface mass densities. A third galaxy, RX J1420-B is an outlier and is dealt with in detail, below.} and are shown in Figure~\ref{fig:compact}. Fourteen hosts have $\sigma_v$ measurements from the Portsmouth pipeline, and a subsample of ten have both types of measurements. The last sample is marked in Figure~\ref{fig:compact} with purple squares. We address the rest of the TDEs in Table~\ref{table:compactness} at the end of this section.

Some of the TDE hosts in Table~\ref{table:compactness} have velocity dispersion values lower than the typical SDSS instrumental resolution of 60 km~s$^{-1}$ \citep{2013MNRAS.431.1383T}. However, we choose to keep these objects in our sample. Recently, \citet{2017MNRAS.471.1694W} measured velocity dispersions for 12 TDE hosts. Six of these (ASASSN14li, ASASSN14ae, PTF09ge, iPTF15af, D23-H1, and TDE1) also have Portsmouth velocity dispersions, which are consistent, even when the Portsmouth values are formally lower than the SDSS instrumental resolution. The host galaxy of PS16dtm, which has a Portsmouth velocity dispersion of $45\pm13~{\rm km~s^{-1}}$, also had spectra taken by \citet{2011ApJ...739...28X} with the Keck Echelette Spectrograph and Imager, as well as the Magellan Echelette spectrograph, which have instrumental dispersions of $22$ and $26~{\rm km~s^{-1}}$, respectively. \citet{2011ApJ...739...28X} measure a velocity dispersion of $45\pm3~{\rm km~s^{-1}}$, consistent with the Portsmouth value.

\floattable
\begin{deluxetable}{lCCCCccc}
 \tablecaption{TDE host properties used in Figure~\ref{fig:compact} \label{table:compactness}}
 \tablehead{
  \colhead{TDE Host} & \colhead{${\rm log}(M_\star/{\rm M_\sun})$\tablenotemark{a}} & \colhead{$r_{50}$\tablenotemark{b}} & \colhead{${\rm log}(\Sigma_{M_\star})$\tablenotemark{c}} & \colhead{$\sigma_v$\tablenotemark{d}} & \colhead{$\sigma_v$ source} & \colhead{${\rm T_{TDE}}$\tablenotemark{e}} & \colhead{${\rm T_{Gal}}$\tablenotemark{f}} \\
  \colhead{} & \colhead{} & \colhead{(arcsec (kpc))} & \colhead{} & \colhead{(km s$^{-1}$)} & \colhead{} & \colhead{} & \colhead{}
 }
 \decimals
 \startdata
  \multicolumn{7}{c}{TDE hosts with S\'{e}rsic half-light radii} \\
  ASASSN14li & 9.3^{+0.1}_{-0.1}     &  1.0~(0.4)  & 10.1^{+0.2}_{-0.2} & 63 \pm 3   & SDSS (Portsmouth) & X-ray+H/He          & QBS \\
  SDSS J1201 & 10.61^{+0.08}_{-0.16} &  1.4~(3.6)  & 9.5^{+0.2}_{-0.3} & \cdots     & $\cdots$          & Likely X-ray        & Q \\ 
  SDSS J1323 & 10.38^{+0.06}_{-0.07} &  1.3~(2.7)  & 9.5^{+0.2}_{-0.2} & 75 \pm 10  & SDSS (Portsmouth) & Likely X-ray        & Q \\ 
  RX J1420-A & 10.53^{+0.07}_{-0.07} &  0.8~(2.5)  & 9.7^{+0.2}_{-0.2} & 131 \pm 13 & SDSS (Portsmouth) & Possible X-ray      & Q \\
  RX J1420-B\tablenotemark{g} &  8.55^{+0.24}_{-0.26} &  1.2~(3.6)  & 7.4^{+0.3}_{-0.3} & 168 \pm 52 & SDSS (Portsmouth) & Possible X-ray & SF \\
  SDSS J0159 & 10.37^{+0.11}_{-0.06} &  0.3~(>1.7) & <9.9              & 128 \pm 17 & SDSS (Portsmouth) & Possible X-ray      & SF \\
  RBS 1032   &  9.19^{+0.15}_{-0.16} &  1.4~(0.7)  & 9.5^{+0.2}_{-0.2} & 36 \pm 9   & SDSS (Portsmouth) & Possible X-ray      & QBS \\
  NGC 5905   & 10.83^{+0.22}_{-0.06} & 11.7~(2.2)  & 10.1^{+0.3}_{-0.2} & \cdots     & $\cdots$          & Possible X-ray      & SF \\
  ASASSN14ae &  9.73^{+0.13}_{-0.13} &  1.7~(1.3)  & 9.5^{+0.2}_{-0.2} & 41 \pm 6   & SDSS (Portsmouth) & Veiled+H/He         & QBS \\
  PTF09ge    &  9.87^{+0.13}_{-0.17} &  1.9~(2.3)  & 9.2^{+0.2}_{-0.3} & 59 \pm 9   & SDSS (Portsmouth) & Veiled+H/He         & Q \\
  SDSS J0748 & 10.18^{+0.06}_{-0.09} &  1.7~(2.3)  & 9.5^{+0.2}_{-0.2} & 126 \pm 7  & SDSS (Portsmouth) & Veiled+coronal+H/He & SF \\
  SDSS J0952 & 10.37^{+0.06}_{-0.07} &  0.5~(>1.0) & <10.4              & \cdots     & $\cdots$          & Veiled+coronal      & SF/AGN \\
  SDSS J1342 &  9.64^{+0.23}_{-0.07} &  1.3~(0.9)  & 9.7^{+0.3}_{-0.2} & 72 \pm 6   & SDSS (Portsmouth) & Veiled+coronal      & SF \\
  SDSS J1350 &  9.94^{+0.17}_{-0.20} &  1.4~(2.0)  & 9.3^{+0.3}_{-0.3} & \cdots     & $\cdots$          & Veiled+coronal      & SF/AGN \\
  iPTF15af    & 10.31^{+0.08}_{-0.10} &  1.9~(2.6)  & 9.5^{+0.2}_{-0.2} & 98 \pm 11  & SDSS (Portsmouth) & Veiled+H/He         & QBS \\
  PS16dtm    &  9.77^{+0.11}_{-0.13} &  0.9~(1.5)  & 9.4^{+0.2}_{-0.2} & 45 \pm 13  & SDSS (Portsmouth) & Veiled              & SF/AGN \\
  \hline
  \multicolumn{7}{c}{TDE hosts with Petrosian half-light radii\tablenotemark{h}} \\
  3XMM J1521 & 10.17^{+0.11}_{-0.20} & 0.8~(2.1)   & 9.5^{+0.2}_{-0.3} & 62 \pm 14  & SDSS (Portsmouth) & Likely X-ray   & Q \\
  D3-13     &  10.7                  & 0.7~(7.2)   & 9.4^{+0.2}_{-0.2} & 133 \pm 6  & \citet{2017MNRAS.471.1694W} & Possible X-ray & Q \\
  iPTF16fnl &   9.59^{+0.20}_{-0.25} & 3.3~(1.1)   & 9.6^{+0.3}_{-0.3} & 55 \pm 2   & \citet{2017MNRAS.471.1694W} & Veiled+H/He    & QBS \\  
  iPTF16axa &  10.25^{+0.05}_{-0.08} & 1.4~(3.4)   & 9.4^{+0.2}_{-0.2} & 82 \pm 3   & \citet{2017MNRAS.471.1694W} & Veiled         & Q \\
  PTF09djl  &   9.91^{+0.13}_{-0.17} & 0.9~(3.8)   & 9.1^{+0.3}_{-0.3} & 64 \pm 7   & \citet{2017MNRAS.471.1694W} & Veiled+H/He    & QBS \\
  PTF09axc  &   9.84^{+0.06}_{-0.09} & 1.1~(2.8)   & 9.2^{+0.2}_{-0.2} & 60 \pm 4   & \citet{2017MNRAS.471.1694W} & Veiled+H/He    & QBS \\
  PS1-10jh  &    9.2^{+0.3}_{-0.3}   & 0.8~(3.1)   & 8.7^{+0.3}_{-0.4} & 65 \pm 3   & \citet{2017MNRAS.471.1694W} & Veiled+H/He    & Q \\
  D23-H1\tablenotemark{i}    &  10.08^{+0.15}_{-0.07} & 0.9~(4.0)   & 9.2^{+0.2}_{-0.2} & 86 \pm 14  & SDSS (Portsmouth) & Veiled         & SF \\
  TDE1\tablenotemark{i}      &  10.08^{+0.08}_{-0.12} & 0.9~(2.8)   & 9.5^{+0.2}_{-0.2} & 137 \pm 12 & SDSS (Portsmouth) & Veiled         & Q \\
 \enddata
 \tablenotetext{a}{Stellar masses computed from SDSS cmodel magnitudes using the {\sc LePhare} code. The masses for SDSS J0159 and SDSS J1350 were computed by the Portsmouth and {\sc Galspec} pipelines, respectively. For these galaxies, the measured masses were reduced by 0.13 dex to reflect the average systematic offset between the results of {\sc LePhare} and the other pipelines.}
 \tablenotetext{b}{$r^\prime$-band S\'{e}rsic half-light radius in arcsec, as it appears in the NYU value-added catalog, and in kpc after conversion based on the SDSS redshift (except for SDSS J1201, which has a non-SDSS redshift, and NGC 5905, which is $\approx39$ Mpc away). Following \citet{2016ApJ...821..101Z}, we adopt an uncertainty of $0.1$ dex on these values.}
 \tablenotetext{c}{Surface stellar mass density, computed as $\Sigma_{M_\star}={\rm log}[(M_\star/r_{50}^2)/({\rm M_\sun/kpc^2})]$.}
 \tablenotetext{d}{Stellar velocity dispersion, as measured by the Portsmouth pipeline.}
 \tablenotetext{e}{TDE type, according to A17.}
 \tablenotetext{f}{Host-galaxy type, where quiescent (Q) and star-forming (SF) galaxies are defined as having H$\alpha$ EW$<3$~\AA\ and H$\alpha$ EW$\geq 3$~\AA, respectively. Quiescent Balmer-strong galaxies (QBS) have H$\alpha$ EW$<3$~\AA\ and H$\delta_\text{A}-\sigma(\text{H}\delta_\text{A})$ Lick index $>1.31$~\AA. Galaxies with H$\alpha$ EW$\geq 3$~\AA\ and evidence of AGN activity are labeled as SF/AGN (see Section~\ref{subsec:agn}).}
 \tablenotetext{g}{We treat RX J1420-B as an outlier and exclude it from all calculations and statistical tests; see Section~\ref{subsec:rxj1420}.}
 \tablenotetext{h}{TDE hosts with $r^\prime$-band Petrosian (instead of S\'{e}rsic) half-light radii. Stellar masses were derived using {\sc LePhare}, except for D3-13, where the mass was derived by \citet{2017arXiv170703458V} using {\sc kcorrect}.}
 \tablenotetext{i}{The host galaxies of D23-H1 and TDE1 also have \citet{2017MNRAS.471.1694W} velocity dispersions: $77\pm18$ and $126\pm7~{\rm km~s^{-1}}$, respectively, consistent with their Portsmouth values.}
\end{deluxetable}

We control for galaxy stellar mass by testing how limiting the stellar mass range of the volume-weighted control sample changes the distributions of $\Sigma_{M_\star}$ and $\sigma_v$. We use several stellar mass ranges: (1) the complete galaxy sample, without any cut (shown in Figure~\ref{fig:compact}); (2) galaxies with $\text{log}(M_\star/{\rm M_\sun})>8$, which include all galaxies that are expected to have SMBHs massive enough to create TDEs; (3) galaxies with $\text{log}(M_\star/{\rm M_\sun})>9$, as all our TDE hosts have masses larger than this lower limit; (4) galaxies with $\text{log}(M_\star/{\rm M_\sun})>9.5$, at which point we begin to exclude TDE host galaxies; and (5) galaxies with $\text{log}(M_\star/{\rm M_\sun})<11$, as more massive galaxies, which harbor SMBHs with Schwarzshild radii that extend beyond the tidal radii, are not expected to harbor TDEs. Table~\ref{table:overdensity_mass} and Figure~\ref{fig:appendix} show the results of these tests, which we discuss below.\footnote{The last stellar mass cut, $\text{log}(M_\star/{\rm M_\sun})<11$, produces nearly identical results as the control sample with no stellar mass cut, and so is not shown in Table~\ref{table:overdensity_mass}.} 

Figure~\ref{fig:compact} shows that TDE host galaxies are found in a narrow band of stellar surface mass density with a median of ${\rm log}(\Sigma_{M_\star})=9.5^{+0.4}_{-0.1}$, where the uncertainties reflect the 16th and 84th percentiles of the distribution of 13 TDE hosts with secure $\Sigma_{M_\star}$ measurements. All thirteen of these TDE hosts have redshifts $z<0.2$, similar to the galaxies in the control sample. 

We test the null hypothesis that the distribution of TDE hosts traces that of the control galaxy population by calculating the probability of obtaining $x_\text{obs}$ or fewer TDE hosts with $\Sigma_{M_\star}$ values below some critical value (either the median $\Sigma_{M_\star}$ of the TDE hosts or a lower limit on $\Sigma_{M_\star}$ below which there are no TDE hosts in our sample). For this test, we use the binomial distribution:
\begin{equation}\label{eq:binom}
 P(x | n,p) = {n \choose x}p^x(1-p)^{(n-x)},
\end{equation}
where $n$ is the number of TDE hosts in our sample, and $p$ and $x$ are the fraction of the control galaxy population and number of TDE hosts, respectively, with stellar surface mass densities smaller than the critical value. We express the resultant $p$-value in terms of the Gaussian distribution's standard deviation, $\sigma$, and claim a result as significant if the null hypothesis is rejected at a $>3\sigma$ significance.

When we use the median $\Sigma_{M_\star}$ value, $n=13$, $p=0.88$, and $x_\text{obs} = 3$, and we sum Equation~\ref{eq:binom} for $x = 0,~1,~2,~3$. In this case, the null hypothesis that the stellar surface mass density of the TDE hosts traces that of the control sample is rejected at $>5\sigma$ significance. For the more conservative lower limit, chosen so that $x_\text{obs} = 0$, $p=0.71$, the probability of obtaining $x_\text{obs}$ or fewer hosts below this lower limit is just Equation~\ref{eq:binom} with $x = 0$. The null hypothesis is still rejected at $>5\sigma$.

Additionally, a Mann-Whitney $U$ test rejects the null hypothesis that the distributions of TDE host and control-galaxy $\Sigma_{M_\star}$ have the same medians with a $p$-value of $<0.01$.

Table~\ref{table:overdensity_mass} and the right-hand panel of Figure~\ref{fig:compact} show that this effect is driven by star-forming TDE hosts, which have significantly higher $\Sigma_{M_\star}$ values than the star-forming control sample. In comparison, quiescent TDE hosts have $\Sigma_{M_\star}$ values more similar to the quiescent control galaxies (i.e., the same binomial test yields only a $>2\sigma$ offset), which have higher $\Sigma_{M_\star}$ values to begin with.

These trends grow weaker as we limit the control sample to successively more massive galaxies. When using either the median or lower limit $\Sigma_{M_\star}$ values noted above, the binomial test provides insignificant ($<3\sigma$) results for the overall control sample at stellar mass cuts of $>10^{9.5}~{\rm M_\sun}$, and for the star-forming galaxy subsample at $>10^9~{\rm M_\sun}$. Likewise, the $p$ value of the Mann-Whitney $U$, for all galaxies and star-forming galaxies specifically, rises to $>0.01$ when we limit the stellar mass of the control sample to $>10^{9.5}~{\rm M_\sun}$. This happens because as we exclude less-massive galaxies, we remove galaxies with lower stellar surface mass densities from the control sample.

If there is a selection effect that somehow disfavors detecting TDEs in star-forming galaxies with low stellar surface mass densities, it might affect our conclusions that TDE hosts prefer galaxies with $\text{log}(\Sigma_{M_\star})>9$ values. Yet this is unlikely for several reasons. First, star-forming galaxies with low $\Sigma_{M_\star}$ values are generally less massive, and thus less luminous, as well as less dusty \citep{2013ApJ...763...92Z}, making it easier to detect TDEs in them. Second, if veiled TDEs, i.e., those TDEs without detected X-ray emission as opposed to those merely unobserved in X-rays, have high extinction, we would expect to see even more TDEs in lower-mass, less dusty galaxies.

Finally, we examine the possible effects of the optical coronal lines present in some of the spectra of the veiled TDE sample. If the TDEs were already present when their images were taken by the SDSS, it could lead to an underestimation of their half-light radii and thus an overestimation of $\Sigma_{M_\star}$. The images of SDSS J1342 and J1350 were taken one and two years prior to their spectra, respectively, but it is unknown when the TDEs occurred. SDSS J0748, on the other hand, was imaged only four months before its spectrum was taken, and the spectrum exhibits broad H and He features in addition to the coronal lines, indicating that the TDE had occurred in the recent past. As some TDEs (e.g., ASASSN14li; \citealt{2016MNRAS.455.2918H}) have exhibited broad H and He features months after they were first detected, it is possible that the TDE in SDSS J0748 was already present when the SDSS image was taken. Removing it from the sample, however, does not change our results.

The 14 TDE hosts with Portsmouth velocity dispersions have a median value of $\text{log}(\sigma_v)=1.9^{+0.2}_{-0.2}$. Quiescent and star-forming TDE hosts have similar values: $1.8^{+0.3}_{-0.2}$ and $1.9^{+0.2}_{-0.2}$, respectively. Both the binomial test and Mann-Whitney $U$ test cannot rule out the null hypothesis that the velocity dispersions of the TDE hosts are randomly drawn from the underlying control populations. This is true whether we use the median $\sigma_v$ or a more conservative upper limit of $\text{log}(\sigma_v)>2.2$ (the upper range of $\sigma_v$ values for the TDE hosts).

\floattable
\begin{deluxetable}{llCCCCCC}
  \tablecaption{Significance of TDE host high stellar surface mass density and low velocity dispersion\label{table:overdensity_mass}}
  \tablehead{
   \colhead{Stellar mass cut} & \colhead{Galaxy type} & \colhead{$f(\Sigma_{M_\star})<\Sigma^{\rm TDE}_{\rm med}~(\Sigma^{\rm TDE}_{\rm min})$\tablenotemark{a}} & \colhead{Significance\tablenotemark{b}} & \colhead{$p_{U}$\tablenotemark{c}} & \colhead{$f(\sigma_v>\sigma^{\rm TDE}_{\rm med}~(\sigma^{\rm TDE}_{\rm max}))$\tablenotemark{d}} & \colhead{Significance\tablenotemark{e}} & \colhead{$p_{U}$\tablenotemark{f}} \\
  }
  \decimals
  \startdata
  No stellar mass cut             & All          & 0.88~(0.71) & 5\sigma~(5\sigma) & <0.01 & 0.48~(0.09) & 1\sigma~(1\sigma) & 0.26 \\
                                  & Star-forming & 0.97~(0.83) & 3\sigma~(3\sigma) & <0.01 & 0.28~(0.06) & 1\sigma~(1\sigma) & 0.12 \\
                                  & Quiescent    & 0.50~(0.22) & 2\sigma~(1\sigma) & 0.62  & 0.68~(0.23) & 1\sigma~(1\sigma) & 0.16 \\
  $M_\star>10^8~{\rm M_\sun}$     & All          & 0.85~(0.63) & 4\sigma~(4\sigma) & <0.01 & 0.48~(0.08) & 1\sigma~(1\sigma) & 0.26 \\
                                  & Star-forming & 0.96~(0.76) & 3\sigma~(3\sigma) & <0.01 & 0.24~(0.04) & 1\sigma~(1\sigma) & 0.08 \\
                                  & Quiescent    & 0.50~(0.22) & 2\sigma~(1\sigma) & 0.63  & 0.68~(0.23) & 2\sigma~(1\sigma) & 0.15 \\
  $M_\star>10^9~{\rm M_\sun}$     & All          & 0.74~(0.42) & 3\sigma~(3\sigma) & <0.01 & 0.58~(0.11) & 1\sigma~(1\sigma) & 0.83 \\
                                  & Star-forming & 0.91~(0.58) & 2\sigma~(2\sigma) & <0.01 & 0.27~(0.01) & 1\sigma~(1\sigma) & 0.10 \\
                                  & Quiescent    & 0.42~(0.11) & 1\sigma~(1\sigma) & 0.84  & 0.78~(0.28) & 2\sigma~(1\sigma) & 0.03 \\
  $M_\star>10^{9.5}~{\rm M_\sun}$ & All          & 0.64~(0.29) & 2\sigma~(2\sigma) & 0.07  & 0.74~(0.16) & 1\sigma~(1\sigma) & 0.53 \\
                                  & Star-forming & 0.88~(0.46) & 2\sigma~(2\sigma) & <0.01 & 0.38~(0.02) & 1\sigma~(1\sigma) & 0.28 \\
                                  & Quiescent    & 0.33~(0.05) & 1\sigma~(1\sigma) & 0.14  & 0.91~(0.34) & 2\sigma~(1\sigma) & 0.01 \\                  
  \enddata
  \tablenotetext{a}{Fraction of volume-weighted galaxy sample below the median $\Sigma_{M_\star}$ of TDE hosts, $\text{log}(\Sigma^{\rm TDE}_{\rm med})=9.5$ for both quiescent and star-forming galaxies, or the lower range of TDE host $\Sigma_{M_\star}$ values, $\text{log}(\Sigma^{\rm TDE}_{\rm min})=9$, in parentheses.}
  \tablenotetext{b}{Significance of binomial test for each $\Sigma_{M_\star}$ value given the cuts on stellar mass and galaxy type (all, quiescent, or star-forming).}
  \tablenotetext{c}{$p$-value of Mann-Whitney $U$ test that compares the medians of the TDE host and control $\Sigma_{M_\star}$ distributions.}
  \tablenotetext{d}{Fraction of volume-weighted galaxy sample above the median $\sigma_v$ of TDE hosts, $\text{log}(\sigma^{\rm TDE}_{\rm med})=1.8$ for quiescent and $\text{log}(\sigma^{\rm TDE}_{\rm med})=1.9$ for star-forming galaxies, or the upper range of TDE host $\sigma_v$ values, $\text{log}(\sigma^{\rm TDE}_{\rm max})=2.2$, in parentheses.}
  \tablenotetext{e}{Significance of finding no TDEs above each $\sigma_v$ value given the cuts on stellar mass and galaxy type (all, quiescent, or star-forming).}
  \tablenotetext{f}{$p$-value of Mann-Whitney $U$ test that compares the medians of the TDE host and control $\sigma_v$ distributions.}
\end{deluxetable}

Thus, we do not claim a detection of an offset between the velocity dispersions of TDE hosts and those of the general galaxy population. However, the observation that quiescent TDE hosts have lower velocity dispersions, at least when compared to the peak of the quiescent galaxy control sample, is suggestive, for two reasons: first, more compact galaxies tend to have higher---not lower---velocity dispersions, as shown by \citet{2017ApJ...841...32Z}; and second, theoretical formulations of the TDE rate claim an inverse dependence on velocity dispersion.

Table~\ref{table:subsamples} once more shows that our results do not depend significantly on the choice of TDE type used in the analysis. The significance of the higher density of the TDE hosts drops as the size of the TDE sample decreases, but only drops to $>2\sigma$ when we confine the analysis to three X-ray and likely X-ray TDEs. Likewise, all TDE types have hosts within the same range of velocity dispersion values, but the significance of TDE hosts having a lower median velocity dispersion than the overall control sample drops to $>2\sigma$ earlier, when the sample is limited to $<9$ events.

Figure~\ref{fig:compact} shows that all types of TDEs considered in this work occupy the same $\Sigma_{M_\star}$--$\sigma_v$ phase space. This clustering is also true regardless of whether the hosts of these TDEs are star-forming, quiescent, or quiescent Balmer-strong. 

Depending on the choice of stellar mass cut on the background galaxy sample (no cut, as well as $\text{log}(M_\star/{\rm M_\sun})>8$, $9$, and $9.5$), the phase-space occupied by the TDE hosts in Figure~\ref{fig:compact}, bound by ${\rm log}(\Sigma_{M_\star})>9$ and ${\rm log}(\sigma_v)<2.2$, accounts for 24\%, 31\%, 48\%, and 56\% of the overall volume-weighted galaxy sample. The same binomial test used above shows that the clustering of TDE hosts in this portion of phase space is significant at $5\sigma$, $5\sigma$, $3\sigma$, and $3\sigma$, respectively.   

While performing the analysis described here, \citet{2017MNRAS.471.1694W} published a sample of velocity dispersions for 12 TDEs. Eight of those have SDSS imaging but do not appear in the NYU value-added catalog, in some cases because their spectra are not from the SDSS legacy survey (e.g., TDE1 and D23-H1). We can use these data to test our results that TDE hosts prefer galaxies with high $\Sigma_{M_\star}$ values. To do so, we substitute SDSS $r^{\prime}$-band Petrosian half-light radii for the S\'{e}rsic half-light radii from the NYU catalog,\footnote{On average, the Petrosian $r_{50}$ are larger than the S\'{e}rsic $R_{50}$ by $\sim0.3^{\prime\prime}$, which we take into account in our calculations.} and measure $M_\star$ using {\sc LePhare} and the galaxies' SDSS cmodel magnitudes (Table~\ref{table:compactness}).\footnote{{\sc LePhare} failed to derive a stellar mass for D3-13. Instead, we use the value measured by \citet{2017arXiv170703458V} using {\sc kcorrect} \citep{2007AJ....133..734B}.} To this sample, we add 3XMM J1521, which has a Portsmouth velocity dispersion but no S\'{e}rsic half-light radius. The host-galaxy type is derived either from Table~\ref{table:host_props} or from the TDE discovery papers (quiescent for both D3-13 and iPTF16axa; \citealt{2008ApJ...676..944G,2017ApJ...842...29H}). The star-forming nature of the host of D23-H1 is apparent from its BOSS spectrum. D3-13 and iPTF16axa are not included in Table~\ref{table:host_props} as we do not have their host-galaxy spectra. They were classified as possible X-ray and veiled TDEs, respectively, by A17.

Because the properties of these TDE hosts are derived differently than for the SDSS control sample and for the hosts used in our analysis, we do not include them in our formal statistical analysis. However, as shown in Figure~\ref{fig:compact}, taking them into account would only strengthen our claims, as these nine TDE hosts fall in the same $\Sigma_{M_\star}$--$\sigma_v$ phase space identified earlier. The only TDE host lying outside of this phase-space is that of PS1-10jh, but even its stellar surface mass density, given the uncertainty, is consistent with the rest of the sample.

\subsection{TDE hosts with AGNs}
\label{subsec:agn}

Several of the TDE hosts in our sample are known, or suspected, to harbor AGNs. These galaxies could potentially bias our results for two reasons: First, quiescent galaxies with AGNs that exhibit broad H$\alpha$ emission lines could be mistaken for star-forming galaxies, as our only criterion for such a classification is an H$\alpha$ EW$\geq3~{\rm \AA}$. Second, at a given stellar mass, AGN host galaxies will appear more compact than other galaxies due to the bright AGN in their cores.

As we describe in Section~\ref{subsec:gals}, we remove AGNs from the SDSS control sample by excluding galaxies with {\sc Galspec} BPT classifications of composite, AGN, or low-S/N LINER. \citet{2017ApJ...835..176F} found LINER-like emission in ASASSN-14li as well as most of the veiled TDEs they studied. However, the ionization source was ambiguous, and could be related to evolved stars (e.g., \citealt{2012ApJ...747...61Y}) or merger shocks \citep{2015ApJS..221...28R,2016ApJS..224...38A}.

Our TDE host sample includes five galaxies with strong evidence for AGNs: ASASSN14li (\citealt{2016ApJ...830L..32P}, along with an AGN classification from {\sc Galspec}), F01004 \citep{2017NatAs...1E..61T}, PS16dtm \citep{2017ApJ...843..106B}, SDSS J0952, and SDSS J1350. The last two are classified by {\sc Galspec} as composite and AGN, respectively.

Five more TDE hosts have conflicting evidence of AGN activity, or evidence that the AGN does not dominate the light of the galaxy. PTF09ge is classified by {\sc Galspec} as a low-S/N LINER, but a close inspection of its SDSS spectrum reveals it to be a quiescent galaxy where an AGN, if present, does not dominate over the light of the galaxy. \citet{2014ApJ...793...38A} noted that the X-ray emission and [O III]/H$\beta$ emission-line ratio of the host of PTF09axc was consistent with a very weak AGN. \citet{2013ApJ...763...84B} discovered radio emission from SDSS  J142025.18+533354.9, a galaxy offset from both RX J1420A and RX J1420B, which were identified by \citet{2000A&A...362L..25G} as the potential host galaxies of the RX J1420 flare. \citet{2007AA...462L..49E} note that 2MASX J0249 is classified as a Seyfert $1.9$ galaxy according to its H$\alpha$ line, but that it fails the [N II] $6583$ \AA/H$\alpha$ AGN diagnostic. Ground-based spectra of NGC 5905 identify it as a starburst galaxy \citep{1995ApJS...98..477H,1999A&A...343..775K}; only a \textit{Hubble Space Telescope} $0.1^{\prime\prime}$ STIS slit is narrow enough to filter out the stellar content of the nucleus and reveal low-luminosity Seyfert 2 emission-line ratios \citep{2003ApJ...592...42G}.

In all of the statistical analyses conducted in this work, excluding the TDE hosts listed above has no significant impact on our results. Specifically, of the five TDE hosts listed as star-forming in Table~\ref{table:compactness} and used in the binomial test, two have AGN activity (PS16dtm and SDSS J1350). As noted above, NGC 5905 is a starbursting galaxy on kpc scales, and is treated here as star-forming. The last two galaxies, namely SDSS J0748 and SDSS J1342 are definitively classified as star-forming (by their colors, specific star-formation rates, and BPT diagnostics). The binomial test continues to reject the null hypothesis that the star-forming TDE hosts have similar stellar surface mass densities as the control star-forming galaxies at high significance even when the number of star-forming TDE hosts is reduced to these last three.

\subsection{The case of RX J1420}
\label{subsec:rxj1420}

RX J1420.4+5334, referred to here as RX J1420, was reported by \citet{2000A&A...362L..25G} as an X-ray source that varied by a factor of $\gtrsim150$ in flux between template ROSAT observation and the ROSAT All-Sky Survey in 1990. Following A17, we treat this object as a likely TDE candidate. \citet{2000A&A...362L..25G} identified two possible host galaxies in optical images within the $10^{\prime\prime}$-radius error circle of ROSAT's High Resolution Imager. Following \citet{2000A&A...362L..25G}, we label these hosts RX J1420-A and RX J1420-B. Both galaxies were later targeted by SDSS for spectroscopy, from which their global properties have been measured here. 

Figure~\ref{fig:compact} shows that RX J1420-A has properties consistent with the rest of the TDE host galaxies in our sample, while RX J1420-B is a stark outlier. Thus, we propose that RX J1420-A is the actual host galaxy of this TDE and suggest that the galaxy properties we focus on here, namely stellar surface mass density and velocity dispersion, can be used to identify the host galaxies of other TDEs in similar cases. Appropriately, throughout this work we have excluded RX J1420-B from calculations and statistical tests.


\section{The TDE rate}
\label{sec:rate}

After establishing that TDEs prefer host galaxies with high surface mass densities (most strongly in star-forming galaxies and, at a lesser significance, in quiescent galaxies) and, perhaps, low velocity dispersions (to some extent in quiescent galaxies but not in star-forming galaxies), and based on theoretical assumptions that the TDE rate should have some dependence (usually expressed as a power law) on the density and velocity dispersion of the stars in the loss cone of the SMBH, in this section we describe a statistical model for the TDE occurrence rate as a function of these global host galaxy properties. We formulate this dependence as
\begin{equation}\label{eq:rate}
 R_\text{TDE} \propto \Sigma_{M_\star}^\alpha \times \sigma_v^\beta.
\end{equation}

\subsection{Statistical model}
\label{subsec:model}

Let $N(\bm{y} | \mu, \Sigma)$ generically indicate a multivariate Gaussian probability density in $y$ with mean $\bm{\mu}$ and covariance $\bm{\Sigma}$. The model consists of the following ingredients.
\begin{enumerate}
 \item Let $\bm{x} = (\log_{10} \Sigma_{M_\star}, \log_{10} \sigma_v)^T$ be the vector of the latent (true, underlying) log values of the stellar mass density and the velocity dispersion of a galaxy. Let $\bm{\hat{x}}$ be the estimated values of these quantities, which differ from the latent values $\bm{x}$ by measurement error. We assume Gaussian errors,\footnote{This is an acceptable assumption, given that the errorbars on $\Sigma_{M_\star}$ and $\sigma_v$ reported in Table~\ref{table:compactness} are mostly symmetric.} so the measurement likelihood function is $P( \bm{\hat{x}} |\, \bm{x}) = N( \bm{\hat{x}} |\, \bm{x}, \bm{W})$. For simplicity, we assume the measurement covariance matrix $\bm{W}$ is diagonal with known variances.
 
 \item Using a large sample of background galaxies (that did not host TDEs), we estimate the density $f_\text{gal}(\bm{x})$ of galaxies in the $\bm{x}$-plane. We do this non-parametrically by applying 2D kernel density estimation (KDE) to the volume-weighted sample of background galaxies described in Section~\ref{subsec:gals}. Because the measurement errors are small relative to the intrinsic scatter of the background galaxies in the $\bm{x}$-plane, they can be ignored in this step when constructing the KDE for $f_\text{gal}(\bm{x})$.
 
 \item We assume the enhancement of the TDE occurrence rate is a power-law function of the galaxy properties $\bm{x}$:
 \begin{equation}
  \begin{split}
   R_\text{TDE}(\bm{x} | \alpha, \beta) &\propto \Sigma_{M_\star}^{\alpha} \times \sigma_{v}^{\beta} \\
   &\propto 10^{\alpha  \log_{10} \Sigma_{M_\star}} \times 10^{\beta \log_{10} \sigma_{v}}
  \end{split}
 \end{equation}
 where $\alpha, \beta$ are the parameters of interest we want to estimate.
\end{enumerate}
Given these ingredients, the probability density of a TDE occurring in a host galaxy with measured properties $\bm{\hat{x}}$ is proportional to the product of the background galaxy density, the enhancement function, and the measurement likelihood function,
\begin{equation}\label{eqn:lkhd}
P( \bm{\hat{x}} | \, \alpha, \beta) = k(\alpha, \beta) \int P(\bm{\hat{x}} | \bm{x}) \, R_\text{TDE}( \bm{x} | \alpha, \beta) \, f_\text{gal}(\bm{x}) \, d\bm{x}
\end{equation}
with the latent galaxy properties $\bm{x}$ integrated out. The normalization factor is determined by requiring the probability to integrate to one, $\int P( \bm{\hat{x}} | \alpha, \beta) \, d\bm{\hat{x}} = 1$, and is a function of the parameters $\alpha, \beta$:
\begin{equation}
k^{-1}(\alpha, \beta) = \int R_\text{TDE}(\bm{x} | \alpha, \beta) f_\text{gal}(\bm{x}) \, d\bm{x}.
\end{equation}
Now, we can construct the likelihood function of the enhancement parameters $\alpha, \beta$, given measurements of the host properties $\mathcal{D} = \{ \bm{\hat{x}}_i \}$ of $N$ observed TDEs:
\begin{equation}
P(\mathcal{D} | \, \alpha, \beta) = \prod_{i=1}^N P(\bm{\hat{x}}_i | \, \alpha, \beta)
\end{equation}
With flat priors  $P(\alpha, \beta) \propto 1$,\footnote{Where $\alpha, \beta$ are evaluated on a wide grid of values spanning $[-5, 5]\times[-5, 5]$.} the posterior density of the parameters is proportional to this likelihood:
\begin{equation}
P( \alpha, \beta | \, \mathcal{D}) \propto P(\mathcal{D} | \, \alpha, \beta) P(\alpha, \beta) \propto  P(\mathcal{D} | \, \alpha, \beta).
\end{equation}

\subsection{Parameter estimation}
\label{subsec:estimate}

\begin{figure}
 \includegraphics[width=0.47\textwidth]{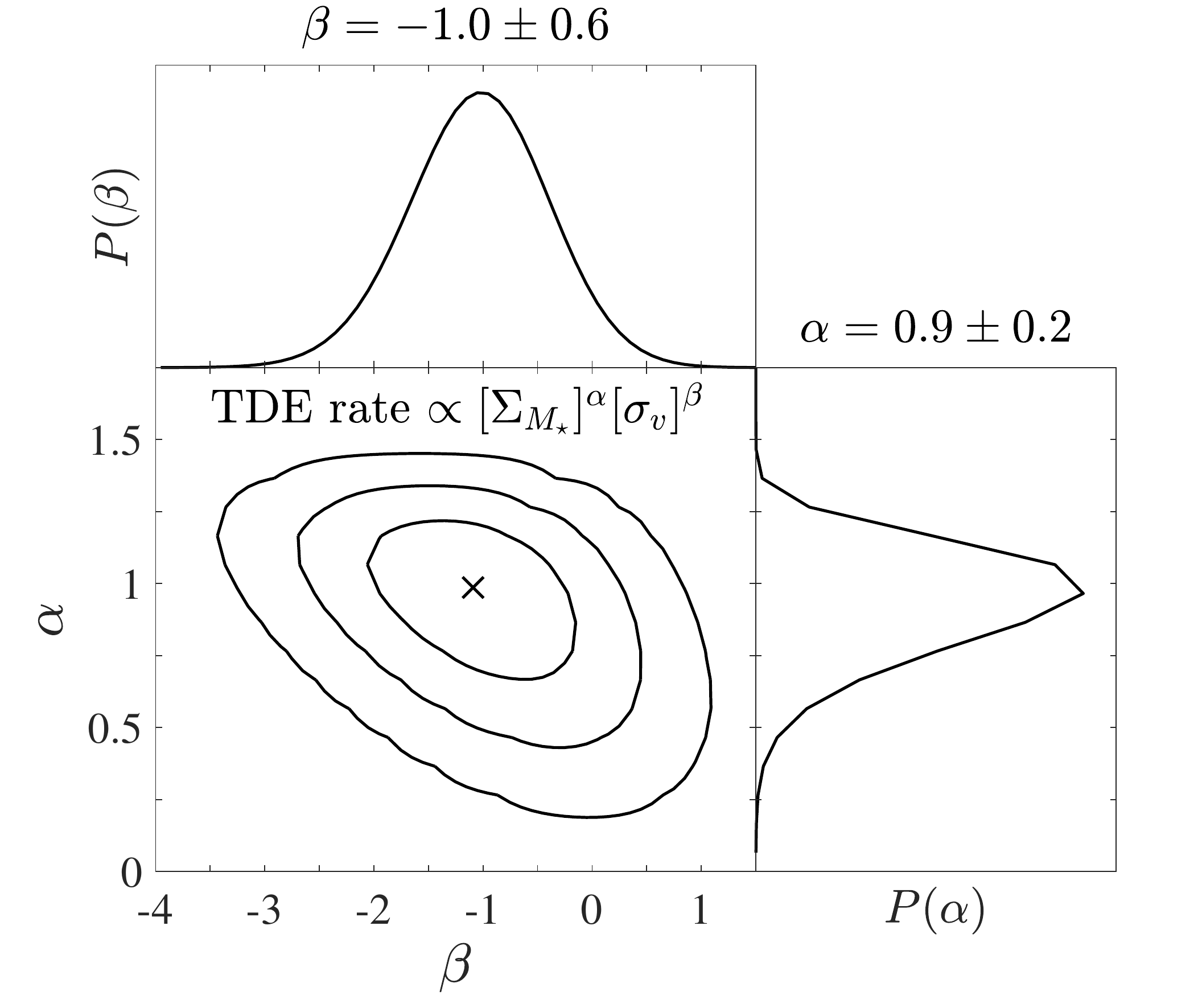}
 \caption{Maximum-likelihood estimates of the exponents $\alpha$ and $\beta$ in the empirical formulation of the TDE rates as $R_{\rm TDE} \propto \Sigma_{M_\star}^\alpha \times \sigma_v^\beta$. The contours in the bottom-left plot contain approximately 68\%, 95\%, and 99.7\% of the highest posterior probability. The bottom-right and upper-left panels show the marginal posteriors of $\alpha$ and $\beta$, respectively. The maximum-likelihood estimate of $\alpha,\beta$ (marked with a cross) is $\hat{\alpha}=0.9\pm0.2$ and $\hat{\beta}=-1.0\pm0.6$. This implies a direct dependence of the TDE rate on global stellar surface mass density and suggests an inverse dependence on the global stellar velocity dispersion.}
 \label{fig:mle}
\end{figure}

We apply our statistical model to the subsample of 10 TDE host galaxies with known $\Sigma_{M_\star}$ and $\sigma_v$ values shown in Figure~\ref{fig:compact} (excluding the outlier RX J1420-B). Our first step is to find the maximum likelihood estimates (MLE) for $\alpha, \beta$ by optimizing the log of this likelihood function. Next, we evaluate this likelihood function on a 2D grid in $\alpha, \beta$ centered on the MLE. We show the likelihood contours in Figure~\ref{fig:mle}. We display the contours containing approximately 68\%, 95\%, and 99.7\% of the highest posterior density as the black curves.  Finally, we marginalize in each direction to obtain marginal posterior densities of each parameter $P(\alpha |\, \mathcal{D})$ and $P(\beta | \, \mathcal{D})$. We compute the posterior mean and standard deviations of each parameter $\hat{\alpha} = 0.9 \pm 0.2$ and  $\hat{\beta} = -1.0 \pm 0.6$. As expected from Section~\ref{subsec:compactness}, we find a significant dependence on surface mass density and a hint of an inverse dependence on velocity dispersion. As more TDEs are discovered and their host galaxies analyzed, our statistical model will be useful for further testing the significance of these trends.

To test the sensitivity of these parameter estimates to the composition of the TDE host sample, we bootstrap resample the TDE hosts. For each bootstrap sample, we use the above procedure to find the parameter estimates. We examine the distribution over the bootstrap samples and find that the mean and standard deviations are $\hat{\alpha}_\text{boot} = 0.88 \pm 0.06$ and $\hat{\beta}_\text{boot} = -0.9 \pm 0.4$. These estimates are consistent with the likelihood analysis above, although the dispersions underestimate the uncertainties. Conservatively, we take the posterior standard deviations of the original sample as our final uncertainties.

In practice, the impact of the measurement uncertainties of $\bm{\hat{x}}$ on the parameter estimates is quite small. This is because the typical uncertainty on $\log_{10}(\Sigma_{M_\star})$ is $\sim 0.20$ and on $\log_{10}(\sigma_v)$ is $\lesssim 0.10$. Since the product of the galaxy density $f_\text{gal}(\bm{x})$ and enhancement function $R_\text{TDE}(\bm{x})$ is smoothly and slowly varying on the scale of the measurement error, it can be approximated as a constant under the integral in Eq.~\ref{eqn:lkhd}, or equivalently, $P(\bm{\hat{x}} | \bm{x})$ can be set equal to a delta function.

As in previous sections, Table~\ref{table:subsamples} shows that the choice of TDE subsample has no effect on the results of our analysis beyond enlarging the uncertainties of the fitted parameters as the size of the samples decrease. Importantly, the direct dependence on surface mass density remains significant. Likewise, applying the stellar mass cuts from Table~\ref{table:overdensity_mass} to the control sample, or limiting the redshift range to $0.01<z<0.1$, does not have a significant impact on $\alpha,\beta$.

In this analysis, we have restricted the TDE host-galaxy sample to those with S\'{e}rsic half-light radii (the top half of Table~\ref{table:compactness}), so that we could compare them directly to the control sample. Adding the nine TDE hosts with Petrosian half-light radii, the majority of which have velocity dispersions measured by \citet{2017MNRAS.471.1694W} instead of the Portsmouth pipeline, results in consistent estimates of $\hat{\alpha} = 0.86 \pm 0.15$ and $\hat{\beta} = -0.8 \pm 0.5$. PS1-10jh, which has a lower $\Sigma_{M_\star}$ value than the rest of the sample, has very little effect on these estimates. Removing it from the sample results in $\hat{\alpha} = 0.89 \pm 0.15$ and $\hat{\beta} = -0.8 \pm 0.5$.

When the TDE and control samples are split between star-forming (three TDE hosts) and quiescent (seven hosts) galaxies, the estimates of $\alpha$ and $\beta$ remain consistent, though with larger statistical uncertainties: $\hat{\alpha}=1.0 \pm 0.3$, $\hat{\beta}=0.5 \pm 1.2$ for star-forming galaxies and $\hat{\alpha}=0.7 \pm 0.3$, $\hat{\beta}=-1.8 \pm 0.8$. As expected from Section~\ref{subsec:compactness}, the dependence on $\Sigma_{M_\star}$ remains significant, while the suggestion of an inverse dependence on $\sigma_v$ remains statistically insignificantly different from zero and is driven by the quiescent TDE hosts.

\section{Discussion}
\label{sec:discuss}

\subsection{Quiescent Balmer strong galaxies and $\Sigma_{M_\star}$}
\label{subsec:PSGdensity}

The $\Sigma_{M_\star}$--$\sigma_v$ phase-space occupied by TDE hosts in Figure~\ref{fig:compact} includes $73$\% of all quiescent Balmer-strong galaxies and $36$\% of the quiescent moderately Balmer-strong galaxies. It is thus tempting to assume that the overabundance of these galaxies among TDE hosts stems from a shared predilection for the high stellar density observed in all TDE hosts. However, the overabundance of high-$\Sigma_{M_\star}$ galaxies among TDE hosts, assuming that they account for only $12$--$30$\% of the galaxy population,\footnote{If we consider the full galaxy sample and the sample with $\text{log}(M_\star/{\rm M_\sun})>8$ in Table~\ref{table:overdensity_mass}.} is a factor of $\sim 3$--$8$, while in this work we find that quiescent Balmer-strong galaxies are overabundant among TDE hosts by factors of at least 10--30. 

By this calculation, it is possible that another physical process, such as the production of a compact SMBH binary in the wake of a merger event (e.g., \citealt{2009ApJ...697L.149C,2011ApJ...729...13C}) is also at play. However, this calculation tacitly assumes that the local stellar surface mass density in the SMBH loss cone is linearly correlated with the global value. If, on the other hand, galaxies with the same global $\Sigma_{M_\star}$ value have a range of sub-pc stellar surface mass densities, and TDEs preferentially occur in galaxies with local densities at the top of this range, our calculation might only represent a lower limit on the effect of stellar density on the TDE rate.

\subsection{$\Sigma_{M_\star}$ and galaxy quenching}
\label{subsec:Sigma1}

There is a growing body of work that posits that before becoming quiescent, star-forming galaxies go through a phase of compaction (e.g., \citealt{2012ApJ...760..131C,2013ApJ...776...63F,2015MNRAS.450.2327Z,2016MNRAS.458..242T,2017ApJ...840...47B}). Interestingly, the critical stellar surface mass density at which this process occurs lies in the range $\text{log}(\Sigma_1)=9$--$9.4$ \citep{2015MNRAS.448..237W}, where $\Sigma_1$ is the stellar surface mass density measured within a radius of 1 kpc around the center of the galaxy. Even though in this work we have measured $\Sigma_{M_\star}$ using either the S\'{e}rsic or Petrosian $r^\prime$-band half-light radii (not strictly within 1 kpc of the core), the median $\text{log}(\Sigma_{M_\star})$ of our TDE hosts, $9.5^{+0.4}_{-0.1}$, is consistent with the above range.

While the physical drivers that change morphology alongside the star-forming properties of galaxies might be different with redshift, and morphological change may not always precede an end to star formation, it is interesting to note that the TDE hosts so far have been either already quiescent galaxies, or star-forming galaxies with unusually high stellar surface densities.

If the TDE rate only depends on the dynamical relaxation of stars around the SMBH, and the effects seen here in the global properties of the TDE hosts are due to a correlation with the smaller-scale stellar properties closest to the SMBH, we should expect that as the TDE sample continues to grow we will begin to discover TDEs in hosts with $\text{log}(\Sigma_{M_\star})<9.5$. If, however, we continue to find that TDE hosts always prefer galaxies with globally-high stellar surface mass densities, this might signal a connection between the process of morphological change in a galaxy and the TDE rate that goes beyond correlations to the conditions nearest to the SMBH.

\subsection{Comparison to other works}
\label{subsec:comp}

\citet{2017ApJ...850...22L}, which appeared on arXiv several days before this work, conducted an independent and complementary analysis of TDE host-galaxy properties. This work considered a smaller sample of TDE host galaxies, which was also matched to the SDSS, but without accounting for the survey's magnitude-limited nature, as we do here.

As in our work, \citet{2017ApJ...850...22L} find that quiescent Balmer-strong galaxies are overrepresented among TDE hosts, though at a formally lower value than found by F16. The overabundance factors they find, based on their smaller sample, are similar to ours. 

Importantly, \citet{2017ApJ...850...22L} concentrate on a complementary set of galaxy properties to those considered here. They find that TDE hosts have high galaxy S\'{e}rsic indices and high bulge-to-total-light ratios, indicating that the cores of TDE hosts are denser than those of other galaxies. This is similar to our finding that TDEs prefer galaxies with high stellar surface mass densities. One difference is that we control for the type of galaxy in which the TDE is found (quiescent or star-forming). As we show above, the determination that TDE hosts have higher stellar densities is driven principally by the star-forming hosts, whereas the stellar densities of the quiescent hosts are more consistent with those of the general quiescent population.

Furthermore, although both works adhere to the A17 classification scheme, we control our analysis for different subsamples of TDE types. We have shown that our results are independent of TDE type.

Finally, we put forward a statistical model that empirically relates the TDE rate to the global galaxy properties we study here, $\Sigma_{M_\star}$ and $\sigma_v$. 

\citet{2017arXiv170703458V} used a subsample of the TDE hosts studied by \citet{2017MNRAS.471.1694W} to reconstruct the TDE luminosity and black-hole mass functions. Using a control sample of synthetic galaxies from the NYU value-added catalog, he shows that, if the TDE rate depends only on the mass of the SMBH, $R_\text{TDE}\propto M_\bullet^\gamma$, then $\gamma$ needs lie in the range $-0.5$ to $0.3$. The empirical rate we derive here is consistent with his measurements, and falls within this range.

\subsection{Comparison to theory}
\label{subsec:theory}

Assuming that the global galaxy properties we use here are directly correlated with the local properties in the vicinity of the SMBH, the values of $\alpha, \beta$ quoted in Section~\ref{sec:rate} imply that the TDE rate is directly proportional to the density---and inversely proportional to the velocity dispersion---of the stars in the loss cone of the SMBH. Based on our TDE sample, only the first of these correlations is statistically significant; a larger sample of TDE host galaxies with both $\Sigma_{M_\star}$ and $\sigma_v$ measurements is required to properly test the second assertion. 

To estimate the rate of disruption, one needs to know the number of stars that orbit the SMBH and the velocity at which they encounter one another. These two quantities are related to each other at the SMBH's sphere of influence, the distance within which the cluster of stars orbiting the SMBH has a mass comparable to the SMBH ($M_\star(r<r_{\rm h})=2M_\bullet$), with a size $r_{\rm h} = G M_\bullet / \sigma^{2}$ \citep{1972ApJ...178..371P}. Its average surface density is then $\Sigma = M_\bullet / \pi r_{\rm h}^{2} = \sigma^{4} / \pi G^{2} M_\bullet$, where $\Sigma$ and $\sigma$ denote the \emph{local} stellar surface mass density and velocity dispersion of the stars surrounding the SMBH. The velocity dispersion, as a property of the entire galaxy, has been found empirically to be correlated with $M_\bullet$, and this allows us to relate $\Sigma$ and $\sigma$.

With all other variables fixed, two-body relaxation yields a disruption rate that scales inversely with $M_\bullet$ and $\sigma^{\eta}$, with the exact value of $\eta$ depending on the cluster's radial density profile but $\eta = 3$ being typical \citep{1999MNRAS.309..447M,2004ApJ...600..149W,2016MNRAS.455..859S}. This suggests that perhaps \emph{higher} velocity dispersions should lead to greater TDE rates. However, if one accounts for the fact that the SMBH mass is itself found to be a strong function of $\sigma$, the net dependence can be inversely proportional to $\sigma$. For different sub-samples of galaxies, a range of $M_\bullet$--$\sigma$ relations $M_\bullet \propto \sigma^{\delta}$ are possible \citep{2013ApJ...764..184M}, with $\Sigma \propto \sigma^{4 - \delta} \propto \sigma^{0}$ for $\delta = 4$ (no dependence) and $\Sigma \propto \sigma^{-2}$ for $\delta = 6$.

Assuming the disruption rate is $R_\text{TDE} \propto M_\bullet^{-\alpha} \sigma^{\eta}$ \citep{2004ApJ...600..149W}, we can now rewrite it in terms of $\Sigma$ and $\sigma$ as $R_\text{TDE} \propto \Sigma^{\alpha} \times \sigma^{\eta - 4 \alpha}$. If we take our nominal estimates for $\alpha$ and $\beta$ ($= \eta - 4 \alpha$) and propagate their uncertainties, we find that $\eta = 2.6 \pm 1.0$ and $\alpha = 0.9 \pm 0.2$, both of which are broadly consistent with the \citet{2004ApJ...600..149W} predictions that $\eta = 3.5$ and $\alpha = 1$. This consistency is suggestive that the enhanced TDE rate is driven by a faster dynamical relaxation of the stars surrounding the SMBH, which may be simply a consequence of the black holes being lower in mass (as suggested by the low $\sigma_v$ values measured by \citealt{2017MNRAS.471.1694W} and hinted at in Section~\ref{subsec:compactness}).


\section{Conclusions}
\label{sec:summary}

The rate of tidal disruption events (TDEs) is predicted to depend on the properties of the stars near the SMBH, which are on sub-pc scales and so rarely measurable. Here, we test whether the TDE rate depends on global galaxy properties, which are on kpc scales and directly observable. We concentrate on global stellar surface mass density, $\Sigma_{M_\star}$, and stellar velocity dispersion, $\sigma_v$, which correlate with galaxy properties on small scales. We further test the overabundance of quiescent Balmer-strong galaxies (many of which have post-starburst, aka ``E+A'', spectra) among the host galaxies of TDEs, as first reported by \citet{2014ApJ...793...38A} and F16. 

We assemble a sample of 37 host-galaxies for 35 TDEs identified primarily from X-ray or UV/optical imaging and classified as bona-fide candidates by A17. The host galaxies range in stellar mass from $\text{log}(M_\star/{\rm M_\sun})=8.5$ to $11$ and in redshift from $0.01$ to $0.4$. We measure the strength of the H$\alpha$ and H$\delta$ lines in archival spectra of the host galaxies and compare to similar values measured by the Sloan Digital Sky Survey (SDSS) Galspec pipeline for a volume-weighted control sample. For ten of these hosts, we use SDSS value-added catalogs to measure homogeneously their $\Sigma_{M_\star}$ and $\sigma_v$ values and compare these properties with the control sample. 

Our findings are:

\begin{enumerate}
 \item TDEs are found in star-forming, quiescent, and quiescent Balmer-strong galaxies.
 
 \item Although the TDEs in our sample are found in a range of galaxy types, four (or eight) of the 35 TDEs lie in rare quiescent, Balmer-strong galaxies, depending on the strength of the H$\delta$ absorption line. When compared to their fractions in a volume-weighted SDSS control sample, the quiescent Balmer-strong hosts are overrepresented among the TDE host galaxies by a factor of $35^{+21}_{-17}$ (or $18^{+8}_{-7}$). This overrepresentation is lower than that found by F16, but formally consistent within the uncertainties. This overabundance does not depend on the TDE class; the values are similar for TDEs with and without known X-ray emission.
 
 \item For the TDE hosts with homogeneous measurements, $\Sigma_{M_\star}$ ranges over $10^9$--$10^{10}~{\rm M_\sun / kpc^2}$ and $\sigma_v$ ranges over $\sim40$--$140~{\rm km~s^{-1}}$. These stellar surface mass densities are higher on average than the volume-weighted control sample of SDSS galaxies with similar redshifts and stellar masses. This difference arises for two reasons: (1) most of the TDE hosts in this subsample are quiescent galaxies, which tend to have higher $\Sigma_{M_\star}$ values than the star-forming galaxies that dominate the SDSS control sample, and (2) the star-forming TDE hosts have higher average $\Sigma_{M_\star}$ values than the SDSS star-forming control galaxies. There is also a (statistically insignificant) suggestion that quiescent TDE hosts have lower velocity dispersions than the control quiescent galaxy sample.
 
 \item The higher-than-normal global stellar density of star-forming TDE hosts (and perhaps of quiescent TDE hosts as well) and the suggestion of lower global velocity dispersions (at least in quiescent TDE hosts) suggests that these global properties can act as proxies for the sub-pc-scale properties of the loss cone surrounding the SMBH. Following theoretical predictions that the TDE rate should depend on the density and velocity dispersion of the stars in the loss cone, we suggest an empirical formulation, $R_\text{TDE} \propto \Sigma_{M_\star}^\alpha \times \sigma_v^\beta$. Applying a statistical model to the TDE hosts and the SDSS volume-weighted control sample, we estimate $\hat{\alpha}=0.9\pm0.2$ and $\hat{\beta}=-1.0\pm0.6$. This significant, roughly linear dependence on $\Sigma_{M_\star}$, coupled with a suggestion of an inverse, linear (but not statistically significant) dependence on $\sigma_v$, is broadly consistent with the TDE rate being tied to the dynamical relaxation of stars around the SMBH.
\end{enumerate}

While a larger sample of TDEs is required to further test this picture, our work here suggests that some global properties of TDE host galaxies may be used to constrain stellar properties in the vicinity of the SMBH. Furthermore, by separating our sample of TDE hosts into star-forming and quiescent galaxies, we make it possible for future theoretical works to undertake more detailed comparisons between the theoretical and observed TDE rates.


\section*{Acknowledgments}

We thank Brad Cenko, Ryan Chornock, Suvi Gezari, Tom W. Holoien, Andrew Levan, Paulina Lira, Peter Maksym, Richard Saxton, and Thomas Wevers for sharing their data with us. We also thank Iair Arcavi, Edo Berger, Daniel Eisenstein, Jose L. Prieto, Nicholas Stone, Sandro Tacchella, Benny Trakhtenbrot, Sjoert van Velzen, and the anonymous referee for helpful discussions and comments.

O.G. is supported by an NSF Astronomy and Astrophysics Fellowship under award AST-1602595. KDF is supported by Hubble Fellowship Grant HST-HF2-51391.001-A, provided by NASA through a grant from the Space Telescope Science Institute, which is operated by the Association of Universities for Research in Astronomy, Incorporated, under NASA contract NAS5-26555. K.S.M. was supported at Harvard by NSF grants AST-1211196 and AST-1516854. A.I.Z. acknowledges funding from NSF grant AST-0908280 and NASA grant ADP-NNX10AE88G. 

This research has made use of NASA's Astrophysics Data System and the NASA/IPAC Extragalactic Database (NED), which is operated by the Jet Propulsion Laboratory, California Institute of Technology, under contract with NASA, as well as the \textit{Open TDE Catalog} (\url{https://tde.space/}), which was produced using {\it AstroCats}.

Funding for the SDSS and SDSS-II has been provided by the Alfred P. Sloan Foundation, the Participating Institutions, the National Science Foundation, the U.S. Department of Energy, the National Aeronautics and Space Administration, the Japanese Monbukagakusho, the Max Planck Society, and the Higher Education Funding Council for England. The SDSS Web Site is \url{http://www.sdss.org/}.

The SDSS is managed by the Astrophysical Research Consortium for the Participating Institutions. The Participating Institutions are the American Museum of Natural History, Astrophysical Institute Potsdam, University of Basel, University of Cambridge, Case Western Reserve University, University of Chicago, Drexel University, Fermilab, the Institute for Advanced Study, the Japan Participation Group, Johns Hopkins University, the Joint Institute for Nuclear Astrophysics, the Kavli Institute for Particle Astrophysics and Cosmology, the Korean Scientist Group, the Chinese Academy of Sciences (LAMOST), Los Alamos National Laboratory, the Max-Planck-Institute for Astronomy (MPIA), the Max-Planck-Institute for Astrophysics (MPA), New Mexico State University, Ohio State University, University of Pittsburgh, University of Portsmouth, Princeton University, the United States Naval Observatory, and the University of Washington.

Funding for SDSS-III has been provided by the Alfred P. Sloan Foundation, the Participating Institutions, the National Science Foundation, and the U.S. Department of Energy Office of Science. The SDSS-III web site is \url{http://www.sdss3.org/}.

SDSS-III is managed by the Astrophysical Research Consortium for the Participating Institutions of the SDSS-III Collaboration including the University of Arizona, the Brazilian Participation Group, Brookhaven National Laboratory, Carnegie Mellon University, University of Florida, the French Participation Group, the German Participation Group, Harvard University, the Instituto de Astrofisica de Canarias, the Michigan State/Notre Dame/JINA Participation Group, Johns Hopkins University, Lawrence Berkeley National Laboratory, Max Planck Institute for Astrophysics, Max Planck Institute for Extraterrestrial Physics, New Mexico State University, New York University, Ohio State University, Pennsylvania State University, University of Portsmouth, Princeton University, the Spanish Participation Group, University of Tokyo, University of Utah, Vanderbilt University, University of Virginia, University of Washington, and Yale University.


\software{Matlab,WebPlotDigitizer}


\appendix

\section{Stellar mass cuts and the surface mass density of TDE hosts}
\label{sec:appendixA}

\begin{figure*}
 \centering
 \begin{tabular}{cc}
  \includegraphics[width=0.47\textwidth]{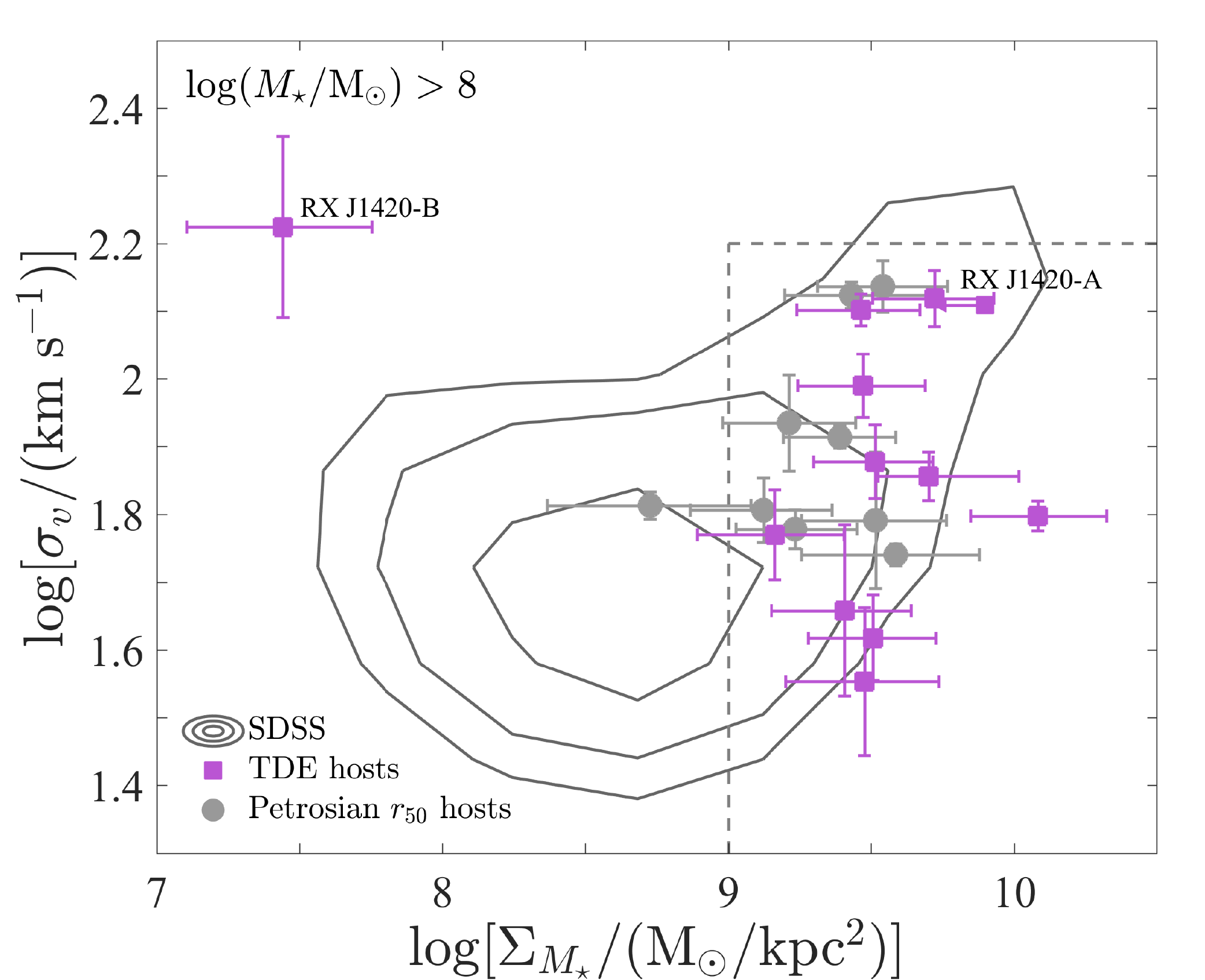} & \includegraphics[width=0.47\textwidth]{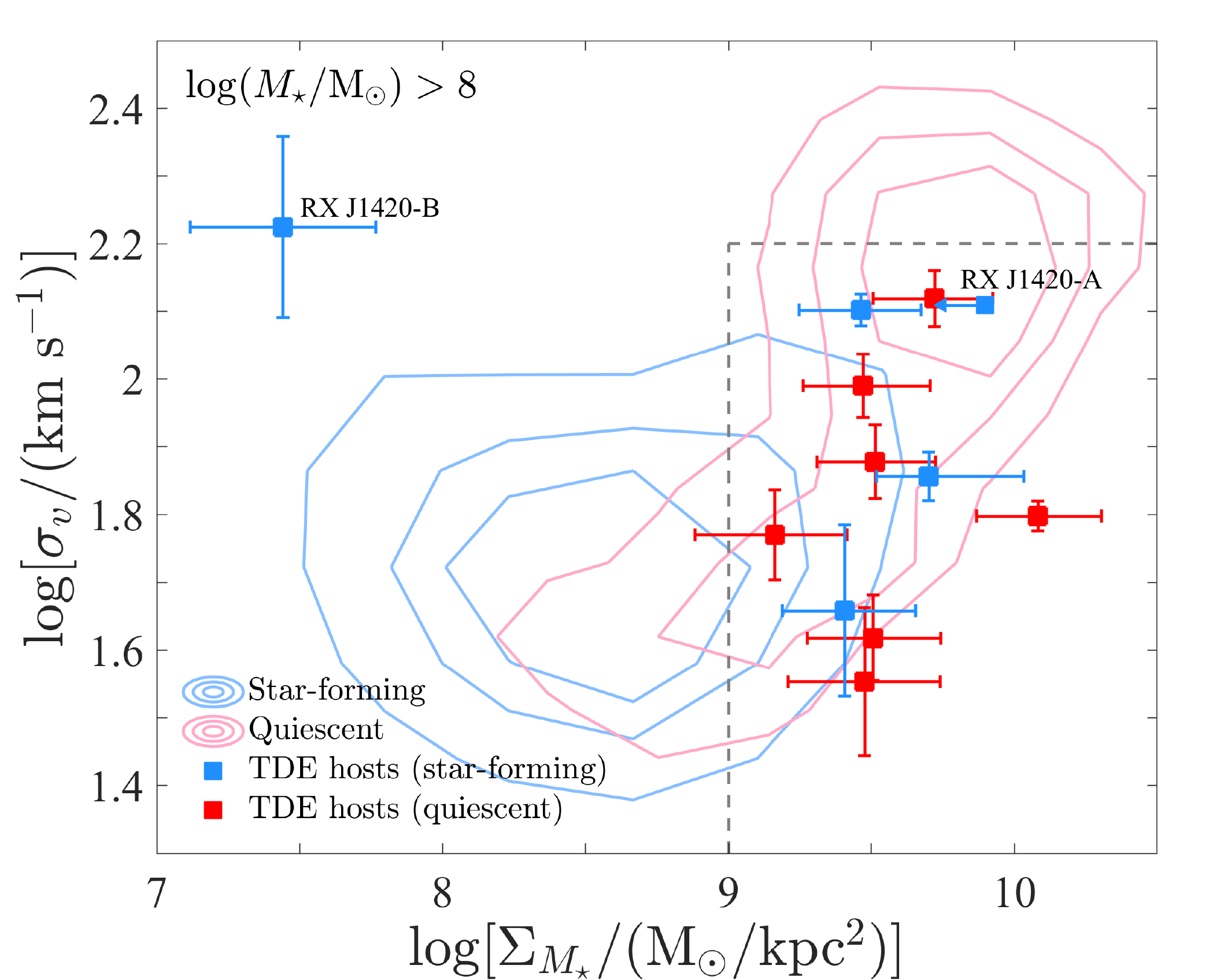} \\
  \includegraphics[width=0.47\textwidth]{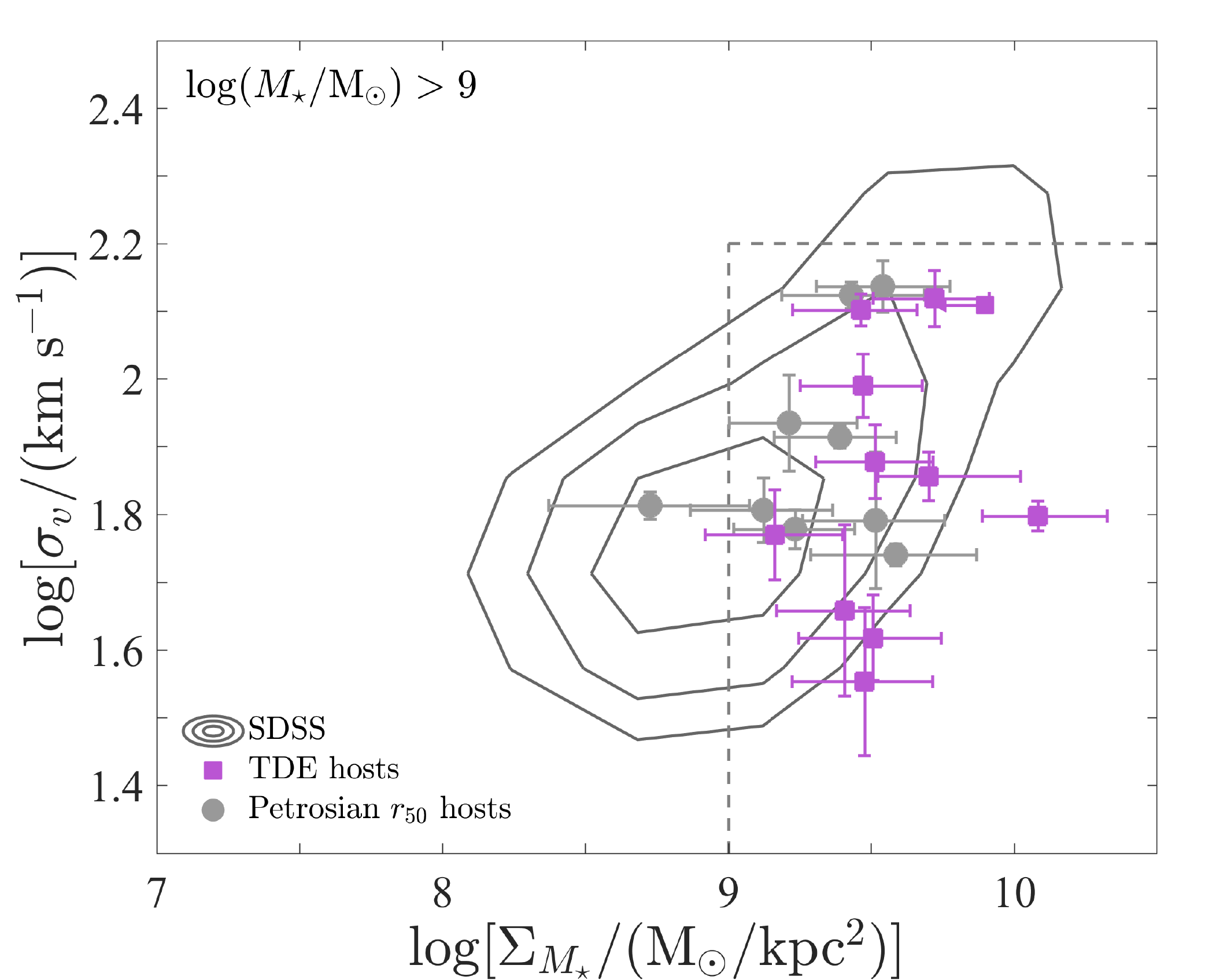} & \includegraphics[width=0.47\textwidth]{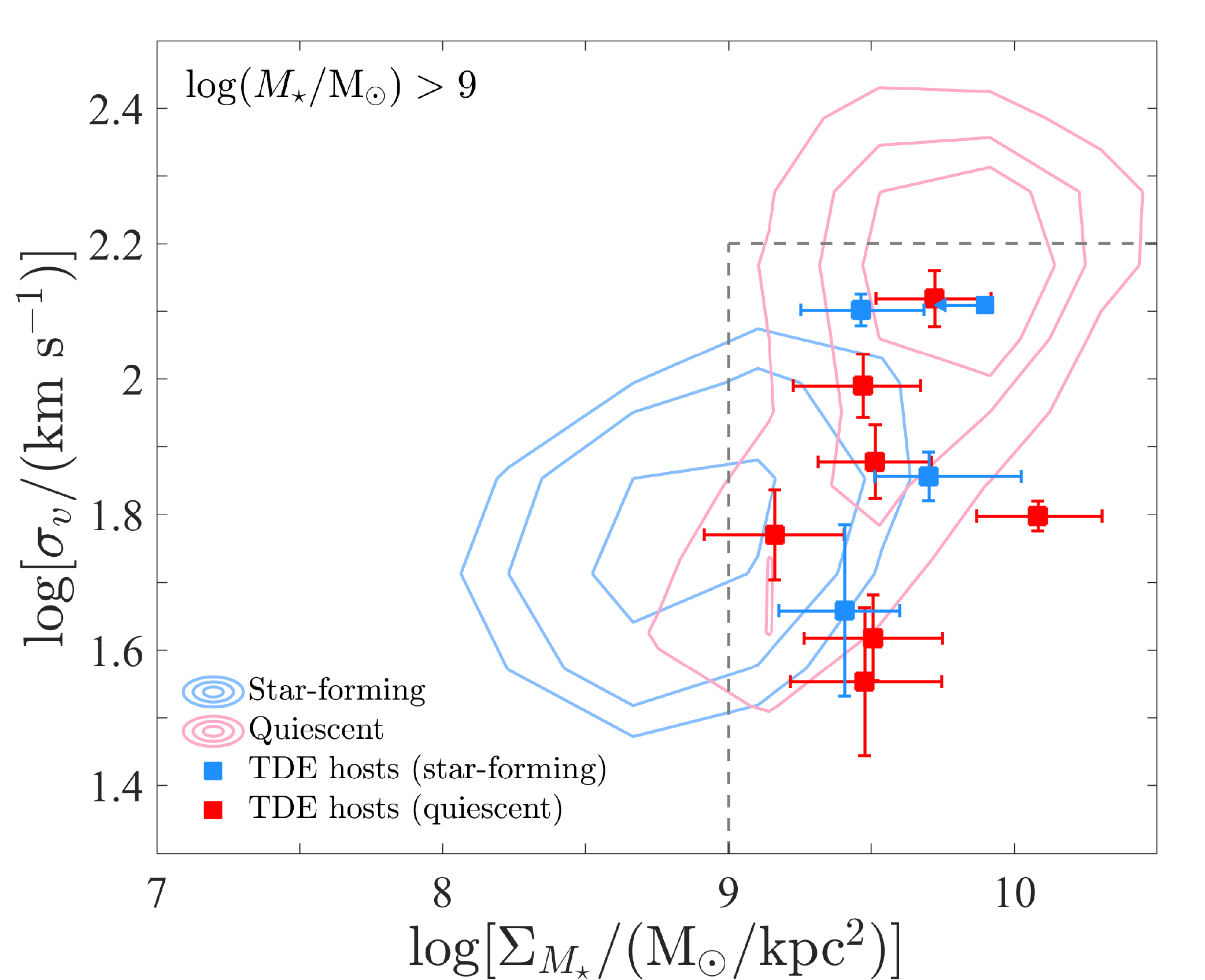} \\
  \includegraphics[width=0.47\textwidth]{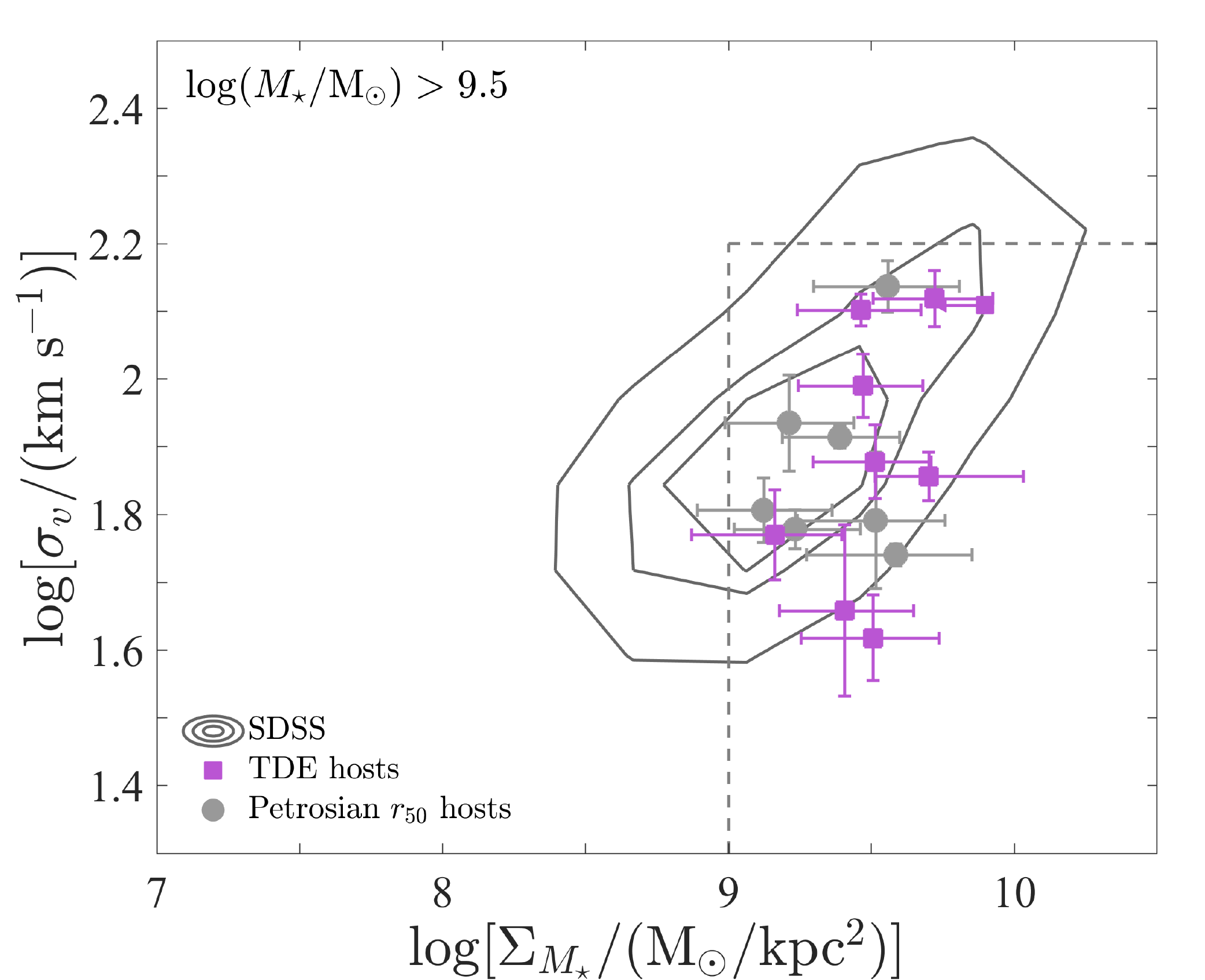} & \includegraphics[width=0.47\textwidth]{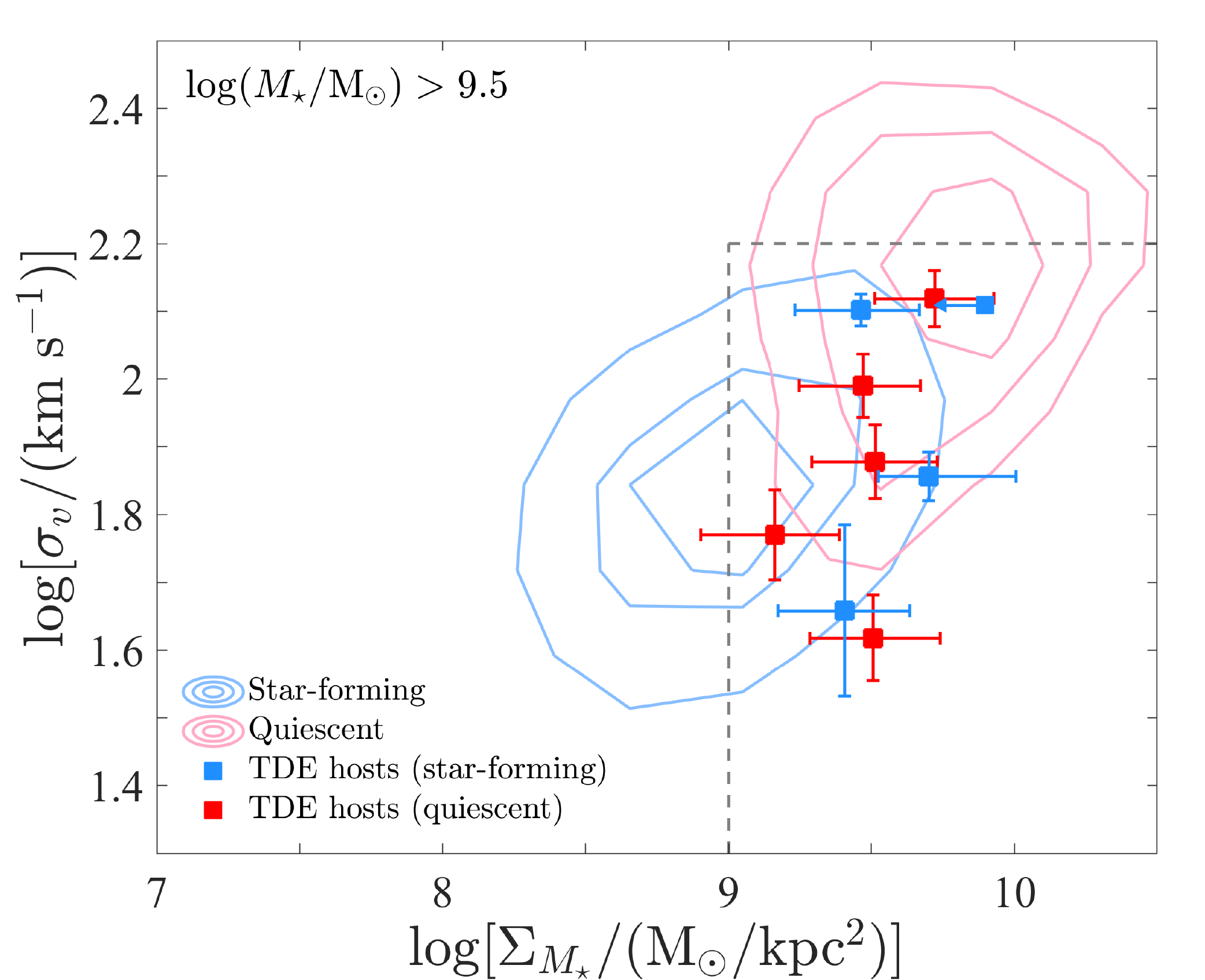} \\
 \end{tabular}
 \caption{Stellar velocity dispersion ($\sigma_v$) vs. stellar surface mass density ($\Sigma_{M_\star}$) of TDE host galaxies with SDSS spectroscopy. Contours and symbols as in Figure~\ref{fig:compact}. The different rows show the effect of limiting the mass range of the background galaxy sample. As detailed in Table~\ref{table:overdensity_mass}, whether we require $\text{log}(M_\star/{\rm M_\sun})>8$ (upper row), $\text{log}(M_\star/{\rm M_\sun})>9$ (center row), or $\text{log}(M_\star/{\rm M_\sun})>9.5$ (bottom row), we find that TDE hosts have significantly higher stellar surface mass densities than the general galaxy population. This trend remains significant in star-forming galaxies, but not in quiescent ones.}
 \label{fig:appendix}
\end{figure*}

In Section~\ref{subsec:compactness}, we describe how limiting the volume-weighted galaxy sample to galaxies with stellar masses $\text{log}(M_\star/{\rm M_\sun})>8$ or $>9$ has no effect on the significance of the higher stellar surface mass density of TDE host galaxies. Here, we show how these stellar mass cuts affect the background distribution of $\Sigma_{M_\star}$ and $\sigma_v$ shown in Figure~\ref{fig:compact}. As expected, limiting the stellar mass range of the control sample shifts it to higher $\Sigma_{M_\star}$ values. This shift, however, is not strong enough to invalidate the significance of the higher stellar surface mass density of the TDE host galaxies.



\begin{thebibliography}{}
\expandafter\ifx\csname natexlab\endcsname\relax\def\natexlab#1{#1}\fi

\bibitem[{{Alam} {et~al.}(2015){Alam}, {Albareti}, {Allende Prieto}, {Anders},
  {Anderson}, {Anderton}, {Andrews}, {Armengaud}, {Aubourg}, {Bailey}, \&
  et~al.}]{2015ApJS..219...12A}
{Alam}, S., {Albareti}, F.~D., {Allende Prieto}, C., {et~al.} 2015, \apjs, 219,
  12

\bibitem[{{Alatalo} {et~al.}(2016){Alatalo}, {Cales}, {Rich}, {Appleton},
  {Kewley}, {Lacy}, {Lanz}, {Medling}, \& {Nyland}}]{2016ApJS..224...38A}
{Alatalo}, K., {Cales}, S.~L., {Rich}, J.~A., {et~al.} 2016, \apjs, 224, 38

\bibitem[{{Alexander}(2012)}]{2012EPJWC..3905001A}
{Alexander}, T. 2012, in European Physical Journal Web of Conferences, Vol.~39,
  European Physical Journal Web of Conferences, 05001

\bibitem[{{Arcavi} {et~al.}(2014){Arcavi}, {Gal-Yam}, {Sullivan}, {Pan},
  {Cenko}, {Horesh}, {Ofek}, {De Cia}, {Yan}, {Yang}, {Howell}, {Tal},
  {Kulkarni}, {Tendulkar}, {Tang}, {Xu}, {Sternberg}, {Cohen}, {Bloom},
  {Nugent}, {Kasliwal}, {Perley}, {Quimby}, {Miller}, {Theissen}, \&
  {Laher}}]{2014ApJ...793...38A}
{Arcavi}, I., {Gal-Yam}, A., {Sullivan}, M., {et~al.} 2014, \apj, 793, 38

\bibitem[{{Arnouts} {et~al.}(1999){Arnouts}, {Cristiani}, {Moscardini},
  {Matarrese}, {Lucchin}, {Fontana}, \& {Giallongo}}]{1999MNRAS.310..540A}
{Arnouts}, S., {Cristiani}, S., {Moscardini}, L., {et~al.} 1999, \mnras, 310,
  540

\bibitem[{{Auchettl} {et~al.}(2017{\natexlab{a}}){Auchettl}, {Guillochon}, \&
  {Ramirez-Ruiz}}]{2017ApJ...838..149A}
{Auchettl}, K., {Guillochon}, J., \& {Ramirez-Ruiz}, E. 2017{\natexlab{a}},
  \apj, 838, 149 (A17)

\bibitem[{{Auchettl} {et~al.}(2017{\natexlab{b}}){Auchettl}, {Ramirez-Ruiz}, \&
  {Guillochon}}]{2017arXiv170306141A}
{Auchettl}, K., {Ramirez-Ruiz}, E., \& {Guillochon}, J. 2017{\natexlab{b}},
  ArXiv e-prints, arXiv:1703.06141

\bibitem[{{Bade} {et~al.}(1996){Bade}, {Komossa}, \&
  {Dahlem}}]{1996A&A...309L..35B}
{Bade}, N., {Komossa}, S., \& {Dahlem}, M. 1996, \aap, 309, L35

\bibitem[{{Baldwin} {et~al.}(1981){Baldwin}, {Phillips}, \&
  {Terlevich}}]{1981PASP...93....5B}
{Baldwin}, J.~A., {Phillips}, M.~M., \& {Terlevich}, R. 1981, \pasp, 93, 5

\bibitem[{{Barro} {et~al.}(2017){Barro}, {Faber}, {Koo}, {Dekel}, {Fang},
  {Trump}, {P{\'e}rez-Gonz{\'a}lez}, {Pacifici}, {Primack}, {Somerville},
  {Yan}, {Guo}, {Liu}, {Ceverino}, {Kocevski}, \&
  {McGrath}}]{2017ApJ...840...47B}
{Barro}, G., {Faber}, S.~M., {Koo}, D.~C., {et~al.} 2017, \apj, 840, 47

\bibitem[{{Blanchard} {et~al.}(2017){Blanchard}, {Nicholl}, {Berger},
  {Guillochon}, {Margutti}, {Chornock}, {Alexander}, {Leja}, \&
  {Drout}}]{2017ApJ...843..106B}
{Blanchard}, P.~K., {Nicholl}, M., {Berger}, E., {et~al.} 2017, \apj, 843, 106

\bibitem[{{Blanton} \& {Roweis}(2007)}]{2007AJ....133..734B}
{Blanton}, M.~R., \& {Roweis}, S. 2007, \aj, 133, 734

\bibitem[{{Blanton} {et~al.}(2005){Blanton}, {Schlegel}, {Strauss},
  {Brinkmann}, {Finkbeiner}, {Fukugita}, {Gunn}, {Hogg}, {Ivezi{\'c}}, {Knapp},
  {Lupton}, {Munn}, {Schneider}, {Tegmark}, \& {Zehavi}}]{2005AJ....129.2562B}
{Blanton}, M.~R., {Schlegel}, D.~J., {Strauss}, M.~A., {et~al.} 2005, \aj, 129,
  2562

\bibitem[{{Bower} {et~al.}(2013){Bower}, {Metzger}, {Cenko}, {Silverman}, \&
  {Bloom}}]{2013ApJ...763...84B}
{Bower}, G.~C., {Metzger}, B.~D., {Cenko}, S.~B., {Silverman}, J.~M., \&
  {Bloom}, J.~S. 2013, \apj, 763, 84

\bibitem[{{Brinchmann} {et~al.}(2004){Brinchmann}, {Charlot}, {White},
  {Tremonti}, {Kauffmann}, {Heckman}, \& {Brinkmann}}]{2004MNRAS.351.1151B}
{Brinchmann}, J., {Charlot}, S., {White}, S.~D.~M., {et~al.} 2004, \mnras, 351,
  1151

\bibitem[{{Brown} {et~al.}(2018){Brown}, {Kochanek}, {Holoien}, {Stanek},
  {Auchettl}, {Shappee}, {Prieto}, {Morrell}, {Falco}, {Strader}, {Chomiuk},
  {Post}, {Villanueva}, {Mathur}, {Dong}, {Chen}, \&
  {Bose}}]{2018MNRAS.473.1130B}
{Brown}, J.~S., {Kochanek}, C.~S., {Holoien}, T.~W.-S., {et~al.} 2018, \mnras,
  473, 1130

\bibitem[{{Cappellari} {et~al.}(2006){Cappellari}, {Bacon}, {Bureau}, {Damen},
  {Davies}, {de Zeeuw}, {Emsellem}, {Falc{\'o}n-Barroso}, {Krajnovi{\'c}},
  {Kuntschner}, {McDermid}, {Peletier}, {Sarzi}, {van den Bosch}, \& {van de
  Ven}}]{2006MNRAS.366.1126C}
{Cappellari}, M., {Bacon}, R., {Bureau}, M., {et~al.} 2006, \mnras, 366, 1126

\bibitem[{{Cenko} {et~al.}(2012){Cenko}, {Bloom}, {Kulkarni}, {Strubbe},
  {Miller}, {Butler}, {Quimby}, {Gal-Yam}, {Ofek}, {Quataert}, {Bildsten},
  {Poznanski}, {Perley}, {Morgan}, {Filippenko}, {Frail}, {Arcavi}, {Ben-Ami},
  {Cucchiara}, {Fassnacht}, {Green}, {Hook}, {Howell}, {Lagattuta}, {Law},
  {Kasliwal}, {Nugent}, {Silverman}, {Sullivan}, {Tendulkar}, \&
  {Yaron}}]{2012MNRAS.420.2684C}
{Cenko}, S.~B., {Bloom}, J.~S., {Kulkarni}, S.~R., {et~al.} 2012, \mnras, 420,
  2684

\bibitem[{{Chen} {et~al.}(2009){Chen}, {Madau}, {Sesana}, \&
  {Liu}}]{2009ApJ...697L.149C}
{Chen}, X., {Madau}, P., {Sesana}, A., \& {Liu}, F.~K. 2009, \apjl, 697, L149

\bibitem[{{Chen} {et~al.}(2011){Chen}, {Sesana}, {Madau}, \&
  {Liu}}]{2011ApJ...729...13C}
{Chen}, X., {Sesana}, A., {Madau}, P., \& {Liu}, F.~K. 2011, \apj, 729, 13

\bibitem[{{Cheung} {et~al.}(2012){Cheung}, {Faber}, {Koo}, {Dutton}, {Simard},
  {McGrath}, {Huang}, {Bell}, {Dekel}, {Fang}, {Salim}, {Barro}, {Bundy},
  {Coil}, {Cooper}, {Conselice}, {Davis}, {Dom{\'{\i}}nguez}, {Kassin},
  {Kocevski}, {Koekemoer}, {Lin}, {Lotz}, {Newman}, {Phillips}, {Rosario},
  {Weiner}, \& {Willmer}}]{2012ApJ...760..131C}
{Cheung}, E., {Faber}, S.~M., {Koo}, D.~C., {et~al.} 2012, \apj, 760, 131

\bibitem[{{Chornock} {et~al.}(2014){Chornock}, {Berger}, {Gezari}, {Zauderer},
  {Rest}, {Chomiuk}, {Kamble}, {Soderberg}, {Czekala}, {Dittmann}, {Drout},
  {Foley}, {Fong}, {Huber}, {Kirshner}, {Lawrence}, {Lunnan}, {Marion},
  {Narayan}, {Riess}, {Roth}, {Sanders}, {Scolnic}, {Smartt}, {Smith},
  {Stubbs}, {Tonry}, {Burgett}, {Chambers}, {Flewelling}, {Hodapp}, {Kaiser},
  {Magnier}, {Martin}, {Neill}, {Price}, \& {Wainscoat}}]{2014ApJ...780...44C}
{Chornock}, R., {Berger}, E., {Gezari}, S., {et~al.} 2014, \apj, 780, 44

\bibitem[{{Dawson} {et~al.}(2013){Dawson}, {Schlegel}, {Ahn}, {Anderson},
  {Aubourg}, {Bailey}, {Barkhouser}, {Bautista}, {Beifiori}, {Berlind},
  {Bhardwaj}, {Bizyaev}, {Blake}, {Blanton}, {Blomqvist}, {Bolton}, {Borde},
  {Bovy}, {Brandt}, {Brewington}, {Brinkmann}, {Brown}, {Brownstein}, {Bundy},
  {Busca}, {Carithers}, {Carnero}, {Carr}, {Chen}, {Comparat}, {Connolly},
  {Cope}, {Croft}, {Cuesta}, {da Costa}, {Davenport}, {Delubac}, {de Putter},
  {Dhital}, {Ealet}, {Ebelke}, {Eisenstein}, {Escoffier}, {Fan}, {Filiz Ak},
  {Finley}, {Font-Ribera}, {G{\'e}nova-Santos}, {Gunn}, {Guo}, {Haggard},
  {Hall}, {Hamilton}, {Harris}, {Harris}, {Ho}, {Hogg}, {Holder}, {Honscheid},
  {Huehnerhoff}, {Jordan}, {Jordan}, {Kauffmann}, {Kazin}, {Kirkby}, {Klaene},
  {Kneib}, {Le Goff}, {Lee}, {Long}, {Loomis}, {Lundgren}, {Lupton}, {Maia},
  {Makler}, {Malanushenko}, {Malanushenko}, {Mandelbaum}, {Manera}, {Maraston},
  {Margala}, {Masters}, {McBride}, {McDonald}, {McGreer}, {McMahon}, {Mena},
  {Miralda-Escud{\'e}}, {Montero-Dorta}, {Montesano}, {Muna}, {Myers},
  {Naugle}, {Nichol}, {Noterdaeme}, {Nuza}, {Olmstead}, {Oravetz}, {Oravetz},
  {Owen}, {Padmanabhan}, {Palanque-Delabrouille}, {Pan}, {Parejko},
  {P{\^a}ris}, {Percival}, {P{\'e}rez-Fournon}, {P{\'e}rez-R{\`a}fols},
  {Petitjean}, {Pfaffenberger}, {Pforr}, {Pieri}, {Prada}, {Price-Whelan},
  {Raddick}, {Rebolo}, {Rich}, {Richards}, {Rockosi}, {Roe}, {Ross}, {Ross},
  {Rossi}, {Rubi{\~n}o-Martin}, {Samushia}, {S{\'a}nchez}, {Sayres}, {Schmidt},
  {Schneider}, {Sc{\'o}ccola}, {Seo}, {Shelden}, {Sheldon}, {Shen}, {Shu},
  {Slosar}, {Smee}, {Snedden}, {Stauffer}, {Steele}, {Strauss}, {Streblyanska},
  {Suzuki}, {Swanson}, {Tal}, {Tanaka}, {Thomas}, {Tinker}, {Tojeiro},
  {Tremonti}, {Vargas Maga{\~n}a}, {Verde}, {Viel}, {Wake}, {Watson}, {Weaver},
  {Weinberg}, {Weiner}, {West}, {White}, {Wood-Vasey}, {Yeche}, {Zehavi},
  {Zhao}, \& {Zheng}}]{Dawson2012BOSS}
{Dawson}, K.~S., {Schlegel}, D.~J., {Ahn}, C.~P., {et~al.} 2013, \aj, 145, 10

\bibitem[{{Doi} {et~al.}(2010){Doi}, {Tanaka}, {Fukugita}, {Gunn}, {Yasuda},
  {Ivezi{\'c}}, {Brinkmann}, {de Haars}, {Kleinman}, {Krzesinski}, \& {French
  Leger}}]{2010AJ....139.1628D}
{Doi}, M., {Tanaka}, M., {Fukugita}, M., {et~al.} 2010, \aj, 139, 1628

\bibitem[{{Dong} {et~al.}(2016){Dong}, {Shappee}, {Prieto}, {Jha}, {Stanek},
  {Holoien}, {Kochanek}, {Thompson}, {Morrell}, {Thompson}, {Basu}, {Beacom},
  {Bersier}, {Brimacombe}, {Brown}, {Bufano}, {Chen}, {Conseil}, {Danilet},
  {Falco}, {Grupe}, {Kiyota}, {Masi}, {Nicholls}, {Olivares E.}, {Pignata},
  {Pojmanski}, {Simonian}, {Szczygiel}, \& {Wo{\'z}niak}}]{2016Sci...351..257D}
{Dong}, S., {Shappee}, B.~J., {Prieto}, J.~L., {et~al.} 2016, Science, 351, 257

\bibitem[{{Donley} {et~al.}(2002){Donley}, {Brandt}, {Eracleous}, \&
  {Boller}}]{2002AJ....124.1308D}
{Donley}, J.~L., {Brandt}, W.~N., {Eracleous}, M., \& {Boller}, T. 2002, \aj,
  124, 1308

\bibitem[{{Dressler} \& {Gunn}(1983)}]{1983ApJ...270....7D}
{Dressler}, A., \& {Gunn}, J.~E. 1983, \apj, 270, 7

\bibitem[{{Dressler} {et~al.}(1999){Dressler}, {Smail}, {Poggianti}, {Butcher},
  {Couch}, {Ellis}, \& {Oemler}}]{1999ApJS..122...51D}
{Dressler}, A., {Smail}, I., {Poggianti}, B.~M., {et~al.} 1999, \apjs, 122, 51

\bibitem[{{Elbaz} {et~al.}(2011){Elbaz}, {Dickinson}, {Hwang},
  {D{\'{\i}}az-Santos}, {Magdis}, {Magnelli}, {Le Borgne}, {Galliano},
  {Pannella}, {Chanial}, {Armus}, {Charmandaris}, {Daddi}, {Aussel}, {Popesso},
  {Kartaltepe}, {Altieri}, {Valtchanov}, {Coia}, {Dannerbauer}, {Dasyra},
  {Leiton}, {Mazzarella}, {Alexander}, {Buat}, {Burgarella}, {Chary}, {Gilli},
  {Ivison}, {Juneau}, {Le Floc'h}, {Lutz}, {Morrison}, {Mullaney}, {Murphy},
  {Pope}, {Scott}, {Brodwin}, {Calzetti}, {Cesarsky}, {Charlot}, {Dole},
  {Eisenhardt}, {Ferguson}, {F{\"o}rster Schreiber}, {Frayer}, {Giavalisco},
  {Huynh}, {Koekemoer}, {Papovich}, {Reddy}, {Surace}, {Teplitz}, {Yun}, \&
  {Wilson}}]{2011A&A...533A.119E}
{Elbaz}, D., {Dickinson}, M., {Hwang}, H.~S., {et~al.} 2011, \aap, 533, A119

\bibitem[{{Esquej} {et~al.}(2007){Esquej}, {Saxton}, {Freyberg}, {Read},
  {Altieri}, {Sanchez-Portal}, \& {Hasinger}}]{2007AA...462L..49E}
{Esquej}, P., {Saxton}, R.~D., {Freyberg}, M.~J., {et~al.} 2007, \aap, 462, L49

\bibitem[{{Esquej} {et~al.}(2008){Esquej}, {Saxton}, {Komossa}, {Read},
  {Freyberg}, {Hasinger}, {Garc{\'{\i}}a-Hern{\'a}ndez}, {Lu}, {Rodriguez
  Zaur{\'{\i}}n}, {S{\'a}nchez-Portal}, \& {Zhou}}]{2008A&A...489..543E}
{Esquej}, P., {Saxton}, R.~D., {Komossa}, S., {et~al.} 2008, \aap, 489, 543

\bibitem[{{Evans} \& {Kochanek}(1989)}]{1989ApJ...346L..13E}
{Evans}, C.~R., \& {Kochanek}, C.~S. 1989, \apjl, 346, L13

\bibitem[{{Fang} {et~al.}(2013){Fang}, {Faber}, {Koo}, \&
  {Dekel}}]{2013ApJ...776...63F}
{Fang}, J.~J., {Faber}, S.~M., {Koo}, D.~C., \& {Dekel}, A. 2013, \apj, 776, 63

\bibitem[{{French} {et~al.}(2016){French}, {Arcavi}, \&
  {Zabludoff}}]{2016ApJ...818L..21F}
{French}, K.~D., {Arcavi}, I., \& {Zabludoff}, A. 2016, \apjl, 818, L21 (F16)

\bibitem[{{French} {et~al.}(2017){French}, {Arcavi}, \&
  {Zabludoff}}]{2017ApJ...835..176F}
---. 2017, \apj, 835, 176

\bibitem[{{Gebhardt} {et~al.}(2000){Gebhardt}, {Bender}, {Bower}, {Dressler},
  {Faber}, {Filippenko}, {Green}, {Grillmair}, {Ho}, {Kormendy}, {Lauer},
  {Magorrian}, {Pinkney}, {Richstone}, \& {Tremaine}}]{2000ApJ...539L..13G}
{Gebhardt}, K., {Bender}, R., {Bower}, G., {et~al.} 2000, \apjl, 539, L13

\bibitem[{{Gezari} {et~al.}(2003){Gezari}, {Halpern}, {Komossa}, {Grupe}, \&
  {Leighly}}]{2003ApJ...592...42G}
{Gezari}, S., {Halpern}, J.~P., {Komossa}, S., {Grupe}, D., \& {Leighly}, K.~M.
  2003, \apj, 592, 42

\bibitem[{{Gezari} {et~al.}(2008){Gezari}, {Basa}, {Martin}, {Bazin},
  {Forster}, {Milliard}, {Halpern}, {Friedman}, {Morrissey}, {Neff},
  {Schiminovich}, {Seibert}, {Small}, \& {Wyder}}]{2008ApJ...676..944G}
{Gezari}, S., {Basa}, S., {Martin}, D.~C., {et~al.} 2008, \apj, 676, 944

\bibitem[{{Gezari} {et~al.}(2012){Gezari}, {Chornock}, {Rest}, {Huber},
  {Forster}, {Berger}, {Challis}, {Neill}, {Martin}, {Heckman}, {Lawrence},
  {Norman}, {Narayan}, {Foley}, {Marion}, {Scolnic}, {Chomiuk}, {Soderberg},
  {Smith}, {Kirshner}, {Riess}, {Smartt}, {Stubbs}, {Tonry}, {Wood-Vasey},
  {Burgett}, {Chambers}, {Grav}, {Heasley}, {Kaiser}, {Kudritzki}, {Magnier},
  {Morgan}, \& {Price}}]{Gezari2012}
{Gezari}, S., {Chornock}, R., {Rest}, A., {et~al.} 2012, \nat, 485, 217

\bibitem[{{Graham}(2016)}]{2016ASSL..418..263G}
{Graham}, A.~W. 2016, Galactic Bulges, 418, 263

\bibitem[{{Graur} {et~al.}(2017){Graur}, {Bianco}, {Huang}, {Modjaz},
  {Shivvers}, {Filippenko}, {Li}, \& {Eldridge}}]{2017ApJ...837..120G}
{Graur}, O., {Bianco}, F.~B., {Huang}, S., {et~al.} 2017, \apj, 837, 120

\bibitem[{{Graur} {et~al.}(2015){Graur}, {Bianco}, \&
  {Modjaz}}]{2015MNRAS.450..905G}
{Graur}, O., {Bianco}, F.~B., \& {Modjaz}, M. 2015, \mnras, 450, 905

\bibitem[{{Greiner} {et~al.}(2000){Greiner}, {Schwarz}, {Zharikov}, \&
  {Orio}}]{2000A&A...362L..25G}
{Greiner}, J., {Schwarz}, R., {Zharikov}, S., \& {Orio}, M. 2000, \aap, 362,
  L25

\bibitem[{{Grupe} {et~al.}(1995){Grupe}, {Beuermann}, {Mannheim}, {Bade},
  {Thomas}, {de Martino}, \& {Schwope}}]{1995A&A...299L...5G}
{Grupe}, D., {Beuermann}, K., {Mannheim}, K., {et~al.} 1995, \aap, 299, L5

\bibitem[{{Grupe} {et~al.}(1999){Grupe}, {Thomas}, \&
  {Leighly}}]{1999AA...350L..31G}
{Grupe}, D., {Thomas}, H.-C., \& {Leighly}, K.~M. 1999, \aap, 350, L31

\bibitem[{{Hills}(1975)}]{1975Natur.254..295H}
{Hills}, J.~G. 1975, \nat, 254, 295

\bibitem[{{Ho} {et~al.}(1995){Ho}, {Filippenko}, \&
  {Sargent}}]{1995ApJS...98..477H}
{Ho}, L.~C., {Filippenko}, A.~V., \& {Sargent}, W.~L. 1995, \apjs, 98, 477

\bibitem[{{Holoien} {et~al.}(2016{\natexlab{a}}){Holoien}, {Kochanek},
  {Prieto}, {Grupe}, {Chen}, {Godoy-Rivera}, {Stanek}, {Shappee}, {Dong},
  {Brown}, {Basu}, {Beacom}, {Bersier}, {Brimacombe}, {Carlson}, {Falco},
  {Johnston}, {Madore}, {Pojmanski}, \& {Seibert}}]{2016MNRAS.463.3813H}
{Holoien}, T.~W.-S., {Kochanek}, C.~S., {Prieto}, J.~L., {et~al.}
  2016{\natexlab{a}}, \mnras, 463, 3813

\bibitem[{{Holoien} {et~al.}(2016{\natexlab{b}}){Holoien}, {Kochanek},
  {Prieto}, {Stanek}, {Dong}, {Shappee}, {Grupe}, {Brown}, {Basu}, {Beacom},
  {Bersier}, {Brimacombe}, {Danilet}, {Falco}, {Guo}, {Jose}, {Herczeg},
  {Long}, {Pojmanski}, {Simonian}, {Szczygie{\l}}, {Thompson}, {Thorstensen},
  {Wagner}, \& {Wo{\'z}niak}}]{2016MNRAS.455.2918H}
---. 2016{\natexlab{b}}, \mnras, 455, 2918

\bibitem[{{Holoien} {et~al.}(2017){Holoien}, {Stanek}, {Kochanek}, {Shappee},
  {Prieto}, {Brimacombe}, {Bersier}, {Bishop}, {Dong}, {Brown}, {Danilet},
  {Simonian}, {Basu}, {Beacom}, {Falco}, {Pojmanski}, {Skowron}, {Wo{\'z}niak},
  {{\'A}vila}, {Conseil}, {Contreras}, {Cruz}, {Fern{\'a}ndez}, {Koff}, {Guo},
  {Herczeg}, {Hissong}, {Hsiao}, {Jose}, {Kiyota}, {Long}, {Monard},
  {Nicholls}, {Nicolas}, \& {Wiethoff}}]{2017MNRAS.464.2672H}
{Holoien}, T.~W.-S., {Stanek}, K.~Z., {Kochanek}, C.~S., {et~al.} 2017, \mnras,
  464, 2672

\bibitem[{{Hung} {et~al.}(2017){Hung}, {Gezari}, {Blagorodnova}, {Roth},
  {Cenko}, {Kulkarni}, {Horesh}, {Arcavi}, {McCully}, {Yan}, {Lunnan},
  {Fremling}, {Cao}, {Nugent}, \& {Wozniak}}]{2017ApJ...842...29H}
{Hung}, T., {Gezari}, S., {Blagorodnova}, N., {et~al.} 2017, \apj, 842, 29

\bibitem[{{Ilbert} {et~al.}(2009){Ilbert}, {Capak}, {Salvato}, {Aussel},
  {McCracken}, {Sanders}, {Scoville}, {Kartaltepe}, {Arnouts}, {Le Floc'h},
  {Mobasher}, {Taniguchi}, {Lamareille}, {Leauthaud}, {Sasaki}, {Thompson},
  {Zamojski}, {Zamorani}, {Bardelli}, {Bolzonella}, {Bongiorno}, {Brusa},
  {Caputi}, {Carollo}, {Contini}, {Cook}, {Coppa}, {Cucciati}, {de la Torre},
  {de Ravel}, {Franzetti}, {Garilli}, {Hasinger}, {Iovino}, {Kampczyk},
  {Kneib}, {Knobel}, {Kovac}, {Le Borgne}, {Le Brun}, {F{\`e}vre}, {Lilly},
  {Looper}, {Maier}, {Mainieri}, {Mellier}, {Mignoli}, {Murayama}, {Pell{\`o}},
  {Peng}, {P{\'e}rez-Montero}, {Renzini}, {Ricciardelli}, {Schiminovich},
  {Scodeggio}, {Shioya}, {Silverman}, {Surace}, {Tanaka}, {Tasca}, {Tresse},
  {Vergani}, \& {Zucca}}]{2009ApJ...690.1236I}
{Ilbert}, O., {Capak}, P., {Salvato}, M., {et~al.} 2009, \apj, 690, 1236

\bibitem[{{Kaiser} {et~al.}(2002){Kaiser}, {Aussel}, {Burke}, {Boesgaard},
  {Chambers}, {Chun}, {Heasley}, {Hodapp}, {Hunt}, {Jedicke}, {Jewitt},
  {Kudritzki}, {Luppino}, {Maberry}, {Magnier}, {Monet}, {Onaka}, {Pickles},
  {Rhoads}, {Simon}, {Szalay}, {Szapudi}, {Tholen}, {Tonry}, {Waterson}, \&
  {Wick}}]{2002SPIE.4836..154K}
{Kaiser}, N., {Aussel}, H., {Burke}, B.~E., {et~al.} 2002, in \procspie, Vol.
  4836, Survey and Other Telescope Technologies and Discoveries, ed. J.~A.
  {Tyson} \& S.~{Wolff}, 154--164

\bibitem[{{Kauffmann} {et~al.}(2003){Kauffmann}, {Heckman}, {Tremonti},
  {Brinchmann}, {Charlot}, {White}, {Ridgway}, {Brinkmann}, {Fukugita}, {Hall},
  {Ivezi{\'c}}, {Richards}, \& {Schneider}}]{2003MNRAS.346.1055K}
{Kauffmann}, G., {Heckman}, T.~M., {Tremonti}, C., {et~al.} 2003, \mnras, 346,
  1055

\bibitem[{{Kesden}(2012)}]{2012PhRvD..85b4037K}
{Kesden}, M. 2012, \prd, 85, 024037

\bibitem[{{Komossa} \& {Bade}(1999)}]{1999A&A...343..775K}
{Komossa}, S., \& {Bade}, N. 1999, \aap, 343, 775

\bibitem[{{Komossa} \& {Greiner}(1999)}]{1999AA...349L..45K}
{Komossa}, S., \& {Greiner}, J. 1999, \aap, 349, L45

\bibitem[{{Komossa} {et~al.}(2009){Komossa}, {Zhou}, {Rau}, {Dopita},
  {Gal-Yam}, {Greiner}, {Zuther}, {Salvato}, {Xu}, {Lu}, {Saxton}, \&
  {Ajello}}]{2009ApJ...701..105K}
{Komossa}, S., {Zhou}, H., {Rau}, A., {et~al.} 2009, \apj, 701, 105

\bibitem[{{Kormendy} \& {Richstone}(1995)}]{1995ARA&A..33..581K}
{Kormendy}, J., \& {Richstone}, D. 1995, \araa, 33, 581

\bibitem[{{Lacy} {et~al.}(1982){Lacy}, {Townes}, \&
  {Hollenbach}}]{1982ApJ...262..120L}
{Lacy}, J.~H., {Townes}, C.~H., \& {Hollenbach}, D.~J. 1982, \apj, 262, 120

\bibitem[{{Law-Smith} {et~al.}(2017){Law-Smith}, {Ramirez-Ruiz}, {Ellison}, \&
  {Foley}}]{2017ApJ...850...22L}
{Law-Smith}, J., {Ramirez-Ruiz}, E., {Ellison}, S.~L., \& {Foley}, R.~J. 2017,
  \apj, 850, 22

\bibitem[{{Leloudas} {et~al.}(2016){Leloudas}, {Fraser}, {Stone}, {van Velzen},
  {Jonker}, {Arcavi}, {Fremling}, {Maund}, {Smartt}, {Kr{\`i}hler},
  {Miller-Jones}, {Vreeswijk}, {Gal-Yam}, {Mazzali}, {De Cia}, {Howell},
  {Inserra}, {Patat}, {de Ugarte Postigo}, {Yaron}, {Ashall}, {Bar},
  {Campbell}, {Chen}, {Childress}, {Elias-Rosa}, {Harmanen}, {Hosseinzadeh},
  {Johansson}, {Kangas}, {Kankare}, {Kim}, {Kuncarayakti}, {Lyman}, {Magee},
  {Maguire}, {Malesani}, {Mattila}, {McCully}, {Nicholl}, {Prentice},
  {Romero-Ca{\~n}izales}, {Schulze}, {Smith}, {Sollerman}, {Sullivan},
  {Tucker}, {Valenti}, {Wheeler}, \& {Young}}]{2016NatAs...1E...2L}
{Leloudas}, G., {Fraser}, M., {Stone}, N.~C., {et~al.} 2016, Nature Astronomy,
  1, 0002

\bibitem[{{Levan} {et~al.}(2011){Levan}, {Tanvir}, {Cenko}, {Perley},
  {Wiersema}, {Bloom}, {Fruchter}, {Postigo}, {O'Brien}, {Butler}, {van der
  Horst}, {Leloudas}, {Morgan}, {Misra}, {Bower}, {Farihi}, {Tunnicliffe},
  {Modjaz}, {Silverman}, {Hjorth}, {Th{\"o}ne}, {Cucchiara}, {Cer{\'o}n},
  {Castro-Tirado}, {Arnold}, {Bremer}, {Brodie}, {Carroll}, {Cooper}, {Curran},
  {Cutri}, {Ehle}, {Forbes}, {Fynbo}, {Gorosabel}, {Graham}, {Hoffman},
  {Guziy}, {Jakobsson}, {Kamble}, {Kerr}, {Kasliwal}, {Kouveliotou},
  {Kocevski}, {Law}, {Nugent}, {Ofek}, {Poznanski}, {Quimby}, {Rol},
  {Romanowsky}, {S{\'a}nchez-Ram{\'{\i}}rez}, {Schulze}, {Singh}, {van
  Spaandonk}, {Starling}, {Strom}, {Tello}, {Vaduvescu}, {Wheatley}, {Wijers},
  {Winters}, \& {Xu}}]{2011Sci...333..199L}
{Levan}, A.~J., {Tanvir}, N.~R., {Cenko}, S.~B., {et~al.} 2011, Science, 333,
  199

\bibitem[{{Lin} {et~al.}(2017){Lin}, {Guillochon}, {Komossa}, {Ramirez-Ruiz},
  {Irwin}, {Maksym}, {Grupe}, {Godet}, {Webb}, {Barret}, {Zauderer}, {Duc},
  {Carrasco}, \& {Gwyn}}]{2017NatAs...1E..33L}
{Lin}, D., {Guillochon}, J., {Komossa}, S., {et~al.} 2017, Nature Astronomy, 1,
  0033

\bibitem[{{Liu} {et~al.}(2017){Liu}, {Modjaz}, \&
  {Bianco}}]{2017ApJ...845...85L}
{Liu}, Y.-Q., {Modjaz}, M., \& {Bianco}, F.~B. 2017, \apj, 845, 85

\bibitem[{{Magorrian} \& {Tremaine}(1999)}]{1999MNRAS.309..447M}
{Magorrian}, J., \& {Tremaine}, S. 1999, \mnras, 309, 447

\bibitem[{{Magorrian} {et~al.}(1998){Magorrian}, {Tremaine}, {Richstone},
  {Bender}, {Bower}, {Dressler}, {Faber}, {Gebhardt}, {Green}, {Grillmair},
  {Kormendy}, \& {Lauer}}]{1998AJ....115.2285M}
{Magorrian}, J., {Tremaine}, S., {Richstone}, D., {et~al.} 1998, \aj, 115, 2285

\bibitem[{{Maksym} {et~al.}(2010){Maksym}, {Ulmer}, \&
  {Eracleous}}]{2010ApJ...722.1035M}
{Maksym}, W.~P., {Ulmer}, M.~P., \& {Eracleous}, M. 2010, \apj, 722, 1035

\bibitem[{{Margutti} {et~al.}(2017){Margutti}, {Metzger}, {Chornock},
  {Milisavljevic}, {Berger}, {Blanchard}, {Guidorzi}, {Migliori}, {Kamble},
  {Lunnan}, {Nicholl}, {Coppejans}, {Dall'Osso}, {Drout}, {Perna}, \&
  {Sbarufatti}}]{2017ApJ...836...25M}
{Margutti}, R., {Metzger}, B.~D., {Chornock}, R., {et~al.} 2017, \apj, 836, 25

\bibitem[{{McConnell} \& {Ma}(2013)}]{2013ApJ...764..184M}
{McConnell}, N.~J., \& {Ma}, C.-P. 2013, \apj, 764, 184

\bibitem[{{Peebles}(1972)}]{1972ApJ...178..371P}
{Peebles}, P.~J.~E. 1972, \apj, 178, 371

\bibitem[{{Phinney}(1989)}]{1989Natur.340..595P}
{Phinney}, E.~S. 1989, \nat, 340, 595

\bibitem[{{Prieto} {et~al.}(2016){Prieto}, {Kr{\"u}hler}, {Anderson},
  {Galbany}, {Kochanek}, {Aquino}, {Brown}, {Dong}, {F{\"o}rster}, {Holoien},
  {Kuncarayakti}, {Maureira}, {Rosales-Ortega}, {S{\'a}nchez}, {Shappee}, \&
  {Stanek}}]{2016ApJ...830L..32P}
{Prieto}, J.~L., {Kr{\"u}hler}, T., {Anderson}, J.~P., {et~al.} 2016, \apjl,
  830, L32

\bibitem[{{Rau} {et~al.}(2009){Rau}, {Kulkarni}, {Law}, {Bloom}, {Ciardi},
  {Djorgovski}, {Fox}, {Gal-Yam}, {Grillmair}, {Kasliwal}, {Nugent}, {Ofek},
  {Quimby}, {Reach}, {Shara}, {Bildsten}, {Cenko}, {Drake}, {Filippenko},
  {Helfand}, {Helou}, {Howell}, {Poznanski}, \&
  {Sullivan}}]{2009PASP..121.1334R}
{Rau}, A., {Kulkarni}, S.~R., {Law}, N.~M., {et~al.} 2009, \pasp, 121, 1334

\bibitem[{{Rees}(1988)}]{1988Natur.333..523R}
{Rees}, M.~J. 1988, \nat, 333, 523

\bibitem[{{Rich} {et~al.}(2015){Rich}, {Kewley}, \&
  {Dopita}}]{2015ApJS..221...28R}
{Rich}, J.~A., {Kewley}, L.~J., \& {Dopita}, M.~A. 2015, \apjs, 221, 28

\bibitem[{{Rodr{\'{\i}}guez Zaur{\'{\i}}n} {et~al.}(2009){Rodr{\'{\i}}guez
  Zaur{\'{\i}}n}, {Tadhunter}, \& {Gonz{\'a}lez Delgado}}]{2009MNRAS.400.1139R}
{Rodr{\'{\i}}guez Zaur{\'{\i}}n}, J., {Tadhunter}, C.~N., \& {Gonz{\'a}lez
  Delgado}, R.~M. 2009, \mnras, 400, 1139

\bibitem[{{Saxton} {et~al.}(2012){Saxton}, {Read}, {Esquej}, {Komossa},
  {Dougherty}, {Rodriguez-Pascual}, \& {Barrado}}]{2012AA...541A.106S}
{Saxton}, R.~D., {Read}, A.~M., {Esquej}, P., {et~al.} 2012, \aap, 541, A106

\bibitem[{{Saxton} {et~al.}(2017){Saxton}, {Read}, {Komossa}, {Lira},
  {Alexander}, \& {Wieringa}}]{2017AA...598A..29S}
{Saxton}, R.~D., {Read}, A.~M., {Komossa}, S., {et~al.} 2017, \aap, 598, A29

\bibitem[{{Schlafly} \& {Finkbeiner}(2011)}]{2011ApJ...737..103S}
{Schlafly}, E.~F., \& {Finkbeiner}, D.~P. 2011, \apj, 737, 103

\bibitem[{{Schmidt}(1968)}]{1968ApJ...151..393S}
{Schmidt}, M. 1968, \apj, 151, 393

\bibitem[{{Smee} {et~al.}(2013){Smee}, {Gunn}, {Uomoto}, {Roe}, {Schlegel},
  {Rockosi}, {Carr}, {Leger}, {Dawson}, {Olmstead}, {Brinkmann}, {Owen},
  {Barkhouser}, {Honscheid}, {Harding}, {Long}, {Lupton}, {Loomis}, {Anderson},
  {Annis}, {Bernardi}, {Bhardwaj}, {Bizyaev}, {Bolton}, {Brewington}, {Briggs},
  {Burles}, {Burns}, {Castander}, {Connolly}, {Davenport}, {Ebelke}, {Epps},
  {Feldman}, {Friedman}, {Frieman}, {Heckman}, {Hull}, {Knapp}, {Lawrence},
  {Loveday}, {Mannery}, {Malanushenko}, {Malanushenko}, {Merrelli}, {Muna},
  {Newman}, {Nichol}, {Oravetz}, {Pan}, {Pope}, {Ricketts}, {Shelden},
  {Sandford}, {Siegmund}, {Simmons}, {Smith}, {Snedden}, {Schneider},
  {SubbaRao}, {Tremonti}, {Waddell}, \& {York}}]{2013AJ....146...32S}
{Smee}, S.~A., {Gunn}, J.~E., {Uomoto}, A., {et~al.} 2013, \aj, 146, 32

\bibitem[{{Stone} \& {Metzger}(2016)}]{2016MNRAS.455..859S}
{Stone}, N.~C., \& {Metzger}, B.~D. 2016, \mnras, 455, 859

\bibitem[{{Stone} \& {van Velzen}(2016)}]{2016ApJ...825L..14S}
{Stone}, N.~C., \& {van Velzen}, S. 2016, \apjl, 825, L14

\bibitem[{{Stoughton} {et~al.}(2002){Stoughton}, {Lupton}, {Bernardi},
  {Blanton}, {Burles}, {Castander}, {Connolly}, {Eisenstein}, {Frieman},
  {Hennessy}, {Hindsley}, {Ivezi{\'c}}, {Kent}, {Kunszt}, {Lee}, {Meiksin},
  {Munn}, {Newberg}, {Nichol}, {Nicinski}, {Pier}, {Richards}, {Richmond},
  {Schlegel}, {Smith}, {Strauss}, {SubbaRao}, {Szalay}, {Thakar}, {Tucker},
  {Vanden Berk}, {Yanny}, {Adelman}, {Anderson}, {Anderson}, {Annis},
  {Bahcall}, {Bakken}, {Bartelmann}, {Bastian}, {Bauer}, {Berman},
  {B{\"o}hringer}, {Boroski}, {Bracker}, {Briegel}, {Briggs}, {Brinkmann},
  {Brunner}, {Carey}, {Carr}, {Chen}, {Christian}, {Colestock}, {Crocker},
  {Csabai}, {Czarapata}, {Dalcanton}, {Davidsen}, {Davis}, {Dehnen},
  {Dodelson}, {Doi}, {Dombeck}, {Donahue}, {Ellman}, {Elms}, {Evans}, {Eyer},
  {Fan}, {Federwitz}, {Friedman}, {Fukugita}, {Gal}, {Gillespie}, {Glazebrook},
  {Gray}, {Grebel}, {Greenawalt}, {Greene}, {Gunn}, {de Haas}, {Haiman},
  {Haldeman}, {Hall}, {Hamabe}, {Hansen}, {Harris}, {Harris}, {Harvanek},
  {Hawley}, {Hayes}, {Heckman}, {Helmi}, {Henden}, {Hogan}, {Hogg}, {Holmgren},
  {Holtzman}, {Huang}, {Hull}, {Ichikawa}, {Ichikawa}, {Johnston}, {Kauffmann},
  {Kim}, {Kimball}, {Kinney}, {Klaene}, {Kleinman}, {Klypin}, {Knapp},
  {Korienek}, {Krolik}, {Kron}, {Krzesi{\'n}ski}, {Lamb}, {Leger},
  {Limmongkol}, {Lindenmeyer}, {Long}, {Loomis}, {Loveday}, {MacKinnon},
  {Mannery}, {Mantsch}, {Margon}, {McGehee}, {McKay}, {McLean}, {Menou},
  {Merelli}, {Mo}, {Monet}, {Nakamura}, {Narayanan}, {Nash}, {Neilsen},
  {Newman}, {Nitta}, {Odenkirchen}, {Okada}, {Okamura}, {Ostriker}, {Owen},
  {Pauls}, {Peoples}, {Peterson}, {Petravick}, {Pope}, {Pordes}, {Postman},
  {Prosapio}, {Quinn}, {Rechenmacher}, {Rivetta}, {Rix}, {Rockosi}, {Rosner},
  {Ruthmansdorfer}, {Sandford}, {Schneider}, {Scranton}, {Sekiguchi}, {Sergey},
  {Sheth}, {Shimasaku}, {Smee}, {Snedden}, {Stebbins}, {Stubbs}, {Szapudi},
  {Szkody}, {Szokoly}, {Tabachnik}, {Tsvetanov}, {Uomoto}, {Vogeley}, {Voges},
  {Waddell}, {Walterbos}, {Wang}, {Watanabe}, {Weinberg}, {White}, {White},
  {Wilhite}, {Wolfe}, {Yasuda}, {York}, {Zehavi}, \&
  {Zheng}}]{2002AJ....123..485S}
{Stoughton}, C., {Lupton}, R.~H., {Bernardi}, M., {et~al.} 2002, \aj, 123, 485

\bibitem[{{Swinbank} {et~al.}(2012){Swinbank}, {Balogh}, {Bower}, {Zabludoff},
  {Lucey}, {McGee}, {Miller}, \& {Nichol}}]{2012MNRAS.420..672S}
{Swinbank}, A.~M., {Balogh}, M.~L., {Bower}, R.~G., {et~al.} 2012, \mnras, 420,
  672

\bibitem[{{Tacchella} {et~al.}(2016){Tacchella}, {Dekel}, {Carollo},
  {Ceverino}, {DeGraf}, {Lapiner}, {Mandelker}, \&
  {Primack}}]{2016MNRAS.458..242T}
{Tacchella}, S., {Dekel}, A., {Carollo}, C.~M., {et~al.} 2016, \mnras, 458, 242

\bibitem[{{Tadhunter} {et~al.}(2017){Tadhunter}, {Spence}, {Rose}, {Mullaney},
  \& {Crowther}}]{2017NatAs...1E..61T}
{Tadhunter}, C., {Spence}, R., {Rose}, M., {Mullaney}, J., \& {Crowther}, P.
  2017, Nature Astronomy, 1, 0061

\bibitem[{{Thomas} {et~al.}(2013){Thomas}, {Steele}, {Maraston}, {Johansson},
  {Beifiori}, {Pforr}, {Str{\"o}mb{\"a}ck}, {Tremonti}, {Wake}, {Bizyaev},
  {Bolton}, {Brewington}, {Brownstein}, {Comparat}, {Kneib}, {Malanushenko},
  {Malanushenko}, {Oravetz}, {Pan}, {Parejko}, {Schneider}, {Shelden},
  {Simmons}, {Snedden}, {Tanaka}, {Weaver}, \& {Yan}}]{2013MNRAS.431.1383T}
{Thomas}, D., {Steele}, O., {Maraston}, C., {et~al.} 2013, \mnras, 431, 1383

\bibitem[{{Tremaine} {et~al.}(2002){Tremaine}, {Gebhardt}, {Bender}, {Bower},
  {Dressler}, {Faber}, {Filippenko}, {Green}, {Grillmair}, {Ho}, {Kormendy},
  {Lauer}, {Magorrian}, {Pinkney}, \& {Richstone}}]{2002ApJ...574..740T}
{Tremaine}, S., {Gebhardt}, K., {Bender}, R., {et~al.} 2002, \apj, 574, 740

\bibitem[{{Ulmer}(1999)}]{1999ApJ...514..180U}
{Ulmer}, A. 1999, \apj, 514, 180

\bibitem[{{van Velzen}(2017)}]{2017arXiv170703458V}
{van Velzen}, S. 2017, ArXiv e-prints, arXiv:1707.03458

\bibitem[{{van Velzen} \& {Farrar}(2014)}]{2014ApJ...792...53V}
{van Velzen}, S., \& {Farrar}, G.~R. 2014, \apj, 792, 53

\bibitem[{{van Velzen} {et~al.}(2011){van Velzen}, {Farrar}, {Gezari},
  {Morrell}, {Zaritsky}, {{\"O}stman}, {Smith}, {Gelfand}, \&
  {Drake}}]{2011ApJ...741...73V}
{van Velzen}, S., {Farrar}, G.~R., {Gezari}, S., {et~al.} 2011, \apj, 741, 73

\bibitem[{{van Velzen} {et~al.}(2016){van Velzen}, {Anderson}, {Stone},
  {Fraser}, {Wevers}, {Metzger}, {Jonker}, {van der Horst}, {Staley}, {Mendez},
  {Miller-Jones}, {Hodgkin}, {Campbell}, \& {Fender}}]{2016Sci...351...62V}
{van Velzen}, S., {Anderson}, G.~E., {Stone}, N.~C., {et~al.} 2016, Science,
  351, 62

\bibitem[{{Vaughan} {et~al.}(2004){Vaughan}, {Edelson}, \&
  {Warwick}}]{2004MNRAS.349L...1V}
{Vaughan}, S., {Edelson}, R., \& {Warwick}, R.~S. 2004, \mnras, 349, L1

\bibitem[{{Wang} \& {Merritt}(2004)}]{2004ApJ...600..149W}
{Wang}, J., \& {Merritt}, D. 2004, \apj, 600, 149

\bibitem[{{Wang} {et~al.}(2012){Wang}, {Zhou}, {Komossa}, {Wang}, {Yuan}, \&
  {Yang}}]{2012ApJ...749..115W}
{Wang}, T.-G., {Zhou}, H.-Y., {Komossa}, S., {et~al.} 2012, \apj, 749, 115

\bibitem[{{Wang} {et~al.}(2011){Wang}, {Zhou}, {Wang}, {Lu}, \&
  {Xu}}]{2011ApJ...740...85W}
{Wang}, T.-G., {Zhou}, H.-Y., {Wang}, L.-F., {Lu}, H.-L., \& {Xu}, D. 2011,
  \apj, 740, 85

\bibitem[{{Wevers} {et~al.}(2017){Wevers}, {van Velzen}, {Jonker}, {Stone},
  {Hung}, {Onori}, {Gezari}, \& {Blagorodnova}}]{2017MNRAS.471.1694W}
{Wevers}, T., {van Velzen}, S., {Jonker}, P.~G., {et~al.} 2017, \mnras, 471,
  1694

\bibitem[{{Woo} {et~al.}(2015){Woo}, {Dekel}, {Faber}, \&
  {Koo}}]{2015MNRAS.448..237W}
{Woo}, J., {Dekel}, A., {Faber}, S.~M., \& {Koo}, D.~C. 2015, \mnras, 448, 237

\bibitem[{{Worthey} \& {Ottaviani}(1997)}]{1997ApJS..111..377W}
{Worthey}, G., \& {Ottaviani}, D.~L. 1997, \apjs, 111, 377

\bibitem[{{Xiao} {et~al.}(2011){Xiao}, {Barth}, {Greene}, {Ho}, {Bentz},
  {Ludwig}, \& {Jiang}}]{2011ApJ...739...28X}
{Xiao}, T., {Barth}, A.~J., {Greene}, J.~E., {et~al.} 2011, \apj, 739, 28

\bibitem[{{Yan} \& {Blanton}(2012)}]{2012ApJ...747...61Y}
{Yan}, R., \& {Blanton}, M.~R. 2012, \apj, 747, 61

\bibitem[{{Yang} {et~al.}(2013){Yang}, {Wang}, {Ferland}, {Yuan}, {Zhou}, \&
  {Jiang}}]{2013ApJ...774...46Y}
{Yang}, C.-W., {Wang}, T.-G., {Ferland}, G., {et~al.} 2013, \apj, 774, 46

\bibitem[{{Yang} {et~al.}(2006){Yang}, {Tremonti}, {Zabludoff}, \&
  {Zaritsky}}]{2006ApJ...646L..33Y}
{Yang}, Y., {Tremonti}, C.~A., {Zabludoff}, A.~I., \& {Zaritsky}, D. 2006,
  \apjl, 646, L33

\bibitem[{{Yang} {et~al.}(2004){Yang}, {Zabludoff}, {Zaritsky}, {Lauer}, \&
  {Mihos}}]{2004ApJ...607..258Y}
{Yang}, Y., {Zabludoff}, A.~I., {Zaritsky}, D., {Lauer}, T.~R., \& {Mihos},
  J.~C. 2004, \apj, 607, 258

\bibitem[{{Yang} {et~al.}(2008){Yang}, {Zabludoff}, {Zaritsky}, \&
  {Mihos}}]{2008ApJ...688..945Y}
{Yang}, Y., {Zabludoff}, A.~I., {Zaritsky}, D., \& {Mihos}, J.~C. 2008, \apj,
  688, 945

\bibitem[{{York} {et~al.}(2000){York}, {Adelman}, {Anderson}, {Anderson},
  {Annis}, {Bahcall}, {Bakken}, {Barkhouser}, {Bastian}, {Berman}, {Boroski},
  {Bracker}, {Briegel}, {Briggs}, {Brinkmann}, {Brunner}, {Burles}, {Carey},
  {Carr}, {Castander}, {Chen}, {Colestock}, {Connolly}, {Crocker}, {Csabai},
  {Czarapata}, {Davis}, {Doi}, {Dombeck}, {Eisenstein}, {Ellman}, {Elms},
  {Evans}, {Fan}, {Federwitz}, {Fiscelli}, {Friedman}, {Frieman}, {Fukugita},
  {Gillespie}, {Gunn}, {Gurbani}, {de Haas}, {Haldeman}, {Harris}, {Hayes},
  {Heckman}, {Hennessy}, {Hindsley}, {Holm}, {Holmgren}, {Huang}, {Hull},
  {Husby}, {Ichikawa}, {Ichikawa}, {Ivezi{\'c}}, {Kent}, {Kim}, {Kinney},
  {Klaene}, {Kleinman}, {Kleinman}, {Knapp}, {Korienek}, {Kron}, {Kunszt},
  {Lamb}, {Lee}, {Leger}, {Limmongkol}, {Lindenmeyer}, {Long}, {Loomis},
  {Loveday}, {Lucinio}, {Lupton}, {MacKinnon}, {Mannery}, {Mantsch}, {Margon},
  {McGehee}, {McKay}, {Meiksin}, {Merelli}, {Monet}, {Munn}, {Narayanan},
  {Nash}, {Neilsen}, {Neswold}, {Newberg}, {Nichol}, {Nicinski}, {Nonino},
  {Okada}, {Okamura}, {Ostriker}, {Owen}, {Pauls}, {Peoples}, {Peterson},
  {Petravick}, {Pier}, {Pope}, {Pordes}, {Prosapio}, {Rechenmacher}, {Quinn},
  {Richards}, {Richmond}, {Rivetta}, {Rockosi}, {Ruthmansdorfer}, {Sandford},
  {Schlegel}, {Schneider}, {Sekiguchi}, {Sergey}, {Shimasaku}, {Siegmund},
  {Smee}, {Smith}, {Snedden}, {Stone}, {Stoughton}, {Strauss}, {Stubbs},
  {SubbaRao}, {Szalay}, {Szapudi}, {Szokoly}, {Thakar}, {Tremonti}, {Tucker},
  {Uomoto}, {Vanden Berk}, {Vogeley}, {Waddell}, {Wang}, {Watanabe},
  {Weinberg}, {Yanny}, \& {Yasuda}}]{2000AJ....120.1579Y}
{York}, D.~G., {Adelman}, J., {Anderson}, Jr., J.~E., {et~al.} 2000, \aj, 120,
  1579

\bibitem[{{Zabludoff} {et~al.}(1996){Zabludoff}, {Zaritsky}, {Lin}, {Tucker},
  {Hashimoto}, {Shectman}, {Oemler}, \& {Kirshner}}]{1996ApJ...466..104Z}
{Zabludoff}, A.~I., {Zaritsky}, D., {Lin}, H., {et~al.} 1996, \apj, 466, 104

\bibitem[{{Zahid} {et~al.}(2016{\natexlab{a}}){Zahid}, {Baeza Hochmuth},
  {Geller}, {Damjanov}, {Chilingarian}, {Sohn}, {Salmi}, \&
  {Hwang}}]{2016ApJ...831..146Z}
{Zahid}, H.~J., {Baeza Hochmuth}, N., {Geller}, M.~J., {et~al.}
  2016{\natexlab{a}}, \apj, 831, 146

\bibitem[{{Zahid} {et~al.}(2016{\natexlab{b}}){Zahid}, {Damjanov}, {Geller},
  {Hwang}, \& {Fabricant}}]{2016ApJ...821..101Z}
{Zahid}, H.~J., {Damjanov}, I., {Geller}, M.~J., {Hwang}, H.~S., \&
  {Fabricant}, D.~G. 2016{\natexlab{b}}, \apj, 821, 101

\bibitem[{{Zahid} \& {Geller}(2017)}]{2017ApJ...841...32Z}
{Zahid}, H.~J., \& {Geller}, M.~J. 2017, \apj, 841, 32

\bibitem[{{Zahid} {et~al.}(2016{\natexlab{c}}){Zahid}, {Geller}, {Fabricant},
  \& {Hwang}}]{2016ApJ...832..203Z}
{Zahid}, H.~J., {Geller}, M.~J., {Fabricant}, D.~G., \& {Hwang}, H.~S.
  2016{\natexlab{c}}, \apj, 832, 203

\bibitem[{{Zahid} {et~al.}(2013){Zahid}, {Yates}, {Kewley}, \&
  {Kudritzki}}]{2013ApJ...763...92Z}
{Zahid}, H.~J., {Yates}, R.~M., {Kewley}, L.~J., \& {Kudritzki}, R.~P. 2013,
  \apj, 763, 92

\bibitem[{{Zolotov} {et~al.}(2015){Zolotov}, {Dekel}, {Mandelker}, {Tweed},
  {Inoue}, {DeGraf}, {Ceverino}, {Primack}, {Barro}, \&
  {Faber}}]{2015MNRAS.450.2327Z}
{Zolotov}, A., {Dekel}, A., {Mandelker}, N., {et~al.} 2015, \mnras, 450, 2327

\end{thebibliography}
\end{document}